\documentclass[aps,pra,twocolumn,superscriptaddress,floatfix]{revtex4-1}
\usepackage{amsmath}
\usepackage{graphicx}
\usepackage{amsfonts}
\usepackage{color}
\usepackage{upgreek}
\usepackage{mathtools}
\usepackage{hyperref}
\usepackage{comment}

\hypersetup{
    colorlinks = true,
    linkcolor =blue,
	citecolor=blue, 
	urlcolor=blue 
}
\usepackage{array}

\usepackage{braket}
\usepackage{bm}
\usepackage[normalem]{ulem}

\newcommand{\squash}[0]{\hspace{-0.5em}}
\newcommand{\superket}[1]{\text{vec($#1$)}}

\newtheorem{lemma}{Lemma}
\newtheorem{theorem}[lemma]{Theorem}

\begin{document}
\title{Searching for Lindbladians obeying local conservation laws and showing thermalization}
	\author{Devashish Tupkary}
	\email{djtupkary@uwaterloo.ca}
	\affiliation{Institute for Quantum Computing and Department of Physics and Astronomy, University of Waterloo, Waterloo, Ontario, Canada, N2L 3G1}	
	
	\author{Abhishek Dhar}
	\email{abhishek.dhar@icts.res.in}
	\affiliation{International Centre for Theoretical Sciences, Tata Institute of Fundamental Research, Bangalore 560089, India}

	\author{Manas Kulkarni}
	\email{manas.kulkarni@icts.res.in}
	\affiliation{International Centre for Theoretical Sciences, Tata Institute of Fundamental Research, Bangalore 560089, India}
		
	\author{Archak Purkayastha}
	\email{archak.p@phy.iith.ac.in}
        \affiliation{Department of Physics, Indian Institute of Technology, Hyderabad 502285, India}
	\affiliation{Centre for complex quantum systems, Aarhus University, Nordre Ringgade 1, 8000 Aarhus C, Denmark}
 \affiliation{School of Physics, Trinity College Dublin, College Green, Dublin 2, Ireland}

\begin{abstract}
       We investigate the possibility of a Markovian quantum master equation (QME) that consistently describes a finite-dimensional system, a part of which is weakly coupled to a thermal bath. In order to preserve complete positivity and trace, such a QME must be of Lindblad form. For physical consistency, it should additionally preserve local conservation laws and be able to show thermalization. We search of Lindblad equations satisfying these additional criteria. First, we show that the microscopically derived Bloch-Redfield equation (RE) violates complete positivity unless in extremely special cases. We then prove that imposing complete positivity and demanding preservation of local conservation laws enforces the Lindblad operators and the lamb-shift Hamiltonian to be `local', i.e, to be supported only on the part of the system directly coupled to the bath. We then cast the problem of finding `local' Lindblad QME which can show thermalization into a semidefinite program (SDP). We call this the thermalization optimization problem (TOP). For given system parameters and temperature, the solution of the TOP conclusively shows whether the desired type of QME is possible up to a given precision. Whenever possible, it also outputs a form for such a QME. For a XXZ chain of few qubits, fixing a reasonably high precision, we find that such a QME is impossible over a considerably wide parameter regime when only the first qubit is coupled to the bath.  Remarkably, we find that when the first two qubits are attached to the bath, such a QME becomes possible over much of the same paramater regime, including a wide range of temperatures.  
\end{abstract}

\maketitle
	

\section{Introduction}
A small finite-dimensional quantum system, a part of which is weakly coupled to a macroscopic thermal bath, is expected to  thermalize to the temperature of the bath. Describing this dynamics is relevant across various fields in quantum science and technology, including quantum information and thermodynamics \cite{quantum_thermodynamics}, quantum optics \cite{LeHur2016}, quantum chemistry \cite{quantum_chemistry}, engineering \cite{quantum_engineering} and biology \cite{quantum_biology}. In absence of  coupling to the macroscopic thermal bath, the dynamics of the density matrix of the system is governed by the Heisenberg equation of motion. This unitary evolution is Markovian. When coupled to the macroscopic thermal bath, the dynamics becomes non-unitary, described by a quantum master equation (QME) \cite{breuer_book,Rivas_2012, carmichael_book}. We investigate under what conditions it is possible to have a physically consistent Markovian QME describing such dynamics. In order to do so, we are led to introduce the ``thermalization optimization problem'' (TOP). This is a semidefinite program (SDP), the output of which conclusively shows whether, for given system parameters and temperature, such a QME is possible, up to given precision.  Whenever possible, the output also yields one possible form for such a QME. Whenever impossible, it means that, for such parameters, the dynamics cannot be described by any Markovian QME even at weak system-bath coupling, and therefore must have some non-Markovian character. The SDP can be solved using standard packages in high-level computing. We note that, while SDP is widely used in many branches of quantum information and communication \cite{watrous_2018,Siddhu_2022}, and also in quantum chemistry \cite{Mazziotti_2004,Mazziotti_2011}, it has been combined with open quantum system techniques in only few previous works \cite{Wolf_2008,Cubitt_2012,Kiilerich_2018}, in very different contexts.

\begin{figure}
	\includegraphics[width = \linewidth]{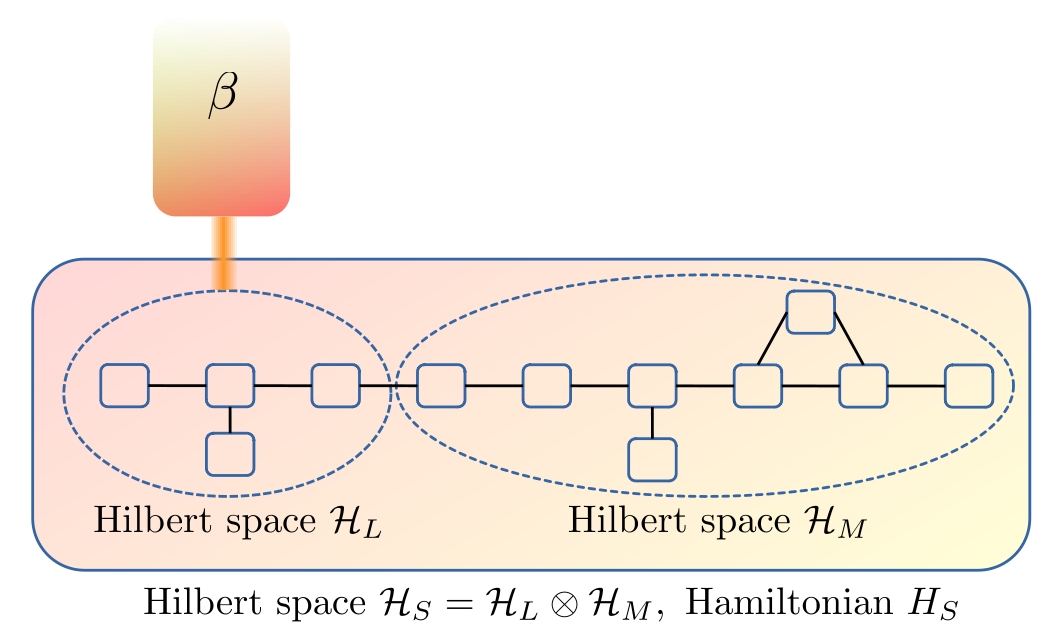}
	\caption{Schematic of the setup we consider: an arbitrary finite dimensional system described by Hamiltonian $H_S$, a part of which is weakly coupled to a thermal bath at inverse temperature $\beta$. The Hilbert space of the system, $\mathcal{H}_S$ is divided into a part $\mathcal{H}_L$ which directly couples to the bath, and the remaining part  $\mathcal{H}_M$.
		\label{fig1}
  }
\end{figure}

It was shown by Gorini, Kossakowski, Sudarshan, and Lindblad (GKSL) \cite{GKS1976, Gorini_1978, lindblad1976} that any QME that preserves complete positivity and trace of the density matrix, and describes Markovian dynamics has to be of the form 
\begin{equation}
\begin{aligned}
	\label{Lindblad_form}
	& \frac{\partial \rho}{\partial t}=i[\rho,H_S + H_{LS}]+\mathcal{D}(\rho), \\
	&\mathcal{D}(\rho) = \sum_{\lambda=1}^{d^2-1}\gamma_{\lambda} \Big( L_{\lambda} \rho L_{\lambda}^\dagger - \frac{1}{2} \{ L_{\lambda}^\dagger L_{\lambda}, \rho   \} \Big),~~\gamma_\lambda \geq 0,
\end{aligned}
\end{equation}
which is commonly called the Lindblad equation. In Eq.~\eqref{Lindblad_form},  $\rho$ is the density matrix of the system, $d$ is the Hilbert space dimension of the system, $H_S$ is the system Hamiltonian, $H_{LS}$ is the Lamb shift Hamiltonian, $L_\lambda$ are the Lindblad operators, $\gamma_\lambda$ are the rates, and $\mathcal{D}(\rho)$ is called the ``dissipator" term. The preservation of complete positivity condition is enforced by demanding $\gamma_\lambda \geq 0$. Lindblad equations have been extensively used in studying both theoretical and experimental setups \cite{breuer_book, carmichael_book,Rivas_2012,Weimer_2021,Plenio_1998,Dalibard1992, Mlmer1993}. 

Given this enormous scope of application, it is of paramount importance to assess the conditions under which such a Markovian description emerges from a more microscopic theory. The standard way to microscopically obtain a Markovian QME is to consider the global Hamiltonian of the system weakly coupled to baths, and to trace out the baths perturbatively to the leading order. Starting with this microscopic viewpoint, it becomes clear that only having an equation in the Lindblad form is not sufficient, there are some additional fundamental requirements for physical consistency \cite{Tupkary_2022}. In particular, one must preserve local conservation laws, and if the steady state is unique, the system is not driven and all baths have same temperature, the system must thermalize to the temperature of the baths. It would be useful to have a QME which, by construction, is of Lindblad form and satisfy these additional requirements. In this paper, we systematically go about searching for such a QME for a setup where a part of the system is coupled to a single bath (see Fig.~\ref{fig1}). This is done in three steps, each step having important consequences: 
\begin{enumerate}
    \item The microscopically derived quantum master equation to the leading order in system-bath coupling is the so-called Bloch-Redfield equation (RE) \cite{breuer_book, redfield_1957,block_1957}. The RE has been shown to preserve local conservation laws and be able to show thermalization \cite{Tupkary_2022}. Here, we provide an explicit, model independent proof that the RE necessarily violates complete positivity unless the Redfield dissipator happens to act ``locally", meaning it is identity on the part of the system that is not directly coupled to the bath. The violation of complete positivity by RE has been previously mostly demonstrated in specific examples \cite{Hartmann_2020_1,Eastham_2016,anderloni_2007,Gaspard_Nagaoka_1999,Kohen_1997,
Gnutzmann_1996,Suarez_1992}. In Ref. \cite{Whitney_2008}, it was shown that the RE would generically violate complete positivity, but the statement on `locality' of the Redfield dissipator was not made. Also, we consider the most general type of system-bath coupling, which was not the case in Ref. \cite{Whitney_2008}.

\item We then prove that enforcing complete positivity condition $\gamma_\lambda \geq 0$ and preservation of local conservation laws necessarily requires the Lindblad operators and the Lamb shift Hamiltonian to be ``local''. That is, they must act only on the part of the system coupled to the bath, and be identity on the part of the system that is not connected to the bath. This directly rules out the possibility of any so-called `global' Lindblad equation, such as the eigenbasis Lindblad equation \cite{breuer_book,carmichael_book}, the Universal Lindblad Equation \cite{ule} to be consistent with local conservation laws.

\item Given the restriction of the dissipator and the Lamb shift Hamiltonian to be ``local', we devise a numerical technique using SDP to check conclusively in a case-by-case basis whether such a QME can show thermalization in a particular situation. We call this the TOP. We use this method to study the case of a XXZ chain of few qubits with a part of it coupled to a bath. If the bath is coupled only to the first qubit, our method conclusively shows that over a large regime of system parameters and temperature, no such QME exists.  However, remarkably if the bath is coupled to two qubits of the chain, up to a chosen precision, our method shows that a Markovian QME respecting all conditions becomes possible over a considerable regime of parameters, including a wide range of temperatures. Note that the RE for the XXZ chain does show thermalization and preserve local conservation laws \cite{Tupkary_2022}, even when only one qubit is attached to a bath. But violates completely positivity.

\end{enumerate}

This work is organized as follows. In Sec.~\ref{sec:model_redfield} we explain the setup studied in this work, derive the Redfield equation for our setup, and show that it will necessarily violate complete positivity, unless the Redfield dissipator happens to act ``locally''. In Sec.~\ref{sec:lindblad} we consider quantum master equations preserving complete positivity and obeying local conservation laws, and show that such equations must have a dissipator and Lamb shift operator that acts only on the part of the system coupled to the bath. In Sec.~\ref{sec:qme_thermal} we discuss the possibility of QMEs respecting complete positivity, local conservation laws and being able to show thermalization. We introduce the TOP, and  use it in the specific case of the few site XXZ chain with one or two sites attached to bath. In Sec.~\ref{sec:summary} we summarize our results, and discuss future directions. Certain details are delegated to the appendices.

\section{The Model and Redfield Description} \label{sec:model_redfield}

Our setup is described schematically in Fig.~\ref{fig1}. The full Hamiltonian be given by 
\begin{equation}
\label{full_H}
H=H_S+\epsilon H_{SB}+H_B,
\end{equation}
where $H_S$ and $H_B$ are the Hamiltonians of the system and bath respectively, $\epsilon \ll 1$ is a small dimensionless parameter controlling system-bath coupling strength, and $H_{SB}$ is the system-bath coupling Hamiltonian. At initial time, the system is considered to be in an arbitrary initial state $\rho(0)$, while the bath is in a thermal state with inverse temperature $\beta$
\begin{align}
\label{initial_state}
    \rho_{\rm tot}(0) = \rho(0) \otimes \rho_B,~~\rho_B = \frac{e^{-\beta H_B}}{Z_B}.
\end{align}
 Starting with this initial state, the whole set-up of the system and the bath is evolved with the full Hamiltonian $H$, and the bath degrees of freedom are traced out to obtain the state of the system, 
\begin{align}
\label{CPTP_map}
    \rho(t) = {\rm Tr}_B\left( e^{-iH t} \rho_{\rm tot} (0)e^{iHt}\right),
\end{align}
where ${\rm Tr}_B(\ldots)$ denotes trace over bath degrees of freedom. The Eq.~\eqref{CPTP_map}, by construction, is a completely positive trace preserving (CPTP) map from $\rho(0)$ to $\rho(t)$ \cite{breuer_book,Rivas_2012}. Without any loss of generality, we assume ${\rm Tr}_B(H_{SB} \rho_B)=0$, where ${\rm Tr}_B(\ldots)$ denotes trace over bath degrees of freedom \cite{breuer_book, Rivas_2012}. The effective equation of motion for the system density matrix written to the leading order in system-bath coupling strength $\epsilon$ is the RE, given by, \cite{breuer_book},
\begin{equation}
    \begin{aligned}
	\label{RE1}
	\frac{\partial\rho}{\partial t}=&i[ \rho(t),H_S] \\
	+& \epsilon^2 \int_0^\infty dt^\prime \operatorname{Tr}_B [H_{SB},[H_{SB}(-t^\prime),\rho(t) \otimes \rho_B]],
    \end{aligned}
\end{equation}
	where 
\begin{equation} 
\label{eq:hsbt}
 H_{SB}(t)=e^{i(H_S+H_B)t}H_{SB}e^{-i(H_S+H_B)t} 
\end{equation}
and $\rho_B$ is the state of the bath. In complete generality, we can write the system-bath coupling Hamiltonian as
\begin{equation}
	\label{couplingH}
	H_{SB} = \sum_{l} (S_l B_l^\dagger + S_l^\dagger B_l),
\end{equation}
where $S_l$ and $B_l$ are operators on the system and bath respectively, and $l$ can be summed over as many indices as required for $H_{SB}$. Using Eq.\eqref{couplingH} in Eq.\eqref{RE1} and simplifying, we have
\begin{equation}
	\begin{aligned}
		\label{RE4}
		\frac{\partial {\rho}}{\partial t}=& \, i[ \rho(t),H_S]  + \epsilon^2 \Big\{& \squash \sum_{l}  \big[S^\dagger_l , S_l^{(1)} \rho(t) \big] - \big[ S^\dagger_l , \rho(t) S_l^{(2)} \big] \\
		&+ \text{H.c} \Big\},
	\end{aligned}
\end{equation}
where 
\begin{equation}
	\begin{aligned}
		\label{RE5}
	S_l^{(1)} & = \sum_m \int_0^\infty dt^\prime \operatorname{Tr} \big[B_l B^\dagger_m (-t^\prime) \rho_B \big] S_m(-t^\prime) \\
	&+\sum_m \int_0^\infty dt^\prime\operatorname{Tr} \big[B_l B_m (-t^\prime) \rho_B \big] S^\dagger_m(-t^\prime)  \\
	S_l^{(2)} & = \sum_m \int_0^\infty dt^\prime \operatorname{Tr} \big[ B^\dagger_m (-t^\prime) B_l \rho_B  \big] S_m(-t^\prime) \\
	&+\sum_m \int_0^\infty dt^\prime\operatorname{Tr} \big[ B_m (-t^\prime)  B_l \rho_B \big] S^\dagger_m(-t^\prime).
	\end{aligned}
\end{equation}
Since the actual microscopic evolution is given by a CPTP map [see Eqs.~\eqref{initial_state},~\eqref{CPTP_map}], one might naively expect that the evolution obtained from the microscopically derived RE respects complete positivity. However, as we prove in the next subsection in generality, unless in extremely special cases, the RE violates complete positivity.

\subsection{Violation of complete positivity in Redfield equation}
\subsubsection{Choosing an operator Basis} \label{subsec:operator}
As shown in Fig.~\ref{fig1}, we consider only a part of the system is coupled to the bath. 
Let us denote $\mathcal{H}_L$ as the Hilbert space of that part of the system that couples to the bath, and let $\mathcal{H}_M$ be the Hilbert space of the remaining part of the system. In mathematical terms, this means that any operator $O_M$ in the Hilbert space $\mathcal{H}_M$ commutes with the system-bath coupling Hamiltonian $H_{SB}$,
\begin{align}
    \label{system-bath-condition}
    [O_M, H_{SB}]=0.
\end{align}
The system Hamiltonian can then be written as
\begin{align}
\label{system_H}
    H_S = H_L + H_{LM} + H_M,
\end{align}
where the Hamiltonian $H_L$ is in Hilbert space $\mathcal{H}_L$, the Hamiltonian $H_M$ is in Hilbert space $\mathcal{H}_M$, and $H_{LM}$ gives the coupling between the two Hilbert spaces. Note that we do not consider this coupling to be small.

Let the dimension of  $\mathcal{H}_L$ and $\mathcal{H}_M$ be $d_L$ and $d_M$ respectively. Then, one can pick an orthonormal basis of operators  $\{f_i\}$ and $\{g_j\}$ on $\mathcal{H}_L$ and $\mathcal{H}_M$ respectively, where $ 1\leq i \leq d_L^2$ and $1 \leq j \leq d_M^2$, and where orthonormality is defined according to the Hilbert Schmidt inner product given by $\braket{A,B} = \text{Tr} [A^\dagger B]$. One can always choose this basis such that  $f_{d_L^2}=I_{L}/ \sqrt{d_L} $ and $g_{d_M^2}=I_M / \sqrt{d_M} $,  where $I_M$ and $I_L$ are the identity operators on those spaces. Such a basis is required by the GKSL theorem \cite{GKS1976,lindblad1976,breuer_book,Rivas_2012}.  Taking the tensor product of these two basis, one can obtain an orthonormal basis $\{F_k\} = \{f_i\} \otimes \{g_j\}$ for operators on $\mathcal{H}_S$, with $F_{d_L^2 d_M^2}=I_S / \sqrt{d} $, where $d=d_Ld_M$ is the dimension of the system Hilbert space. Without loss of generality, the Lindblad equation [Eq.~\eqref{Lindblad_form}] written in this basis is given by
\begin{equation}\label{eq:RE_lindblad1}
		\frac{\partial {\rho}}{\partial t} = i [\rho,H_S+H_{LS}]  + \squash \sum_{\alpha,{\tilde{\alpha} } = 1}^{d^2-1} \Gamma_{\alpha {\tilde{\alpha} }} \Big( F_{{\tilde{\alpha} }} \rho F^\dagger_{\alpha} - \frac{\{ F^\dagger_\alpha F_{\tilde{\alpha} }, \rho \}}{2} \Big),
\end{equation}
where complete positivity of $\rho$ is preserved iff $\Gamma$ is positive semidefinite \cite{breuer_book,Rivas_2012}. Eq.~\eqref{eq:RE_lindblad1} can be turned into Eq.~\eqref{Lindblad_form} by diagonalizing the matrix $\Gamma$. 

The complete positivity of RE can be checked by taking the RE to the same form as Eq.\eqref{eq:RE_lindblad1} and checking if the corresponding $\Gamma$ is positive semidefinite. To do so, let us relabel the indices so that $F_i= f_i \otimes I_M / \sqrt{d_M} $ for $1\leq i \leq d_L^2-1$. This allows us to expand the system operators in Eq.~\eqref{RE4} as, 
\begin{equation}
	\begin{aligned}
	\label{eq:S_expansions}
		S_l&=\sum_{\alpha=1}^{d^2} a_{l \alpha} F_\alpha, \quad S^\dagger_l=\sum_{\alpha=1}^{d^2} a^\prime_{l \alpha} F_\alpha,    \\
		S^{(1)}_l&=\sum_{\alpha=1}^{d^2} b_{l \alpha} F_\alpha, \quad   S^{(1) \dagger}_l=\sum_{\alpha=1}^{d^2} b^\prime_{l \alpha} F_\alpha,  \\
		S^{(2)}_l&=\sum_{\alpha=1}^{d^2} c_{l \alpha} F_\alpha , \quad  S^{(2) \dagger }_l=\sum_{\alpha=1}^{d^2} c^\prime_{l \alpha} F_\alpha, 
	\end{aligned}
\end{equation}
where $a_{l \alpha}=a^\prime_{l \alpha} = 0,  \forall \quad d_L^2 \leq \alpha \leq d^2 -1 $ since $S_l$ and $S_l^\dagger$ are identity on $\mathcal{H}_M$. Substituting Eq.~\eqref{eq:S_expansions} into Eq.~\eqref{RE4}, we obtain
\begin{equation}
	\begin{aligned} 
		\label{RE6}
			& \frac{\partial {\rho}}{\partial t} = i [\rho,H_S] - \epsilon^2 \sum_l \sum_{\alpha,{\tilde{\alpha}}=1}^{d^2} \Big\{ a^*_{l \alpha} b_{l {\tilde{\alpha}}}[F^\dagger_\alpha,F_{\tilde{\alpha}} \rho]  \\ 
			&+ c^{\prime *}_{l \alpha} a^\prime_{l {\tilde{\alpha}}}[\rho F^\dagger_\alpha,F_{\tilde{\alpha}}] 
		+b^*_{l \alpha} a_{l {\tilde{\alpha}}}[\rho F_\alpha^\dagger, F_{\tilde{\alpha}}]+ a^{\prime *}_{l \alpha} c^\prime_{l {\tilde{\alpha}}} [F^\dagger_\alpha,F_{\tilde{\alpha}} \rho] \Big\}.
		\end{aligned}
\end{equation}
Using some straightforward algebra (see Appendix~\ref{appendixa}), Eq.~\eqref{RE6} can be simplified to 
\begin{equation}
	\begin{aligned} \label{RE:lindblad}
			\frac{\partial {\rho}}{\partial t}&= i [\rho,H_S+H_{LS} ] + \sum_{\alpha, {\tilde{\alpha}}=1}^{d^2-1} \Gamma_{\alpha {\tilde{\alpha}}} \bigg( F_{\tilde{\alpha}} \rho F_\alpha^\dagger - \frac{\{ F_\alpha^\dagger F_{\tilde{\alpha}},\rho \}}{2}  \bigg),
	\end{aligned}
\end{equation}
where $\Gamma_{\alpha{\tilde{\alpha}}}$ is a $(d^2-1)\times(d^2-1)$ hermitian matrix given by
\begin{equation} 
	\label{eq:gamma_re}
	\Gamma_{\alpha {\tilde{\alpha}}}=\epsilon^2 \sum_l ( a^*_{l \alpha } b _{l {\tilde{\alpha}}} +c^{\prime *}_{l \alpha } a^\prime _{l {\tilde{\alpha}}} +b^*_{l \alpha } a_{l {\tilde{\alpha}}} +a^{\prime *}_{l \alpha } c^\prime_{l {\tilde{\alpha}}} ),
\end{equation}
where $x^*$ in Eq.~\eqref{eq:gamma_re} denotes the complex conjugate of $x$, and the expression of $H_{LS}$ that appears in Eq.~\eqref{RE:lindblad} can be found in Appendix \ref{appendixa}. Since Eq.~\eqref{eq:gamma_re} is of the form in Eq.~\eqref{eq:RE_lindblad1}, the condition for preserving complete positivity of $\rho$ is equivalent to $\Gamma$ being positive semidefinite. 
 
\subsubsection{The dissipator}
Let us now look at $\Gamma_{\alpha {\tilde{\alpha}}}$ from Eq.~\eqref{eq:gamma_re}. Recall from Sec.~\ref{subsec:operator}, that if $ \alpha \geq d_L^2$, then $a_{l \alpha}=a^\prime_{l \alpha}=0$. Therefore, $\Gamma_{\alpha {\tilde{\alpha}}}=0$ when $\alpha,{\tilde{\alpha}} \geq d_L^2$, and has the following structure:
\begin{equation} \label{gamma_struc}
	\Gamma=\left[ \begin{array}{ c | c  }
		\Gamma_{\alpha,{\tilde{\alpha}} < d_L^2} & \Gamma_{\alpha < d_L^2,{\tilde{\alpha}} \geq d_L^2}\\ 
		\hline
		\Gamma_{\alpha \geq d_L^2, {\tilde{\alpha}} < d_L^2} & 0 \\
	\end{array} \right].
\end{equation}
In general, the off-diagonal blocks $ \Gamma_{\alpha < d_L^2,{\tilde{\alpha}} \geq d_L^2}$, and $\Gamma_{\alpha \geq d_L^2, {\tilde{\alpha}} < d_L^2}$ will not be identically zero. The RE will preserve complete positivity if and only if the matrix $\Gamma$ is positive semidefinite. To move forward we will require the following Lemma concerning positive semidefinite matrices.
\begin{lemma} \label{lemma:positivity}
	Let $M$ be any positive semidefinite matrix, such that $M_{jj} = 0$. Then, $\forall i$ it must be the case that $M_{ij} = M_{ji} = 0$. Thus, if a positive semidefinite matrix has a zero as its $j$th diagonal element, then the entire $j$th row and $j$th column must consist of zeros.
\end{lemma}
\textbf{Proof} : For a matrix $M$ to be positive semidefinite, it is necessary that every $2 \times 2$  matrix ($M^\prime$) of the form
\begin{equation}
	M^\prime=\left[ \begin{array}{ c | c  }
		M_{ii} & M_{ij}\\ 
		\hline
		M_{ji} & M_{jj} \\
	\end{array} \right]
\end{equation}
is positive semidefinite. Otherwise there would exist a vector $\ket{v^\prime}$ such that $\bra{v^\prime} M^\prime \ket{v^\prime} < 0$. Then, there would exist a vector $\ket{v}$ which is non-zero only on the $i$th and $j$th entries, such that $\bra{v} M \ket{v} = \bra{v^\prime} M^\prime \ket{v^\prime}  < 0$. 
If $M_{jj}=0$, the eigenvalues of $M^\prime$ are given by solving the characteristic equation: $\lambda^2-M_{ii} \lambda -M_{ij} M_{ji}=0$. Since $M$ is hermitian, for both eigenvalues to be non-negative, it must be the case that $M_{ij} = M_{ji} = 0$.

Applying Lemma~\ref{lemma:positivity} to Eq.~\eqref{gamma_struc}, we find that $\Gamma$ can be positive semidefinite only if the off-diagonal blocks $\Gamma_{\alpha < d_L^2,{\tilde{\alpha}} \geq d_L^2}$ and $	\Gamma_{\alpha \geq d_L^2, {\tilde{\alpha}} < d_L^2} $ are zero. Since this is not generically true for RE, we have shown that the RE equation for any generic setup where the bath only couples to a part of the system violates complete positivity. An important exception here are situations where the RE happens to be such that the off-diagonal blocks in $\Gamma$ are identically zero. In such cases, the RE dissipator consists of operators of the form $f_i \otimes I_M / \sqrt{d_M}$, and acts only the part of the system connected to the bath. However, such situations can be expected to arise in only some extremely special cases. Even in such situations, RE may or may not preserve complete positivity, depending on $\Gamma_{\alpha < d_L^2, {\tilde{\alpha}}< d_L^2}$.   Thus we have rigorously shown that for setups where the bath couples to only a part of the system, the RE will violate complete positivity unless its dissipator is ``local'' and acts only on the part of the system that is connected to the bath. We also see from Eq.~\eqref{eq:gamma_re}, that both the positive and negative eigenvalues of the $\Gamma$ matrix can be the same order in system-bath coupling, i.e $O(\epsilon^2)$. 

We stress that although it is known that RE does not conserve complete positivity, most previous works show this via specific examples (for instance, \cite{Hartmann_2020_1,Eastham_2016, Suarez_1992, Spohn_1980}). A model-independent proof that RE generically violates complete positivity can be found in Ref. \cite{Whitney_2008}, for the case where the system-bath coupling consists of a single term of the form $H_{SB} = \epsilon S B$, which is not the most general form of system-bath coupling. Our proof is completely general, and highlights the role of locality of the system-bath coupling and the Redfield dissipator, which Ref. \cite{Whitney_2008} does not.  

 In Appendix~\ref{appendixb}, we give a concrete example of a two-qubit XXZ system with the first qubit connected to the bath. We indeed find that the matrix $\Gamma$ for this specific example has the expected structure from Eq.~\eqref{gamma_struc}, and negative eigenvalues. 

A natural question that arises, is how one could recover complete positivity of RE via a suitable approximation to the RE. In order to recover complete positivity, one must change $\Gamma_{\alpha {\tilde{\alpha}}}$ to some $\widetilde{\Gamma}_{\alpha {\tilde{\alpha}}}$ that is positive semidefinite. By the above discussion, for any ${\tilde{\alpha}} \geq d_L^2$, this will require either (i) making $\widetilde{\Gamma}_{\alpha \alpha}>0$, $\forall \alpha>d_L^2$  or 
(ii) making $\widetilde{\Gamma}_{\alpha {\tilde{\alpha}}}=0$  $\forall \alpha<d_L^2$. We will see in the next section (Sec.~\ref{sec:lindblad}) that any equation with  (i), will violate local conservation laws.

\section{Lindblad descriptions obeying local conservation laws} \label{sec:lindblad}

\subsection{Local conservation laws} \label{subsec:conservation}

Let us make precise what we mean by preservation of local conservation laws. For our setup, since the bath only acts on $\mathcal{H}_L$ part of the system, any operator on $\mathcal{H}_M$ commutes with the system-bath coupling Hamiltonian $H_{SB}$ (see Eq.~\eqref{system-bath-condition}, Fig.~\ref{fig1}). So, writing down the Heisenberg equation of motion with respect to the full Hamiltonian $H$ [Eq. \eqref{full_H}], and using Eq.~\eqref{system-bath-condition}, we see that the dynamical equation for the expectation value of any observable $O_M$ on $\mathcal{H}_M$ is given by
    \begin{equation} \label{cont_1}
    	\frac{d}{dt} \braket{ I_L \otimes O_M}=-i \braket{[I_L\otimes O_M, H_S]}
    \end{equation}
where $\braket{X} = \text{Tr} [X\rho]$. Any effective QME obtained by integrating out the bath should satisfy this property. We call QMEs satisfying this property as ones preserving local conservation laws. The justification for this name becomes clear if we look at an operator in $\mathcal{H}_M$ that would remain conserved if there is coupling with $\mathcal{H}_L$, i.e, if $H_{LM}$ in Eq.\eqref{system_H} is zero. One such operator is the Hamiltonian $H_M$. The dynamical equation for expectation value of $H_M$ gives the energy continuity equation
\begin{align}
    \frac{d}{dt} \braket{ H_M}= J_{L \to M},~~ J_{L \to M}=  -i \braket{[H_M, H_{LM}]}.
\end{align}
Here, $J_{L \to M}$ can be interpreted as the energy current from the region $L$ to the region $M$ (see Fig.~\ref{fig1}). In steady state, the rate of change of any system operator is zero. From above equation, this gives, $J_{L \to M}=0$ in steady state.  Thus, steady state energy current inside the system is zero. This is a statement of local conservation of energy and is one of the fundamental physical requirements for a system coupled to a single bath that follows from the more general requirement Eq.~\eqref{cont_1}. 

Importantly, the RE, i.e, Eq.~\eqref{RE1}, can be shown to satisfy Eq.~\eqref{cont_1} \cite{Tupkary_2022} and thereby preserves local conservation laws. We can write any QME to leading order in system bath coupling as
\begin{equation} \label{eq:qme}
    \frac{\partial {\rho}}{\partial t} = \mathcal{L}_0 (\rho)+\epsilon^2 \mathcal{L}_2(\rho),
\end{equation}
where $\mathcal{L}_0 (\rho)=i[\rho,H_S]$ and $\mathcal{L}_2(\rho)$ contains both the dissipator and the Lamb-shift Hamiltonian. Computing the left hand size of Eq.~\eqref{cont_1} using Eq.~\eqref{eq:qme}, and comparing with the right hand size of Eq.~\eqref{cont_1}, we obtain \cite{Tupkary_2022},
\begin{equation} \label{eq:old_condition}
    \text{Tr}[(I_M \otimes O_M) \mathcal{L}_2(\rho)] = 0.
\end{equation}
This is a necessary condition for satisfying local conservation laws. If we now further restrict the QME to be of Lindblad form, i.e, of the form Eq.~\eqref{Lindblad_form}, thereby respecting complete positivity, we obtain the following theorem. 
\begin{theorem}
\label{theorem}
Any QME of Lindblad form Eq.~\eqref{Lindblad_form} (thereby satisfying complete positivity) that also satisfies local conservation laws must have the Lindblad operators and the Lamb-shift Hamiltonian acting only on the part of the system connected to the bath. That is,  $L_\lambda=L_\lambda^L \otimes I_M$, $H_{LS}=H_{LS}^L \otimes I_M$, where $L_\lambda^L$, $H_{LS}^L$ act only on the Hilbert space $\mathcal{H}_L$ which is coupled to the bath by system-bath coupling Hamiltonian $H_{SB}$, and $I_M$ is the identity on the remaining of the system Hilbert space $\mathcal{H}_M$.
\end{theorem}
In the next subsection, we give the proof of this theorem.


\subsection{Proof of theorem \ref{theorem}}

We start by writing the most general Lindblad equation in the basis of Sec.~\ref{subsec:operator},
\begin{equation} \label{eq:lindblad1}
		\frac{\partial {\rho}}{\partial t} = i [\rho,H_S+\widetilde{H}_{LS}] + \squash \sum_{\alpha,{\tilde{\alpha}} = 1}^{d^2-1}  \widetilde{\Gamma}_{\alpha {\tilde{\alpha}}} \Big( F_{{\tilde{\alpha}}} \rho F^\dagger_{\alpha} - \frac{\{ F^\dagger_\alpha F_{\tilde{\alpha}}, \rho \}}{2} \Big)
\end{equation}
where $\widetilde{H}_{LS}$ is some Lamb shift Hamiltonian, and $\widetilde{\Gamma}$ is a positive semidefinite matrix. As mentioned before, this form can be reduced to the form of Eq.\eqref{Lindblad_form} by transforming to a basis where matrix $\widetilde{\Gamma}$ is diagonal. So, it suffices to work with this form. Since $F_i = f_i \otimes I_M$ for $1  \leq i < d_L^2$, the condition for the Lindblad operators to act only on $\mathcal{H}_L$ then translates to the matrix $\widetilde{\Gamma}$ being of the form
\begin{equation} \label{eq:gamma_struct_lindblad}
	\widetilde{\Gamma}=\left[ \begin{array}{ c | c  }
		\widetilde{\Gamma}_{\alpha,{\tilde{\alpha}} < d_L^2} & 0\\ 
		\hline
	0& 0 \\
	\end{array} \right].
\end{equation}
In the following we prove that in order to preserve local conservation laws the matrix $\widetilde{\Gamma}$ must be of this form.


\subsubsection{The restriction on $\widetilde{\Gamma}$}
Writing down the evolution of any observable $I_L \otimes O_M$ for Eq.~\eqref{eq:lindblad1}, we have
\begin{equation}
	\begin{aligned} \label{eq:evol_eq_lindblad}
	&\frac{d \braket{I_L \otimes O_M}}{dt}=-i \braket{[I_L \otimes O_M,H_S] } -i \braket{[I_L \otimes O_M,\widetilde{H}_{LS} ] } \\
		&+ \sum_{\alpha=1}^{d_L^2-1} \sum_{{\tilde{\alpha}}=d_L^2}^{d^2-1} \frac{\widetilde{\Gamma}_{\alpha {\tilde{\alpha}}}}{2} \braket{[I_L \otimes O_M,F^\dagger_\alpha F_{\tilde{\alpha}}]}  \\
		&+  \sum_{\alpha=d_L^2}^{d^2-1} \sum_{{\tilde{\alpha}}=1}^{d_L^2-1} \frac{\widetilde{\Gamma}_{\alpha {\tilde{\alpha}}} }{2}\braket{[F^\dagger_\alpha F_{\tilde{\alpha}}, I_L \otimes O_M]} \\
		&+ \sum_{\alpha, {\tilde{\alpha}}=d_L^2}^{d^2-1} \Big( \widetilde{\Gamma}_{\alpha {\tilde{\alpha}}}  \Big( \braket{ F^\dagger_\alpha   (I_L \otimes O_M) F_{\tilde{\alpha}}  } - \frac{1}{2} \braket{ (I_L \otimes O_M) F_\alpha^\dagger F_{\tilde{\alpha}}} \\
		&- \frac{1}{2}\braket{F_\alpha^\dagger F_{\tilde{\alpha}} (I_L \otimes O_M) }  \Big)
	\end{aligned}
\end{equation}
where $\braket{X}$ denotes the expectation value of $X$ given by $\text{Tr} [X \rho]$, and we use the fact that $[F_i, I_L \otimes O_M]= 0$ when $i < d_L^2$. We can combine all commutators in above equation into a single commutator with an effective Lamb-shift like Hamiltonian, which we denote $\widetilde{H}^{(2)}_{LS}$. This gives
\begin{equation}
	\begin{aligned} \label{eq:evol_eq_lindblad_2}
	&\frac{d \braket{I_L \otimes O_M}}{dt}=-i \braket{[I_L \otimes O_M,H_S+\widetilde{H}^{(2)}_{LS} ] } \\
		&+ \sum_{\alpha, {\tilde{\alpha}}=d_L^2}^{d^2-1} \widetilde{\Gamma}_{\alpha {\tilde{\alpha}}} \Big( \braket{ F^\dagger_\alpha  (I_L \otimes O_M)  F_{\tilde{\alpha}} } - \frac{1}{2} \braket{ (I_L \otimes O_M) F_\alpha^\dagger F_{\tilde{\alpha}}} \\
		&- \frac{1}{2}\braket{F_\alpha^\dagger F_{\tilde{\alpha}} (I_L \otimes O_M) }  \Big)
	\end{aligned}
\end{equation}
where $\widetilde{H}^{(2)}_{LS}$ is some hermitian operator. Comparing with Eq.~\eqref{cont_1}, and noting that $\operatorname{Tr}[M\rho]=0 \quad \forall \rho$ implies $ M=0$, we obtain the condition for satisfying Eq.~\eqref{cont_1} as
\begin{equation}
\label{cont_conjecture}
	\begin{aligned}
		&-i [I_L \otimes O_M,\widetilde{H}^{(2)}_{LS} ]+ \sum_{\alpha, {\tilde{\alpha}}=d_L^2}^{d^2-1} \widetilde{\Gamma}_{\alpha {\tilde{\alpha}}} \Big( F^\dagger_\alpha  (I_L \otimes O_M) F_{\tilde{\alpha}}  \\
		&-\frac{1}{2} (I_L \otimes O_M) F_\alpha^\dagger F_{\tilde{\alpha}} - \frac{1}{2} F_\alpha^\dagger F_{\tilde{\alpha}} (I_L \otimes O_M)  \Big)=0 \\
		& \hspace{19em} \forall O_M.
	\end{aligned}
\end{equation}
We will now relabel the indices $\alpha, {\tilde{\alpha}}$ for the sake of convenience.  Recall from Sec.~\ref{subsec:operator}, that  $\alpha,{\tilde{\alpha}}$ which appear in the above expression can be equivalently written as $\alpha \rightarrow (\alpha_L, \alpha_M), {\tilde{\alpha}} \rightarrow ({\tilde{\alpha}}_L, {\tilde{\alpha}}_M)$, and vice-versa, where $F_\alpha=f_{\alpha_L}\otimes g_{\alpha_M}$. Therefore, $
\sum_{\alpha = 1}^{d^2-1}$ is equivalent to $\sum_{\alpha_M =1}^{d^2_M-1} \sum_{\alpha_L = 1}^{d_L^2} $. We can also expand $\widetilde{H}^{(2)}_{LS}$ as
\begin{equation} \label{eq:HLS2_expansion}
	\widetilde{H}^{(2)}_{LS}=\sum_{\alpha=1}^{d^2} \nu_\alpha F_\alpha = \sum_{\alpha_L=1}^{d^2_L} \sum_{\alpha_M=1}^{d^2_M} \nu_{\alpha_L,\alpha_M} (f_{\alpha_L} \otimes g_{\alpha_M} ).
\end{equation}
This basis aids in taking a partial trace over the $L$ part of the system. Performing this partial trace and using the orthonormality of $f_i$ operators, we can rewrite Eq.~\eqref{cont_conjecture} as
\begin{equation}
	\begin{aligned}
		\label{cont_conjecture_2}
		&-i \sum_{\alpha_M=1}^{d_M^2} \nu_{d_L^2,\alpha_M}  [O_M,g_{\alpha_M}] + \sum_{\alpha_M, {\tilde{\alpha}}_M=1}^{d_M^2-1} \widetilde{\Lambda}_{\alpha_M,{\tilde{\alpha}}_M} \\
		& \Big(  g_{\alpha_M}^\dagger   O_M  g_{{\tilde{\alpha}}_M} - \frac{1}{2}  O_M  g_{\alpha_M}^\dagger g_{{\tilde{\alpha}}_M}  - \frac{1}{2}  g_{\alpha_M}^\dagger  g_{{\tilde{\alpha}}_M}   O_M  \Big)=0, \\
		& \hspace{19em} \forall O_M.
	\end{aligned}
\end{equation}
where
\begin{equation}
	\begin{aligned}
		\widetilde{\Lambda }_{\alpha_M,{\tilde{\alpha}}_M} &= \sum_{\alpha_L, {\tilde{\alpha}}_L=1}^{d_L^2} \delta_{\alpha_L,{\tilde{\alpha}}_L} \widetilde{\Gamma}_{(\alpha_L, \alpha_M) , ({\tilde{\alpha}}_L, {\tilde{\alpha}}_M)} \\
	\end{aligned}
\end{equation}
 We show in Appendix~\ref{appendixc} that Eq.~\eqref{cont_conjecture_2} implies 
\begin{equation} \label{eq:rigorous_condition}
 \sum_{\alpha_M=1}^{d_M^2-1} \widetilde{\Lambda}_{\alpha_M,\alpha_M} =  \sum_{\alpha_M=1}^{d_M^2-1} \sum_{\alpha_L=1}^{d_L^2}  \widetilde{\Gamma}_{(\alpha_L, \alpha_M) , (\alpha_L ,\alpha_M)}  =0
\end{equation} 
Since we require $\widetilde{\Gamma}$ to be positive semidefinite, it cannot have negative values on the diagonal. Therefore, Eq.~\eqref{eq:rigorous_condition} implies $\widetilde{\Gamma}_{(\alpha_L, \alpha_M) , (\alpha_L, \alpha_M)} = 0$ for $1 \leq \alpha_L \leq d_L^2$ and $1 \leq \alpha_M \leq d_M^2-1$ . Equivalently, $\widetilde{\Gamma}_{\alpha \alpha} = 0$ for $\alpha \geq d_L^2$. Applying Lemma \ref{lemma:positivity} to this case, we see that $\widetilde{\Gamma}_{\alpha, {\tilde{\alpha}}} $ can be non-zero only when both $\alpha, {\tilde{\alpha}} < d_L^2$, and therefore $\widetilde{\Gamma}$ is a matrix of the form in Eq.~\eqref{eq:gamma_struct_lindblad}.
This concludes the first part of the proof.

\subsubsection{The restriction on $\widetilde{H}_{LS}$}
Given this structure for $\widetilde{\Gamma}$ from Eq.~\eqref{eq:gamma_struct_lindblad}, we now investigate the restrictions on the Lamb shift Hamiltonian $\widetilde{H}_{LS}$.
Since $\widetilde{\Gamma}$ has to obey Eq.~\eqref{eq:gamma_struct_lindblad}, we find that $\widetilde{H}_{LS} = \widetilde{H}_{LS}^{(2)} = \sum_{\alpha_L,\alpha_M} \nu_{\alpha_L,\alpha_M} (f_{\alpha_L} \otimes g_{\alpha_M} ) $ in Eq.~\eqref{cont_conjecture}. Then, our condition for satisfying local conservation laws from  Eq.~\eqref{cont_conjecture} is given by, 
\begin{equation}
-i  \sum_{\alpha_L = 1}^{d^2_L} \sum_{\alpha_M=1}^{d^2_M} \nu_{\alpha_L,\alpha_M}   [I_L \otimes O_M,  (f_{\alpha_L} \otimes g_{\alpha_M} ) ] = 0 \quad \forall O_M . 
\end{equation}
which further implies
\begin{equation} \label{eq:hls_condition}
        -i  \sum_{\alpha_L = 1}^{d^2_L} \sum_{\alpha_M=1}^{d^2_M} \nu_{\alpha_L,\alpha_M} f_{\alpha_L} \otimes   [ O_M,   g_{\alpha_M} ] = 0 \quad \forall O_M . 
\end{equation}
Multiplying both sides by $f^\dagger_{\alpha_L} \otimes I_M$, and tracing out the $L$ part of the system, we obtain
\begin{equation}
    -i  \sum_{\alpha_M=1}^{d^2_M} \nu_{\alpha_L,\alpha_M}   [ O_M,   g_{\alpha_M} ] = 0 \quad \forall \alpha_L, O_M .
\end{equation}
This can happen only if 
\begin{equation} 
\label{eq:HLS_identity}
  \sum_{\alpha_M=1}^{d^2_M} \nu_{\alpha_L,\alpha_M}  g_{\alpha_M} \propto I_M \quad \forall \alpha_L
    \end{equation}
Using Eq.~\eqref{eq:HLS_identity} in Eq.~\eqref{eq:HLS2_expansion}, and recalling that $\widetilde{H}_{LS} = \widetilde{H}^{(2)}_{LS}$, we obtain $\widetilde{H}_{LS} =\widetilde{H}^{(L)}_{LS}  \otimes I_M $. This concludes the second part of the proof.

\subsection{Remarks on Theorem~\ref{theorem}}
Theorem~\ref{theorem} says that for a QME to preserve complete positivity and obey local conservation laws, both the Lamb shift Hamitonian and the dissipator must only act on the part of the system connected to the bath. Such a Lindblad equation is often termed a `local Lindblad equation'. Theorem~\ref{theorem} thus says that only local Lindblad equations are consistent with local conservation laws. The RE preserves local conservation laws without having a local dissipator, but it does so at the cost of losing complete positivity. Any global form of Lindblad equation, like the eigenbasis Lindblad equation \cite{carmichael_book,breuer_book} and the universal Lindblad Equation \cite{ule}, violates local conservation laws, while preserving complete positivity. One main reason such global forms of Lindblad equations are often derived under various approximations is that they can be proven to show thermalization. However, general statements related to thermalization in local Lindblad equations are usually difficult to make. In the next section, we discuss a numerical technique which allows to study whether, in a given set-up, a QME consistent with Theorem~\ref{theorem} is possible such that it also shows thermalization.




\section{On the possibility of thermalization with local dissipators} \label{sec:qme_thermal}



\subsection{Condition for satisfying thermalization}

We start by making precise what we mean by thermalization. Going back to the form of the QME in Eq.~\eqref{eq:qme}, the setup is said to show thermalization if 
\begin{align}
\label{eq:thermalization_definition}
    \lim_{\epsilon \to 0} \left(\lim_{t \to \infty} e^{t (\mathcal{L}_0 + \epsilon^2 \mathcal{L}_2)} \rho(0)\right)=\rho_{th},~~\rho_{th} = \frac{e^{-\beta H_S}}{\text{Tr} [e^{-\beta H_S}]},
\end{align}
irrespective of the initial state of the system $\rho(0)$. Physically, it means that if the system is weakly coupled to a thermal bath for a long time, and then the system-bath coupling is slowly switched-off, the state of the system will be the Gibbs state at the temperature of the bath, irrespective of the system's initial state. If there is no explicit time-dependence in the Hamiltonian, as in Eq.~\eqref{full_H}, this statement can be proven starting with the initial state of the full set-up in Eq.~\eqref{initial_state}, and assuming that the steady state is unique \cite{Tupkary_2022}. Given a QME of the form in Eq.\eqref{eq:qme}, in Ref.~\cite{Tupkary_2022}, we showed that the following condition needs to be satisfied in order to guarantee thermalization, 
\begin{equation} \label{eq:thermal_condition}
    \bra{E_i} \mathcal{L}_2 (\rho_{th}) \ket{E_i} = 0  \quad \forall~i.
\end{equation}
where $\ket{E_i}$ is the eigenvector of the system Hamiltonian $H_S$ with eigenvalue $E_i$. The derivation of this condition can also be found in Appendix \ref{appendix_thermal}. Given a system Hamiltonian $H_S$, the inverse temperature $\beta$ of the bath and a partition of the system Hilbert space into the part $\mathcal{H}_L$ which is attached to a bath and the remainder of Hilbert space $H_M$ (see Fig.\ref{fig1}), we would like to find a QME respecting the restrictions in theorem \ref{theorem} and satisfying Eq.\eqref{eq:thermal_condition}. As we will see later by example, such a QME is not guaranteed to be possible. In the next subsection, we provide a numerical way to conclusively check if such a QME is possible in a given setup.

\subsection{The thermalization optimization problem}
\label{subsec:TOP}

For the setup shown in Fig.~\ref{fig1}, the most general form of QME respecting the restrictions in theorem \ref{theorem} for satisfying complete positivity and local conservation laws is
\begin{equation}
	\begin{aligned} \label{eq:local_lindblad_for_TOP}
			&\frac{\partial {\rho}}{\partial t}= i [\rho,H_S+\epsilon^2 H^{(L)}_{LS} \otimes I_M ] + \epsilon^2 \sum_{\alpha_L, {\tilde{\alpha}}_L=1}^{d_L^2-1} \Gamma^{(L)}_{\alpha_L, {\tilde{\alpha}}_L}  \\
			&\bigg( (f_{\tilde{\alpha}_L} \otimes I_M) \rho (f_{\alpha_L} \otimes I_M)^\dagger\\
			&- \frac{\{ (f_{\alpha_L} \otimes I_M)^\dagger (f_{\tilde{\alpha}_L} \otimes I_M),\rho \}}{2}  \bigg).
	\end{aligned}
\end{equation}
Here $H^{(L)}_{LS}$ is a Lamb shift Hamiltonian that acts on the $\mathcal{H}_L$ part of the system, $\Gamma^{(L)}$ is $(d_L^2-1) \otimes (d_L^2-1)$ matrix that must be positive semidefinite, $d_L$ is the dimension of the Hilbert space $\mathcal{H}_L$ that is directly coupled with the bath. We include the factor of $\epsilon^2$ in front of $\Gamma^{(L)}$ and $H^{(L)}_{LS}$ explicitly. The system Hamiltonian $H_S$ and the Hilbert space $\mathcal{H}_L$ are assumed to be given. The task is then to find $H^{(L)}_{LS}$ and $\Gamma^{(L)}$, such that Eq.\eqref{eq:thermal_condition} is satisfied to a given precision, for a given inverse temperature $\beta$. To this end, we introduce the quantity
\begin{equation}  
\label{eq:thermal_condition_tau}
    \tau = \sum_i \left\vert \bra{E_i}  \mathcal{L}_2 (\rho_{th} ) \ket{E_i}\right\vert,
\end{equation}
where $\mathcal{L}_2$ consist of all terms in Eq.~\eqref{eq:local_lindblad_for_TOP} which are multiplied by $\epsilon^2$, i.e all terms except the commutator with the system Hamiltonian. Then, we can cast the task in terms of an optimization problem given by :
\begin{equation} \label{eq:thermal_optimization}
    \begin{aligned}
    \text{minimize : } \quad &\tau ~ \textrm{by varying } H^{(L)}_{LS},~\Gamma^{(L)},  \\
    \text{subject to : } \quad & H^{(L)}_{LS} \text{ is hermitian,} \\
    & \text{Tr}(\Gamma^{(L)}) = 1,~~ \Gamma^{(L)} \geq 0, \\
    \end{aligned}
\end{equation}
where we use $\Gamma^{(L)} \geq 0$ to denote $\Gamma^{(L)}$ being positive semidefinite. The condition $\text{Tr}(\Gamma^{(L)}) = 1$ is imposed to avoid the trivial solution $H^{(L)}_{LS},\Gamma^{(L)}=0$, which trivially gives the global minimum $\tau=0$. Since we want $\Gamma^{(L)}$ to be a non-zero positive semi-definite matrix, it must have a positive trace. We have fixed that trace to one in some arbitrarily chosen energy unit. This does not cause loss of generality since the strength of system-bath coupling is explicitly governed by $\epsilon^2$ in Eq.\eqref{eq:local_lindblad_for_TOP}. We call the optimization problem in Eq.~\eqref{eq:thermal_optimization} the ``thermalization optimization problem''(TOP). Let $\tau_{opt}$ be the optimal value obtained from solving TOP. Given a tolerance $\delta$, 
\begin{align}
\label{eq:tolerance}
\begin{array}{cc}
     \textrm{if } \tau_{opt}<\delta, &\textrm{ the desired QME is possible}, \\
     \textrm{else, } & \textrm{ the desired QME is impossible},
\end{array}
\end{align}
up to the precision $\delta$.

Most interestingly, we find that the TOP in Eq.~\eqref{eq:thermal_optimization} can be written as a SDP. The background and theoretical framework of SDP is discussed in Appendix~\ref{subsec:sdp}. The TOP is proven to be a SDP in Appendix~\ref{subsec:reduction}. In particular, any choice of $H^{(L)}_{LS}$ and $\Gamma^{(L)}$ in Eq.~\eqref{eq:thermal_optimization} can be used to obtain an upper bound on $\tau_{opt}$. Then, the theoretical framework of SDP can be used to construct a ``dual'' problem to the optimization problem in Eq.~\eqref{eq:thermal_optimization}. This dual problem can then be used to obtain a lower bound on $\tau_{opt}$. If the lower bound and upper bound match, then this guarantees that one has found the global optimal value of $\tau$ . Our above described approach is transparent, and simple to use, since the optimization problem in Eq.~\eqref{eq:thermal_optimization} can be directly put into standard packages for disciplined convex optimization like the CVX MATLAB package \cite{cvx}. In particular, CVX itself automatically constructs the dual problem, outputs $\tau_{opt}$, and gives one choice of  $H^{(L)}_{LS}$, and $\Gamma^{(L)}$ which yields the output value of $\tau_{opt}$.  Thus, if $\tau_{opt} < \delta$, it not only says that the desired type of QME is possible but also it outputs one possible candidate for such a QME. If $\tau_{opt} \geq \delta$, the desired type of QME is impossible.

We would like to point out here that, in a microscopic derivation, given the temperature of the baths, $H^{(L)}_{LS}$ and $\Gamma^{(L)}$ would depend only on the bath spectral functions and system-bath coupling Hamiltonian. So, the TOP can be thought of as varying over all possible bath spectral functions and system-bath coupling Hamiltonians to find $\tau_{opt}$. Thus, if $\tau_{opt} \geq \delta$, we can conclusively say that, for the chosen system parameters and temperature, under no choice of bath spectral function and system-bath coupling Hamiltonian, can a Markovian QME be derived which simultaneously satisfies complete positivity, local conservation laws and shows thermalization up to the chosen precision.  In the next subsection, we look at the TOP in an open XXZ qubit chain.

\subsection{Open XXZ qubit chain as an example}

We study the possibility of having a Lindblad description satisfying local conservation laws and showing thermalization in an open  XXZ qubit chain system with some of the qubits attached to baths. The system Hamiltonian for this setup is given by 
    \begin{equation}
        \begin{aligned}
        H_S = \sum_{\ell=1}^N \frac{\omega^{(\ell)}_0}{2} {\sigma}_z^{(\ell)}  -  \sum_ {\ell=1}^{N-1} g_\ell \big(& {\sigma}_x^{(\ell)} {\sigma}_x^{(\ell+1)} + {\sigma}_y^{(\ell)} {\sigma}_y^{(\ell+1)} \\
    &+ \Delta_\ell {\sigma}_z^{(\ell)} {\sigma}_z^{(\ell+1)} \big),
        \end{aligned}
    \end{equation}
    where $\sigma^{(\ell)}_{x,y,z}$ denotes the Pauli matrices acting on the $\ell^{\text{th}}$ qubit, and $\omega^{(\ell)}_0$, $g_\ell$, and $g_\ell \Delta_\ell$ represent the magnetic field, the overall qubit-qubit coupling strength and the anisotropy respectively. The first $N_L$ qubits are attached to a bath, while the remaining $N_M = N-N_L$ qubits are not attached to any bath. We use the formalism of Sec.~\ref{subsec:TOP} to investigate thermalization in this set-up for various values of $N_L$ and $N_M$. In order to do so, we first need to construct the basis for operators in Hilbert space of the first $N_L$ qubits, so that the most general form of the desired QME can be written as in Eq.~\eqref{eq:local_lindblad_for_TOP}. For the $\ell$th qubit, we choose the basis $\{ -\sigma^{(\ell)}_z /\sqrt{2}, \sigma^{(\ell)}_+, \sigma^{(\ell)}_-, I^{(\ell)}_2 / \sqrt{2} \} $, where  $\sigma^{(\ell)}_{+}=(\sigma^{(\ell)}_{x}+i \sigma^{(\ell)}_{y})/2$, $\sigma^{(\ell)}_{-}=(\sigma^{(\ell)}_{x}-i \sigma^{(\ell)}_{y})/2$, and $I^{(\ell)}_2$ is the identity operator for the qubit Hilbert space. The basis for the first $N_L$ qubits is obtained by direct product of the basis of each of the qubits. We construct the TOP [Eq.\eqref{eq:thermal_optimization}] in this basis, which we then directly input in the CVX MATLAB package to obtain $\tau_{opt}$. We set an ad-hoc value of the tolerance $\delta=10^{-6}$ [see Eq.\eqref{eq:tolerance}]. In a typical calculation from a weak system-bath coupling QME, usually the error due to neglecting higher order terms would be larger than such a low value of tolerance.

\begin{figure}
    \centering
    \includegraphics[width = \linewidth]{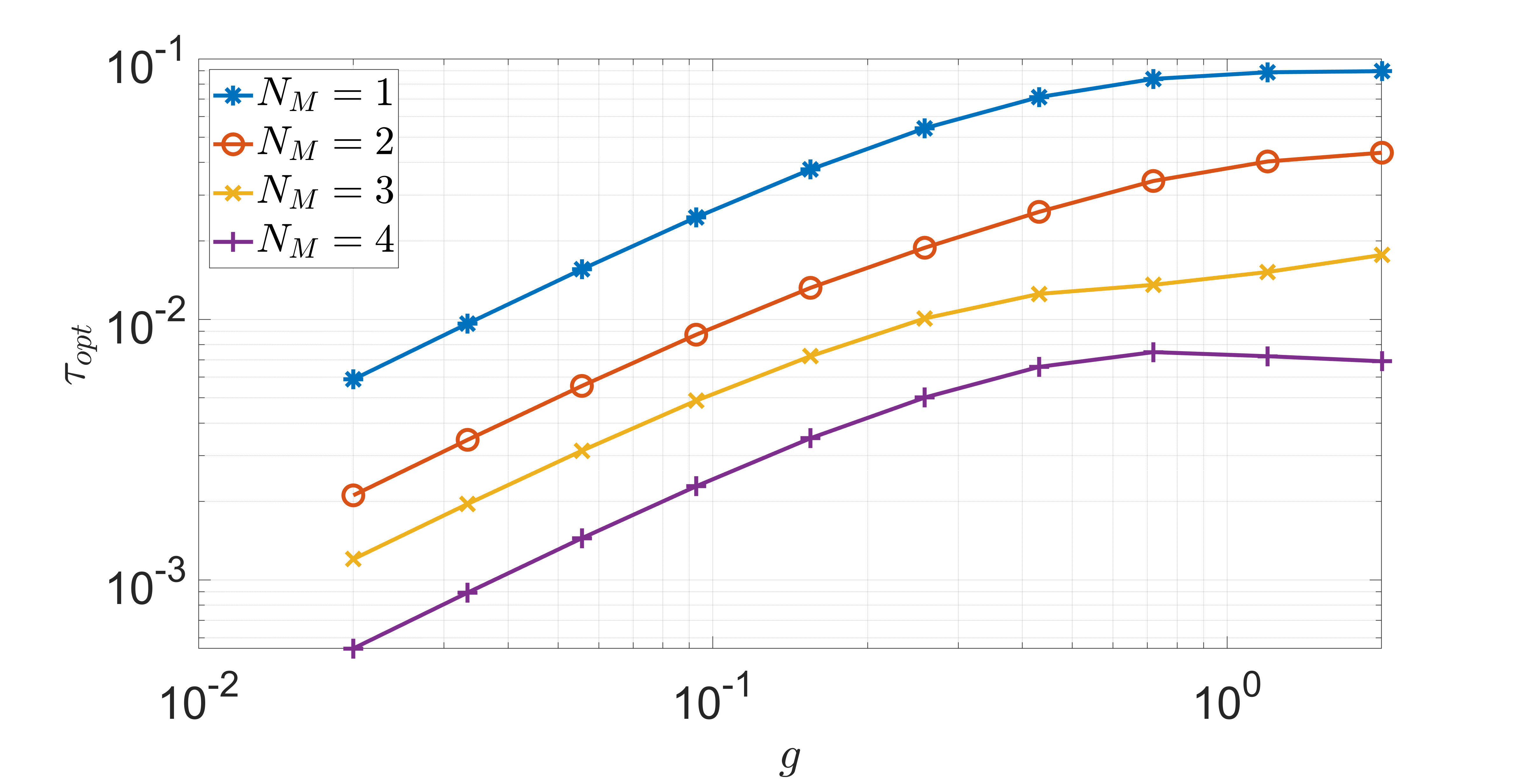}
    \caption{$\tau_{opt}$ vs $g$, for $N_L = 1$, with $\omega^{(\ell)}_0 = 1$, $\Delta_\ell = 1$, $\beta = 1$ and $g_\ell = g$ for all $\ell$. The tolerance chosen is $\delta=10^{-6}$. We find that $\tau_{opt} \gg \delta$, conclusively showing that, for such setups, no QME can simultaneously preserve complete positivity, obey local conservation laws, and show thermalization up to the precision set by the tolerance. }
    \label{fig2}
\end{figure}

\begin{figure}
    \centering
    \includegraphics[width = \linewidth]{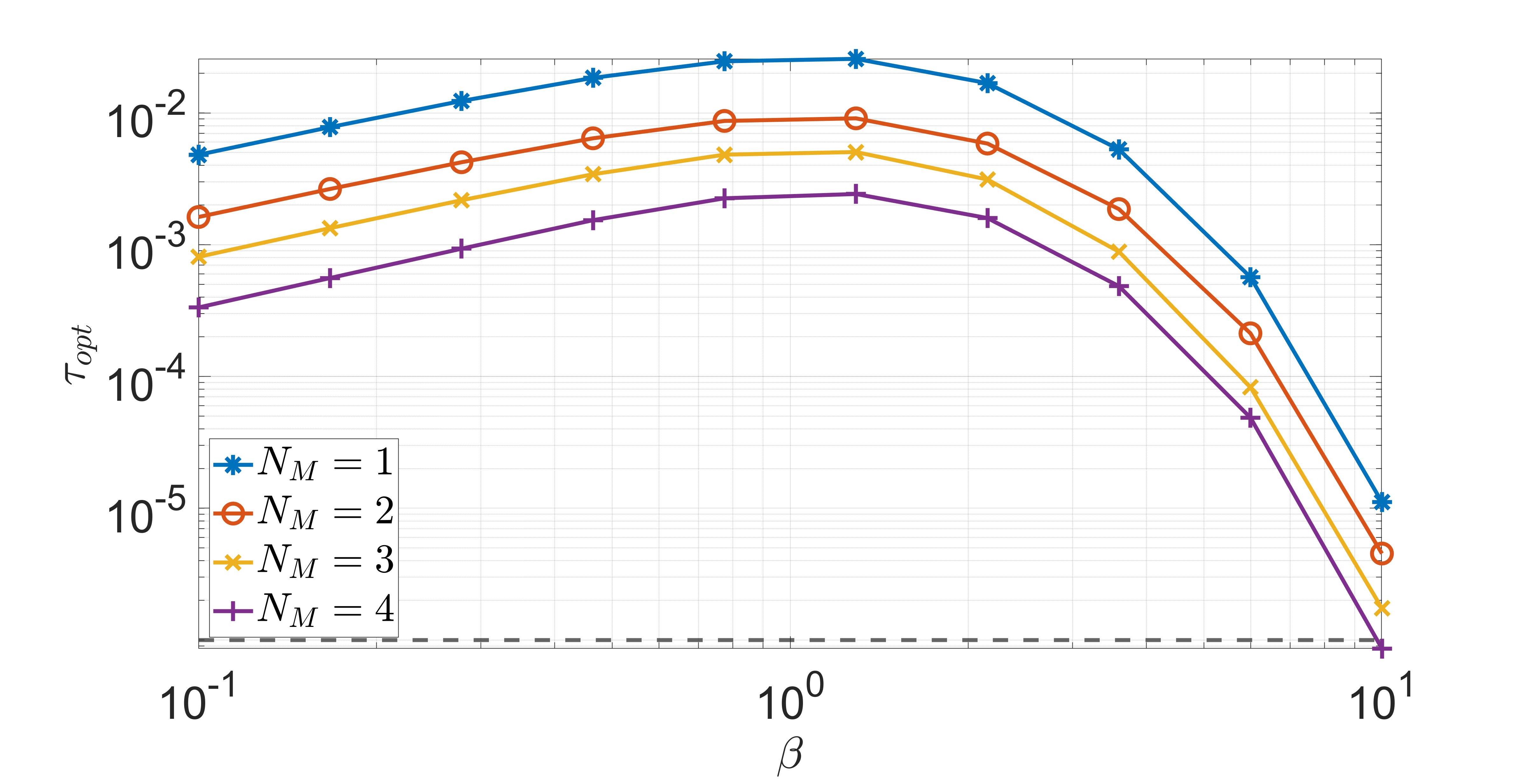}
    \caption{$\tau_{opt}$ vs $\beta$, for $N_L = 1$, with $\omega^{(\ell)}_0 = 1$, $\Delta_\ell = 1$, $g_\ell= 0.1$  for all $\ell$. The tolerance chosen is $\delta=10^{-6}$, and is plotted as the dashed horizontal line. We find that $\tau_{opt} \gg \delta$, indicating that, for such setups, it is not possible to have a QME simultaneously preserving complete positivity, obeying local conservation laws, and showing thermalization up to the precision set by the tolerance. 
    }
    \label{figbeta1}
    \end{figure}

\subsubsection{Single qubit attached to bath}
First we consider the case where the first qubit of the chain is coupled to a bath, so, $N_L = 1$. In Fig.~\eqref{fig2}, we plot $\tau_{opt}$ vs $g$ for $N_M = 1,2,3$, fixing $\omega^{(\ell)}_0 = 1$, $\Delta_\ell = 1$ for all $\ell$ and $\beta= 1$.  We find that $\tau_{opt} \gg \delta$ when $N_L = 1$, and $N_M = 1,2,3$. Thus, no Markovian QME can simultaneously preserve complete positivity, obey local conservation laws, and satisfy thermalization for such setups in the chosen range of parameters. This explicit example also directly rules out the possibility of having a general form of Markovian QME that is guaranteed to meet all the fundamental requirements. We see from Fig.~\eqref{fig2} that $\tau_{opt}$ increases with $g$. This is consistent with previous results showing that such local Lindblad equations become a good description when the coupling between the system qubits are weak. Additionally, we see that for a given value of $g$,  $\tau_{opt}$ decreases as $N_M$ increases. This, interestingly, seems to indicate that for long chains, local Lindblad with dissipator acting on only one qubit might be able to describe the thermal state up to a reasonably good precision. However, more detailed studies are required to make any conclusive statement in this direction.

Intuitively, memory effects, and hence Markovianity of open system dynamics, depend on temperature. So, we might expect that at a different temperature, $\tau_{opt}$ might be smaller than $\delta$.
In Fig.~\eqref{figbeta1}, we plot $\tau_{opt}$ vs $\beta$ for $N_L=1$, fixing $\omega^{(\ell)}_0 = 1$, $\Delta_\ell = 1$, $g_\ell= 0.1$  for all $\ell$. We find that $\tau_{opt} \gg \delta$, for almost the entire range of $\beta$ chosen, showing that for these parameters, no Markovian QME can simultaneously preserve complete positivity, obey local conservation laws, and satisfy thermalization. But, we see some interesting features. Firstly, as before, we see that $\tau_{opt}$ decreases as $N_M$ increases. Secondly, we see that  $\tau_{opt}$ varies non-monotonically with $\beta$. At very low temperatures, $\tau_{opt}$ decreases tends to decay below $\delta$. In Fig.~\eqref{figbeta1}, for $N_M=4$ and $\beta=10$, $\tau_{opt}<\delta$. At high temperatures also, $\tau_{opt}$ decreases.
This suggests that at very low and very high temperatures, it is possible to obtain a local Lindblad equation that shows thermalization. That this is indeed true can be checked independently. At such extremes of temperatures, for at least some choices of baths and system-bath couplings, local Linblad equations can be microscopically derived \cite{Milburn_book1,Becker_2021}.  This explains the non-monotonic dependence of $\tau_{opt}$ on $\beta$.
Since very high and very low temperatures can allow for a local Lindblad description, it is then intuitive that departure from such behavior is maximum when $\beta$ is of the order of system time scales. Indeed it is close to such values of $\beta$, i.e, $\beta \sim \omega_0^{(\ell)}=1$, that the highest value of $\tau_{opt}$ is seen in Fig.~\eqref{figbeta1}.  

We would like to point out here that, $\tau_{opt}\geq \delta$ does not mean the system coupled to a thermal bath is unable to thermalize for such parameters. Unless in extremely special cases, it is almost always possible to find a bath spectral function and system-bath coupling which ensure thermalization. So $\tau_{opt}\geq \delta$ instead means that, for those parameters, the dynamics of approach to thermal state cannot be governed by a completely positive Markovian QME preserving local conservation laws. The dynamics then must have some non-Markovian character. In fact, as we see from above example, $\tau_{opt}$ gives an estimate of how close to Markovianity the open system dynamics can be for the chosen parameters. 

Next, we discuss the case where first two qubits are attached to the bath and highlight the drastic difference observed for similar choice of parameters.

\begin{figure}
    \centering
    \includegraphics[width = \linewidth]{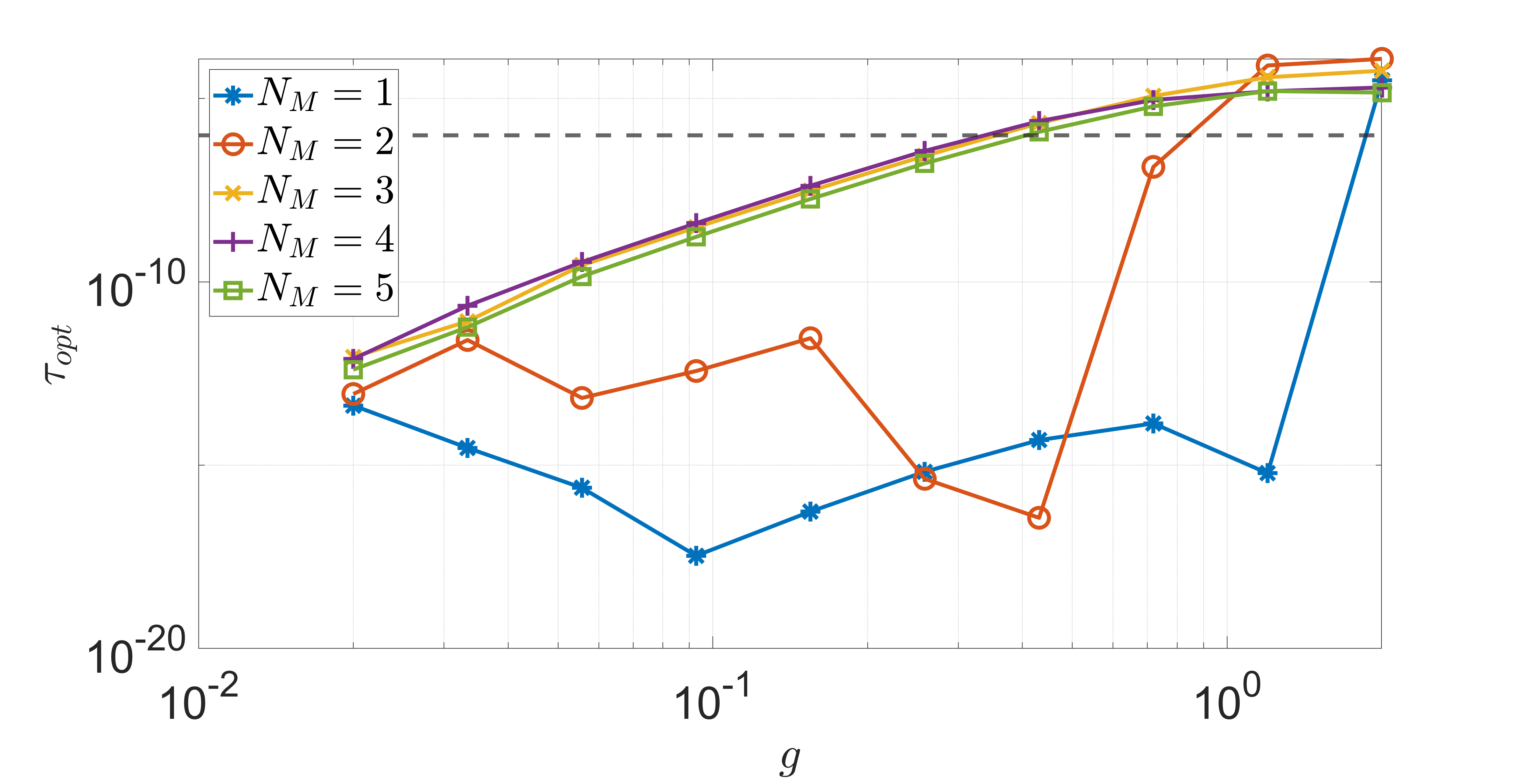}
    \caption{$\tau_{opt}$ vs $g$, for $N_L = 2$, with $\omega^{(\ell)}_0 = 1$, $\Delta_\ell = 1$, $g_\ell = g$ for all $\ell$ and $\beta = 1$. The tolerance chosen is $\delta=10^{-6}$, and is plotted as the dashed horizontal line. We find that $\tau_{opt} \ll \delta$ for smaller values of $g$, indicating that, for such setups, it is possible to have a QME simultaneously preserving complete positivity, obeying local conservation laws, and showing thermalization up to the precision set by the tolerance. 
    }
    \label{fig3}
    \end{figure}

\begin{figure}
    \centering
    \includegraphics[width = \linewidth]{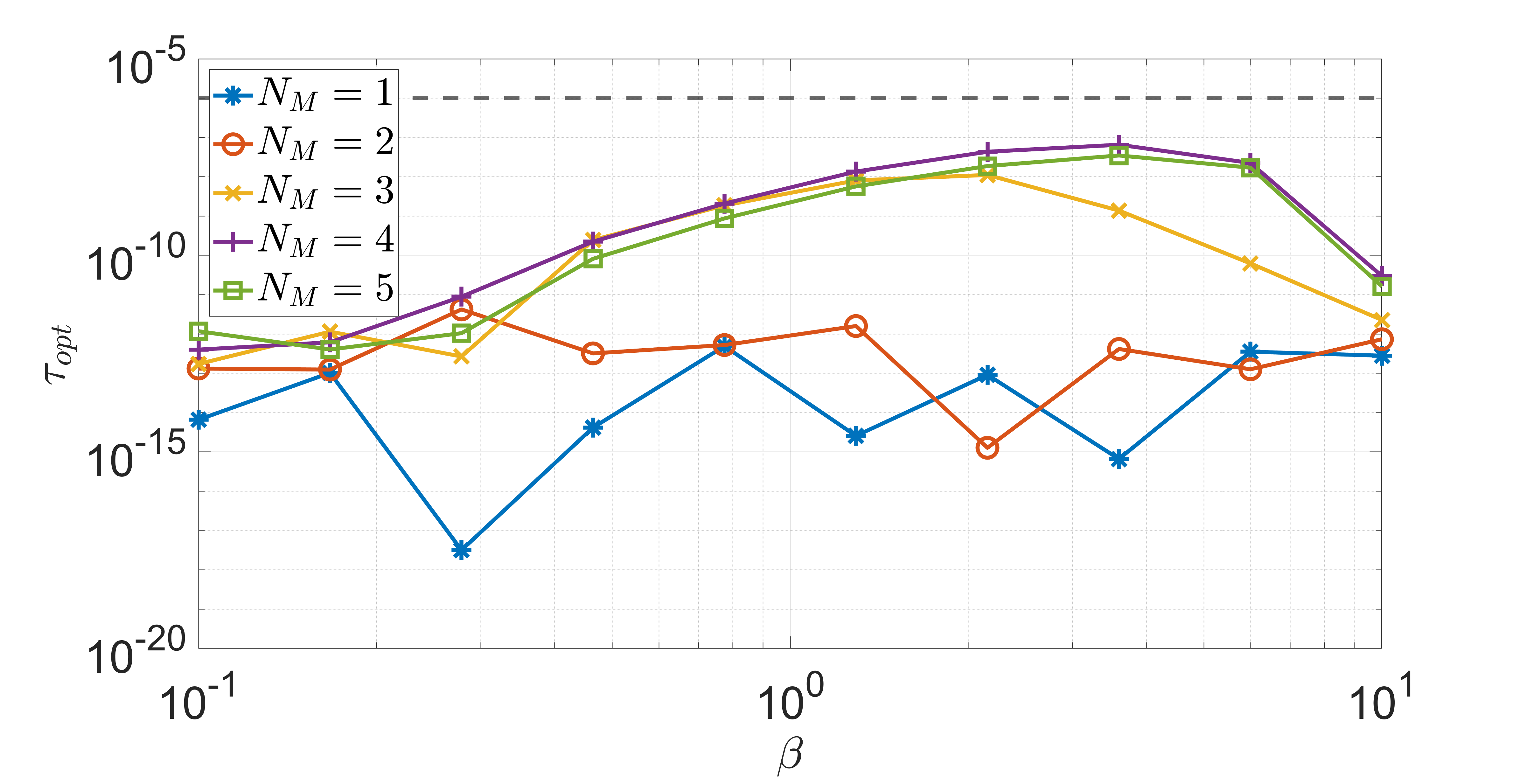}
    \caption{$\tau_{opt}$ vs $\beta$, for $N_L = 2$, with $\omega^{(\ell)}_0 = 1$, $\Delta_\ell = 1$, $g_\ell= 0.1$  for all $\ell$. The tolerance chosen is $\delta=10^{-6}$, and is plotted as the dashed horizontal line. We find that $\tau_{opt} \ll \delta$ for smaller values of $g$, indicating that, for such setups, it is possible to have a QME simultaneously preserving complete positivity, obeying local conservation laws, and showing thermalization up to the precision set by the tolerance. The values in the figure less than $10^{-12}$ are below the numerical precision of CVX Matlab package.
    }
    \label{figbeta2}
    \end{figure}

\begin{table*}[t]
    \centering
    \begin{tabular}{|l|l|l|l|l|l|l|l|l|l|l|l|l|l|l|}
    \hline
        0.00033 & 0 & 0 & -0.00123 & 0 & 0 & 0.00033 & 0 & 0 & 0.00019 & 0 & 0 & 0.00001 & 0 & 0 \\ \hline
        0 & 0.00196 & 0 & 0 & -0.00179 & 0 & 0 & -0.03638 & 0 & 0 & 0 & 0 & 0 & -0.00002 & 0 \\ \hline
        0 & 0 & 0.00081 & 0 & 0 & 0 & 0 & 0 & -0.00081 & 0 & 0 & 0.01307 & 0 & 0 & -0.00021 \\ \hline
        -0.00123 & 0 & 0 & 0.02611 & 0 & 0 & -0.00282 & 0 & 0 & 0.00157 & 0 & 0 & -0.00143 & 0 & 0 \\ \hline
        0 & -0.00179 & 0 & 0 & 0.00172 & 0 & 0 & 0.03491 & 0 & 0 & 0 & 0 & 0 & 0.00011 & 0 \\ \hline
        0 & 0 & 0 & 0 & 0 & 0 & 0 & 0 & 0 & 0 & 0 & 0 & 0 & 0 & 0 \\ \hline
        0.00033 & 0 & 0 & -0.00282 & 0 & 0 & 0.00045 & 0 & 0 & 0.00002 & 0 & 0 & 0.00011 & 0 & 0 \\ \hline
        0 & -0.03638 & 0 & 0 & 0.03491 & 0 & 0 & 0.70685 & 0 & 0 & 0 & 0 & 0 & 0.00209 & 0 \\ \hline
        0 & 0 & -0.00081 & 0 & 0 & 0 & 0 & 0 & 0.00082 & 0 & 0 & -0.01311 & 0 & 0 & 0.00021 \\ \hline
        0.00019 & 0 & 0 & 0.00157 & 0 & 0 & 0.00002 & 0 & 0 & 0.00035 & 0 & 0 & -0.00014 & 0 & 0 \\ \hline
        0 & 0 & 0 & 0 & 0 & 0 & 0 & 0 & 0 & 0 & 0 & 0 & 0 & 0 & 0 \\ \hline
        0 & 0 & 0.01307 & 0 & 0 & 0 & 0 & 0 & -0.01311 & 0 & 0 & 0.26032 & 0 & 0 & -0.00208 \\ \hline
        0.00001 & 0 & 0 & -0.00143 & 0 & 0 & 0.00011 & 0 & 0 & -0.00014 & 0 & 0 & 0.00009 & 0 & 0 \\ \hline
        0 & -0.00002 & 0 & 0 & 0.00011 & 0 & 0 & 0.00209 & 0 & 0 & 0 & 0 & 0 & 0.0001 & 0 \\ \hline
        0 & 0 & -0.00021 & 0 & 0 & 0 & 0 & 0 & 0.00021 & 0 & 0 & -0.00208 & 0 & 0 & 0.00009 \\ \hline
    \end{tabular}
    \caption{\label{Table:Gamma} The $\Gamma^{(L)}$ obtained from CVX for $N_L=2$, $N_M=4$, $\omega_0^{(\ell)} = 1$, $g_\ell = 0.1$, $\Delta_\ell = 1$, $\beta=1$, i.e, the values used to compute Fig.~\ref{figtau}. Every entry is rounded to 5 digits after the decimal point for convenience of representation. The corresponding $H^{(L)}_{LS}$ is given to be zero. This $\Gamma^{(L)}$ satisfies complete positivity, local conservation laws and thermalization up to a precision of $\delta=10^{-6}$ for the given choice of parameters.  }
\end{table*}

\subsubsection{Two qubits attached to bath}

In Fig.~\eqref{fig3}, we plot $\tau_{opt}$ vs $g_\ell = g$ for $N_L = 2$, i.e, first two qubits attached to the bath, and $N_M = 1,2,3,4$, taking fixed values of $\omega^{(\ell)}_0 = 1$, $\Delta_\ell = 1$ and $\beta = 1$. These parameters are the same as in Fig.~\ref{fig2} and $g$ is varied over the same range. Quite remarkably, in stark contrast to Fig.~\ref{fig2},  we find that in this case $\tau_{opt} \ll \delta$ over a considerable range for $g<1$. We then look at the behavior of $\tau_{opt}$ versus $\beta$, fixing $\omega^{(\ell)}_0 = 1$, $\Delta_\ell = 1$, $g_\ell=g=0.1$. This is shown in Fig.~\ref{figbeta2}. For $N_M>N_L$, we again see the non-monotonic behavior. However, in stark contrast to Fig.~\ref{figbeta1}, we find that over the entire chosen range of $\beta$  $\tau_{opt} \ll \delta$. Thus, for $N_L=2$, over a considerable range of parameters, a QME that simultaneously preserves complete positivity, obeys local conservation laws, and satisfies thermalization up to the given precision is possible. This is a highly non-trivial result. 

Previously, local two-qubit Lindblad dissipators have been used to study energy transport in XXZ-type qubit chains (for example, Refs.~\cite{ Prosen_2009,Mendoza-Arenas_2015}). However, those local two-qubit Lindblad operators were constructed so as to thermalize the two qubits only, in absence of coupling to the rest of the chain. Such Lindblad description is not guaranteed to thermalize the whole chain to the given inverse temperature $\beta$ of the bath \cite{thermalization_2010_znidaric}. Our result here shows that it is possible to have a two-qubit local Lindblad description that can thermalize the full chain to the given temperature of the bath to a good approximation. 

As mentioned before, CVX also outputs a possible choice of $\Gamma^{(L)}$ and $H^{(L)}_{LS}$ matrices corresponding to $\tau_{opt}$. So, when $\tau_{opt}< \delta$, we get one possible candidate for the desired type of QME. For our choice of parameters, we find that CVX always outputs $H^{(L)}_{LS}=0$, and a non-trivial value of $\Gamma^{(L)}$ that would be hard to guess otherwise. In Table.~\ref{Table:Gamma}, we demonstrate the $\Gamma^{(L)}$ obtained for $N_L=2$, $N_M=4$, $\omega_0^{(\ell)} = 1$, $g_\ell = 0.1$, $\Delta_\ell = 1$, $\beta=1$. The $\Gamma^{(L)}$ matrix corresponds to the basis of operators $\{F_k\}$ chosen as 
\begin{equation}
F_{k} =  f_{\lceil k/4 \rceil }  \otimes  f_{k(\text{mod } 4)}  \otimes I_M 
\end{equation}
where $\{f_i\} = \{ -\sigma_z / \sqrt{2}, \sigma_-, \sigma_+, I_2 / \sqrt{2} \} $, $\lceil k/4 \rceil$ denotes the nearest integer greater than or equal to $k/4$, and $k(\text{mod } 4)$ denotes the value of $k$ modulo $4$, and $k$ goes from $1$ to $15$. We also note that the exact values of $\Gamma^{(L)}$ and $H^{(L)}_{LS}$ computed by CVX may depend on the exact configuration of the programming environment (such as internal solvers used by CVX). 

\begin{figure}
    \centering
    \includegraphics[width = \linewidth]{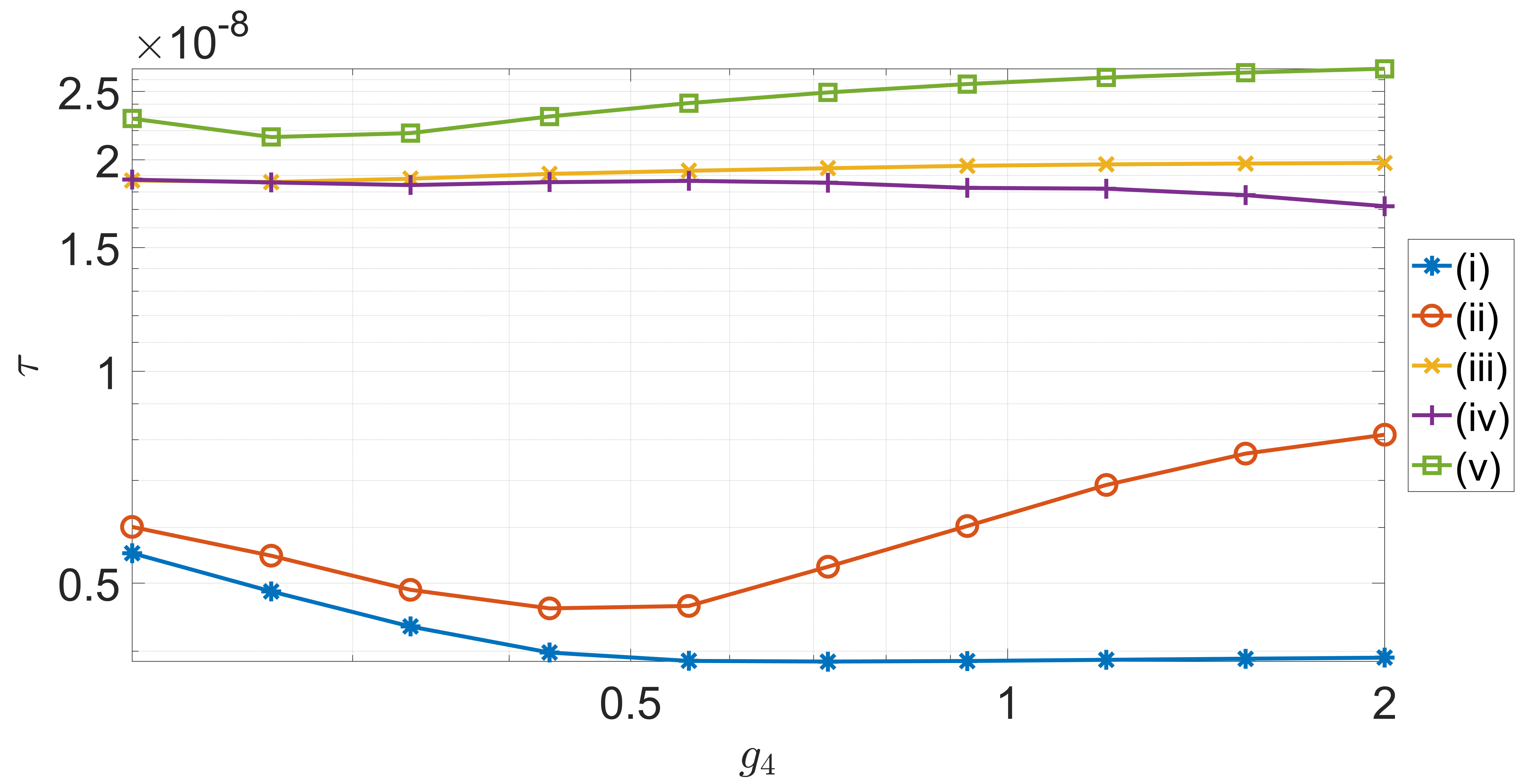}
    \caption{$\tau$ vs $g_4$, for $N_L = 2, N_M=4$, with $\omega^{(\ell)}_0 = 1$, $\Delta_\ell = 1$, $\beta = 1$ and $g_\ell = 0.1$ for all $\ell$ unless otherwise mentioned. $\tau$ is computed from $\Gamma^{(L)}$ and $H^{(L)}_{LS}$ obtained from CVX for $N_L = 2, N_M=4$, with $\omega^{(\ell)}_0 = 1$, $\Delta_\ell = 1$, $\beta = 1$ and $g_\ell = 0.1$ for all $\ell$.
    The modified parameters for the plots are given by (i) (no parameters changed), (ii) $\Delta_3 = 0.4, \Delta_4=1.2$, (iii) $\omega^{(3)}_0 = 1.5, \omega^{(4)}_0 = 1.5, g_5 = 0.3$, (iv)  $\omega^{(3)}_0 = 1.5, \omega^{(4)}_0 = 1.5, g_5 = 0.3, \Delta_4 = 0.5$, (v) $g_3 = 0.3$. We find that $\tau \ll \delta=10^{-6}$ even if parameters are changed for qubits of the system that are not coupled to the baths.
    }
    \label{figtau}
    \end{figure}

For every parameter of the system, there is a different $\tau_{opt}$, with a corresponding value of $\Gamma^{(L)}$ and $H^{(L)}_{LS}$ given by CVX. If we want to explore a large parameter space of the system, it seems that we need a different $\Gamma^{(L)}$ and $H^{(L)}_{LS}$ for each parameter point. Surprisingly, we find that this is not always required. If $\tau_{opt} \ll \delta$ for one set of parameters, we can substantially change parameters of the system far from the qubits attached to baths, and still obtain a value of $\tau \ll \delta$ with the same value of $\Gamma^{(L)}$ and $H^{(L)}_{LS}$. This is shown in Fig.~\eqref{figtau}, where $\tau$ is calculated changing various parameters away from the two qubits coupled with the bath, fixing $H^{(L)}_{LS}=0$ and $\Gamma^{(L)}$ to be the same as in Table.~\ref{Table:Gamma}. Over the entire regime of chosen parameters $\tau \ll \delta$. Note that, in contrast to previous plots, this is not be the optimal value of $\tau$. Nevertheless, if $\tau \ll \delta$, we still get a completely positive Markovian QME preserving local conservation laws and showing thermalization up to the chosen precision. Given $\Gamma^{(L)}$ and $H^{(L)}_{LS}$, it is much easier to just check this rather than finding the optimal value $\tau_{opt}$.

If parameters of the two qubits that are coupled to the bath are changed, we can no longer use the same $\Gamma^{(L)}$ and $H^{(L)}_{LS}$. For example, if we choose the same $\Gamma^{(L)}$ as in Fig.~\ref{figtau}, and change $g_1$ to $0.2$ from $0.1$, we get $\tau  = 0.0014 \gg \delta$.

The above observation suggests that the values of $\Gamma^{(L)}$ and $H^{(L)}_{LS}$ obtained by CVX can be used to define a QME, independent of the parameters in the bulk of the system. This is consistent with underlying picture that each value of $\Gamma^{(L)}$ and $H^{(L)}_{LS}$ corresponds to a different choice of the bath spectral function and the system-bath coupling Hamiltonian. If we change any parameter of the qubits attached to the baths, the change reflects substantially on the system-bath coupling Hamiltonian, so the value of $\tau$ changes drastically from $\tau_{opt}$ obtained with original parameters. If we change any parameter away from the two qubits, the change reflects much less on the system-bath coupling Hamiltonian, causing $\tau$ to be of the same order as the original value of $\tau_{opt}$. This presents an exciting prospect for studying the dynamics of the system-bath setup over a wide range of parameters, including a wide range of temperatures, with physically consistent Markovian QMEs. Such studies may also be possible for long chains, since local Markovian dissipation is favourable for tensor network based numerical techniques. Such dissipation may, also, in principle, be engineered in quantum computing and quantum simulation platforms, like ion traps \cite{Barreiro_2011, Schindler_2013}, Rydberg atoms \cite{Weimer_2010,Nguyen_2018}, superconducting qubits \cite{Garcia_2020} and quantum dots \cite{Kim_2022}.

\section{Summary and outlook} \label{sec:summary}

{\it Searching for a physically consistent Markovian QME ---}
A physically consistent Markovian QME must satisfy complete positivity, obey local conservation laws and be able to show thermalization. In this work, we have systematically gone about searching for such QMEs.  
This is done in three steps, and the result in each step has important consequences. Especially, we are led to introduce the TOP problem, which is an optimization problem for finding a QME with all the above properties up to a given precision. The TOP opens a completely new avenue in the study of dissipative quantum systems.

We consider a finite-dimensional undriven system a part of which is weakly coupled to a thermal bath. The microscopically derived QME written to leading order in system-bath coupling is the RE, which is known to obey local conservation laws and be able to show thermalization \cite{Tupkary_2022}. First, we show in generality that the RE violates complete positivity, unless in extremely special cases. Although there are previous works showing this via specific examples (for instance, \cite{Hartmann_2020_1,Eastham_2016, Suarez_1992, Spohn_1980}), we are unaware of a model independent proof similar to ours. Next, we prove that imposing complete positivity and preservation of local conservation laws enforces the QME to be of `local' form. That is, the Lindblad operators and the Lamb-shift Hamiltonian must have support only on the part of the system directly coupled to the bath, and be identity elsewhere. This rules out the possibility of any `global' forms of Lindblad equations, which are usually constructed to show thermalization, to be consistent with local conservation laws. Then, we ask if a `local' Lindblad equation can be found which is able to show thermalization. We find that, the task of finding such a Lindblad equation can be cast as an optimization problem, which we call TOP. Most interestingly, this optimization problem turns out to be a SDP. For given system and parameters, the SDP can be efficiently solved using high-level programming packages like the CVX Matlab package. The output of the TOP conclusively shows whether the desired type of QME is possible for the chosen system parameters and temperature, up to a chosen precision. For  numerical example, we look at the TOP in a XXZ qubit chain of few sites, fixing a reasonably high precision. When only the first site is coupled to a bath, we find that, unless in extremes of temperatures, it is impossible to find a local Lindblad equation that is capable of showing thermalization up to the chosen precision. 

{\it Discussion in light of various existing forms of QMEs ---}
Various forms of QMEs have been derived and analyzed in the literature under various approximations (for example, \cite{Becker2022, Becker_2021, ule, mccauley2020, Harbola2006, ULE2, Trushechkin_2021, Trushechkin_2016,Kleinherbers_2020,Davidovic_2020,mozgunov2020, kirvsanskas2018,segal_2022,massimo_2017,massimo_2017_2}. Although the above example shows that there is no general form of physically consistent Markovian QME, this does not immediately make them unusable. Instead, it turns out that in each of these forms of QME, some elements of the system density matrix are given correctly, while the others are not \cite{Tupkary_2022}. So, one needs to be careful in interpreting the results from them, always keeping in mind their microscopic derivation and approximations. The RE, despite not being completely positive, is more accurate than all such Lindblad QMEs in most cases. To elucidate how this can happen, imagine that, in a given setup, physically, the population of one energy level, say, $\bra{E_j}\rho \ket{E_j}$, is zero in steady state. The RE might then give a small negative value (say, $\bra{E_j}\rho \ket{E_j}=-10^{-3}$), while any of the Lindblad equations can give a larger positive value, which might be (say, $\bra{E_j}\rho \ket{E_j}=0.1$). Either case is a problem if we want to calculate various kinds of entropies, as often required in quantum information and thermodynamics. In case of RE, unphysical results can often be ruled out by checking the scaling with system-bath coupling \cite{Tupkary_2022, Archak_2020}. This is often more difficult in Lindblad QMEs, where approximations are often less controlled. The state obtained from the recently derived ULE \cite{ule}, which been shown to violate local conservation laws \cite{Tupkary_2022}, can be corrected to obtain results as accurate as the RE \cite{ULE2}. This re-instates the local conservation laws, at the cost of also re-instating the same positivity problem of the density matrix as in RE. In another recent work, a general form of QME has been derived \cite{Becker2022} which is more accurate than RE, even though complete positivity of dynamics is still not guaranteed.

{\it TOP and (non) Markovianity ---}
In the microscopic picture, given the temperature of the bath, the QME is completely defined by the bath spectral functions and the type of system-bath coupling. The TOP can then be thought of as varying over all possible bath spectral functions and types of system-bath couplings to find the closest to satisfying thermalization the local Lindblad equation can be. So, when TOP shows that the desired type of QME is impossible, it means no matter what type of bath is attached and how it is coupled to the system, for the chosen parameters, it is impossible to describe the dynamics via a completely positive Markovian QME satisfying local conservation laws and showing thermalization. The approach to thermal state must then have some non-Markovian character for such system parameters and temperature. The output of TOP, $\tau_{opt}$, shows non-trivial dependence on the system parameters and the temperature. This dependence seems to capture how close to Markovian the dynamics can be for the chosen parameters.

{\it Surprises when two qubits are attached to bath ---}
Surprisingly, we have found that, when first two qubits of the few-site XXZ chain are attached to a bath, solving the TOP shows that it is possible to find Lindbladians obeying  local conservation laws and showing thermalization up to quite high precision. This holds over a considerable range of parameters, including a wide range of temperatures. Notably, in this entire parameter regime, when one qubit was coupled to a bath, such a QME was  impossible. 

Whenever the TOP shows a QME respecting all conditions is possible, standard high-level programming packages used to solve the SDP also outputs one possible form for such a QME. When two qubits are attached to the bath, the form of QME  so obtained, which respects all the requirements, is quite non-trivial and would be hard to guess otherwise. Even more interestingly, we have found that if we take one such QME obtained for one choice of system parameters, and change some system parameters away from two qubits that couple to the bath, the QME still satisfies all the requirements. This opens several exciting possibilities that we describe below.

{\it Future directions ---}
Our results open the exciting possibility of studying the dynamics of approach to thermal state in open quantum many-body systems using physically consistent Markovian QMEs, over a wide range of parameters, including a wide range of temperatures. This is particularly aided by the fact that local Lindblad equations are favourable for tensor network techniques.  Studying such dynamics at finite temperatures is often quite challenging otherwise, requiring simulation of non-Markovian dynamics \cite{Fux_2022,PReB_2021,Lacerda_2022}.  

The TOP lets us find parameters of the system where local Lindblad equations can show thermalization. For two qubits attached to bath, this range of parameters can be considerably large, as we have seen. It may be possible to design such local dissipation in quantum computing and quantum simulation platforms like ion traps \cite{Barreiro_2011, Schindler_2013}, Rydberg atoms \cite{Weimer_2010,Nguyen_2018}, superconducting qubits \cite{Garcia_2020} and quantum dots \cite{Kim_2022}. Especially in ion traps and Rydberg atom platforms, this offers an interesting way to controllably prepare finite temperature states of complex quantum many-body systems in these platforms, which is presently a technological challenge. Usually, one would require global Lindblad dissipators to ensure that a thermal state is prepared. This would be hard to design in quantum simulation platforms if one wants to simulate complex many-body systems. The possibility of having local dissipation confined to two qubits offers a much easier alternative.   

Moreover, as we have seen in the example of XXZ qubit chain, the dependence of the output of the TOP, $\tau_{opt}$, on various parameters of the system already encode rich and interesting physics. For complex quantum many-body systems, one may need more scalable techniques for SDP, which is itself a direction of research in computer science \cite{SDP_scalability_2020}. Using these techniques,  the rich behavior of $\tau_{opt}$ with various parameters can then be studied.

It is therefore clear that our results, especially the introduction of the TOP, leads to new paradigm within the fields of quantum information, computation and technology. Nevertheless, various questions still remain. One main question concerns steady-state coherences \cite{Guarnieri_2018, Archak_2020, Trushechkin_2022, Cresser_2021}. When coupled to a thermal bath at any finite coupling, the system density matrix will have coherences in energy eigenbasis of the system \cite{Cresser_2021, Trushechkin_2022}. These coherences can be important in quantum information and thermodynamics \cite{Streltsov_2017,Francica_2019,Santos_2019,Narasimhachar_2015} and are given correctly to the leading order by the RE \cite{Tupkary_2022,Archak_2020, Cresser_2021}. However, it is not clear that  the steady-state coherences calculated from physically consistent Markovian QME obtained via TOP will be the same as those obtained from RE. Further investigation is required in this respect, which will be carried out in future works.

All code used in this work can be found at \cite{devashishcodegithub}.

\section*{Acknowledgements}
 MK would like to acknowledge support from the project 6004-1 of the Indo-French Centre for the Promotion of Advanced Research (IFCPAR), Ramanujan Fellowship (SB/S2/RJN-114/2016), SERB Early Career Research Award (ECR/2018/002085) and SERB Matrics Grant (MTR/2019/001101) from the Science and Engineering Research Board (SERB), Department of Science and Technology, Government of India. AD and MK acknowledge support of the Department of Atomic Energy, Government of India, under Project No. RTI4001. AP acknowledges funding from the European Research Council (ERC) under the European Unions Horizon 2020 research and innovation program (Grant Agreement No. 758403). A.P also acknowledges funding from the Danish National Research Foundation through the Center of Excellence ``CCQ'' (Grant agreement no.: DNRF156).
 
\appendix
\section{Casting Eq.\eqref{RE6} to Eq.\eqref{RE:lindblad}}
\label{appendixa}
In this appendix we show the steps for taking  Eq.~\eqref{RE6} to the form of Eq.~\eqref{RE:lindblad} which is  more amenable to studying issues related to conservation of complete positivity. We start with Eq.~\eqref{RE6}, which we recall to be 
\begin{equation}
	\begin{aligned} 
		\label{RE6_app}
			& \frac{\partial {\rho}}{\partial t} = i [\rho,H_S] - \epsilon^2 \sum_l \sum_{\alpha,{\tilde{\alpha}}=1}^{d^2} \Big\{ a^*_{l \alpha} b_{l {\tilde{\alpha}}}[F^\dagger_\alpha,F_{\tilde{\alpha}} \rho]  \\ 
			&+ c^{\prime *}_{l \alpha} a^\prime_{l {\tilde{\alpha}}}[\rho F^\dagger_\alpha,F_{\tilde{\alpha}}] 
		+b^*_{l \alpha} a_{l {\tilde{\alpha}}}[\rho F_\alpha^\dagger, F_{\tilde{\alpha}}]+ a^{\prime *}_{l \alpha} c^\prime_{l {\tilde{\alpha}}} [F^\dagger_\alpha,F_{\tilde{\alpha}} \rho] \Big\}.
		\end{aligned}
\end{equation}
This can be rewritten as
\begin{widetext}
\begin{equation}
\label{eq:re_expanded}
	\begin{aligned}
			\frac{\partial {\rho}}{\partial t}&=i [\rho,H_S]+\epsilon^2 \sum_l \sum_{\alpha,{\tilde{\alpha}}=1}^{d^2} 
		\Big\{ a^*_{l \alpha } b _{l {\tilde{\alpha}}} \Big( F_{\tilde{\alpha}} \rho F_\alpha^\dagger - \frac{\{ F_\alpha^\dagger F_{\tilde{\alpha}},\rho \}}{2} - \frac{[F_\alpha^\dagger F_{\tilde{\alpha}},\rho]}{2}\Big) 
		+ c^{\prime *}_{l \alpha } a^\prime _{l {\tilde{\alpha}}} \Big( F_{\tilde{\alpha}} \rho F_\alpha^\dagger - \frac{\{ F_\alpha^\dagger F_{\tilde{\alpha}},\rho \}}{2} + \frac{[F_\alpha^\dagger F_{\tilde{\alpha}},\rho]}{2}\Big)  \\
		&+ b^*_{l \alpha } a_{l {\tilde{\alpha}}} \Big( F_{\tilde{\alpha}} \rho F_\alpha^\dagger - \frac{\{ F_\alpha^\dagger F_{\tilde{\alpha}},\rho \}}{2} + \frac{[F_\alpha^\dagger F_{\tilde{\alpha}},\rho]}{2}\Big) 
		+a^{\prime *}_{l \alpha } c^\prime_{l {\tilde{\alpha}}} \Big( F_{\tilde{\alpha}} \rho F_\alpha^\dagger - \frac{\{ F_\alpha^\dagger F_{\tilde{\alpha}},\rho \}}{2} - \frac{[F_\alpha^\dagger F_{\tilde{\alpha}},\rho]}{2}\Big)
		\Big\},
	\end{aligned}
\end{equation}
\end{widetext}
where, ${A,B}=AB+BA$ is the anti-commutator.
Next, we note that the summation  $\sum_{\alpha,{\tilde{\alpha}}=1}^{d^2} $ in above equation, can be written as 
$
 \sum_{\alpha,{\tilde{\alpha}}=1}^{d^2}   = \sum_{\alpha = \tilde{\alpha} = d^2}+ \sum_{\alpha = 1, \tilde{\alpha} = d^2 }^{d^2-1}   + \sum_{{\tilde{\alpha}}=1, \alpha=d^2}^{d^2-1}  +\sum_{\alpha,{\tilde{\alpha}}=1}^{d^2-1}.   
$
Using this, and the fact that $F_{d^2} = I_S / \sqrt{d}$ commutes with all operators, we combine all commutator terms and write them as as $ i [\rho, H_{S} + H_{LS} ] $ to obtain
\begin{equation} \label{eq:model}
	\begin{aligned} 
			\frac{\partial {\rho}}{\partial t}&= i [\rho,H_S+H_{LS} ] + \hspace{-0.5em} \sum_{\alpha, {\tilde{\alpha}}=1}^{d^2-1} \Gamma_{\alpha {\tilde{\alpha}}} \Big( F_{\tilde{\alpha}} \rho F_\alpha^\dagger - \frac{\{ F_\alpha^\dagger F_{\tilde{\alpha}},\rho \}}{2}  \Big).
	\end{aligned}
\end{equation}
Here 
\begin{equation}
\begin{aligned}
& H_{LS} =  \epsilon^2 \sum_l \bigg\{ \\
& \sum_{\alpha,{\tilde{\alpha}}=1}^{d^2} \Big\{ \frac{a^*_{l \alpha } b _{l {\tilde{\alpha}}} }{2i}  - \frac{c^{\prime *}_{l \alpha } a^\prime _{l {\tilde{\alpha}}} }{2i} -  \frac{b^*_{l \alpha } a_{l {\tilde{\alpha}}}}{2i} + \frac{a^{\prime *}_{l \alpha } c^\prime_{l {\tilde{\alpha}}}}{2i} \Big\} F^\dagger_\alpha F_{\tilde{\alpha}} \\
&+  \sum_{\alpha=1}^{d^2-1} \frac{(a^*_{l \alpha} b_{l,d^2} + c^{\prime *}_{l \alpha} a^\prime_{l d^2} + b^*_{l \alpha} a_{l ,d^2} + a^{\prime *}_{l \alpha} c^\prime_{l,d^2} )}{2 i \sqrt{d} } F^\dagger_\alpha \\
&- \sum_{{\tilde{\alpha}}=1}^{d^2-1} \frac{( a^*_{ l  d^2} b_{l {\tilde{\alpha}} } +  c^{\prime *}_{l,d^2} a^\prime_{l {\tilde{\alpha}}} + b^{*}_{l,d^2} a_{l {\tilde{\alpha}}} + a^{\prime *}_{l , d^2} c^\prime_{l {\tilde{\alpha}}} )}{2 i \sqrt{d} } F_{\tilde{\alpha}} \bigg\}
\end{aligned}
\end{equation}
and 
\begin{equation}
	\Gamma_{\alpha {\tilde{\alpha}}}=\epsilon^2 \sum_l ( a^*_{l \alpha } b _{l {\tilde{\alpha}}} +c^{\prime *}_{l \alpha } a^\prime _{l {\tilde{\alpha}}} +b^*_{l \alpha } a_{l {\tilde{\alpha}}} +a^{\prime *}_{l \alpha } c^\prime_{l {\tilde{\alpha}}} ),
\end{equation}
$\alpha,\tilde{\alpha}$ going from $1$ to $d^2-1$. This is Eq.~\eqref{RE:lindblad} given in the main text.

\section{An example of RE violating complete positivity}
\label{appendixb}
In this section, we will present a simple example of the discussion in Sec.~\ref{sec:model_redfield}. Our setup consists of a two-qubit XXZ qubit chain, where only the first qubit is connected to the bath modelled by an infinite number of bosonic modes. Let $H$ be the Hamiltonian of the full set-up, given by 
\begin{equation}
\label{eq:hfull_app}
H=H_S + \epsilon \, H_{SB} + H_B, 
\end{equation}
where 
\begin{equation}
    \begin{aligned}
H_S &=  \frac{\omega_0}{2} (\sigma_z^{(1)} +\sigma_z^{(2)}) \\
&-   g(\sigma_x^{(1)} \sigma_x^{(2)} + \sigma_y^{(1)} \sigma_y^{(2)} + \Delta \sigma_z^{(1)} \sigma_z^{(2)} ) \\
 H_{SB}  &=\sum_{r=1}^\infty  (\kappa_{ r} \hat{B}^{ \dagger}_r \sigma^{(1)}_{-} + \kappa_{r}^* \hat{B}_r \sigma^{(1)}_+) \\
 H_B&= \sum_{r=1}^\infty \Omega_r \hat{B}^{ \dagger}_r \hat{B}_r 
    \end{aligned}
\end{equation}
where $\sigma^{(\ell)}_{x,y,z}$ denotes the Pauli matrices acting on the $\ell^{\text{th}}$ qubit, $\sigma^{(\ell)}_{+}=(\sigma^{(\ell)}_{x}+i \sigma^{(\ell)}_{y})/2$, $\sigma^{(\ell)}_{-}=(\sigma^{(\ell)}_{x}-i \sigma^{(\ell)}_{y})/2$, $\hat{B}_r$ is bosonic annihilation operator for the $r^{\text{th}}$ mode of the bath. Here, $\omega_0$, $g$, and $g\Delta$ represent the magnetic field, the overall qubit-qubit coupling strength and the anisotropy respectively. 
The RE for this setup can be computed to be \cite{Tupkary_2022}
\begin{equation}
	\begin{aligned}
		\frac{\partial {\rho}}{\partial t}=i[ \rho(t),H_S]  +\epsilon^2   \big\{ &   [S^\dagger , S^{(1)} \rho(t) ] - [S^\dagger , \rho(t) S^{(2)}] \\
	    &+ \text{H.c} \big\}
	\end{aligned}
\end{equation}
with
\begin{equation} \label{eq:Soperators_app}
    \begin{aligned}
        S^\dagger &= \sigma_+^{(1)}, \qquad S = \sigma_-^{(1)} \\
        S^{(1)} &= \sum_{j, k = 1}^4 \ket{E_j} \bra{E_j} \sigma^{(1)}_-  \ket{E_k}\bra{E_k}   D(j,k), \\ 
        S^{(2)} &= \sum_{j,k = 1}^4 \ket{E_j} \bra{E_j} \sigma^{(1)}_-  \ket{E_k} \bra{E_k}   C(j,k)
    \end{aligned},
\end{equation}
and
\begin{equation}
\begin{aligned}
C(j,k) &= \frac{\mathfrak{J}(E_{k j}) n(E_{k j})}{2 } - i \mathcal{P} \int_{0}^{\infty} d \omega \frac{\mathfrak{J}(\omega) n (\omega)}{\omega-E_{k j}},  \\
D(j,k) &= \frac{ e^{\beta(E_{k j}-\mu_{\ell})} \mathfrak{J}(E_{k j}) n(E_{k j})}{2 }\\ &- i \mathcal{P} \int_{0}^{\infty} d \omega \frac{e^{\beta(\omega-\mu)} \mathfrak{J}(\omega) n(\omega)}{\omega-E_{k j}}, \\
\mathfrak{J}(\omega) &=  \sum_{k=1}^{\infty} 2 \pi \left| \kappa_{ k} \right|^2 \delta (\omega-\Omega_k), \\
n(\omega) &=[e^{\beta\omega}- 1]^{-1}.
\label{redfield:constants}
\end{aligned}
\end{equation}
In above, $\mathfrak{J}(\omega)$ is called the bath spectral function. Let us consider bosonic baths described by Ohmic spectral functions with Gaussian cut-offs, $ \mathfrak{J}(\omega)=\omega e^{-(\omega / \omega_c)^2}\Theta(\omega),$ where $\Theta(\omega)$ is the Heaviside step function, and $\omega_c$ is the cut-off frequency.  The above operators can then be computed numerically.

The next step is to choose the basis $f_i$ and $g_j$ for operators on $\mathcal{H}_L$ and $\mathcal{H}_M$. For the general case, one can start with any set of linearly independent operators that forms a basis and includes the identity operator, and then apply the Gram Schmidt orthonormalization procedure to produce an orthonormal basis that includes the normalized identity operator. For our case, one can easily verify that the set $\{ -\sigma^{(i)}_z /\sqrt{2}, \sigma^{(i)}_-, \sigma^{(i)}_+, I^{(i)}_2 / \sqrt{2} \} $ suffices, where $i=1$ for $\{ f_i \}$ and $i=2$ for $\{ g_j\}$, and $I_2$ is the identity operator.

The basis for the full system $\{F_i\}$ can be constructed from the above basis as described in subsection ~\ref{subsec:operator}, and is given by
\begin{equation}
    \begin{aligned}
        F_i &= f_i \otimes \frac{I_2^{(2)}}{2} \qquad \text{(for $i = 1,2,3$)} \\
        F_{3i+j} &= f_i \otimes g_j \qquad \text{(for $i=1,2,3,4$ and $j=1,2,3$)} \\
        F_{16} &= \frac{I_4}{2}
    \end{aligned}
\end{equation}
Any operator $X$ can be expanded in terms of the above basis as $X=\sum_\alpha x_\alpha F_\alpha$,  where $x_\alpha = \braket{F_\alpha,X} = \text{Tr}(F^\dagger_\alpha X)$. Thus, expanding $S,S^\dagger, S^{(1)}, S^{(2)}$ [Eq.~\eqref{eq:Soperators_app}], one can evaluate all the coefficients in Eq.~\eqref{eq:S_expansions}. Finally, one can compute the matrix $\Gamma$ according to Eq.~\eqref{eq:gamma_re}. The matrix $\Gamma$ for this example, with parameters chosen as $g= 0.1, \omega_0 = 1, \omega_c = 10,  \beta=1, \mu=-0.5, \Delta = 1$ is given by

\begin{widetext}
\setcounter{MaxMatrixCols}{20}
\begin{equation} 
 \Gamma = \epsilon^2 
\begin{bmatrix}
        0 & 0 & 0 & 0 & 0 & 0 & 0 & 0 & 0 & 0 & 0 & 0 & 0 & 0 & 0 \\ 
        0 & 0 &1.542+3.428i & 0 & 0 & 0.014+0.047i & 0 & 0 & 0 & 0 & 0 & 0 & 0 & 0 & 0 \\ 
        0 & 1.542-3.428i & 0 & 0 & -0.18-0.007i & 0 & 0.18+0.007i & 0 & 0 & 0 & 0 & 0 & 0 & 0 & 0 \\ 
        0 & 0 & 0 & 0 & 0 & 0 & 0 & 0 & 0 & 0 & 0 & 0 & 0 & 0 & 0 \\
        0 & 0 & -0.18+0.007i & 0 & 0 & 0 & 0 & 0 & 0 & 0 & 0 & 0 & 0 & 0 & 0 \\ 
        0 & 0.014-0.047i & 0 & 0 & 0 & 0 & 0 & 0 & 0 & 0 & 0 & 0 & 0 & 0 & 0 \\ 
        0 & 0 & 0.18-0.007i & 0 & 0 & 0 & 0 & 0 & 0 & 0 & 0 & 0 & 0 & 0 & 0 \\
        0 & 0 & 0 & 0 & 0 & 0 & 0 & 0 & 0 & 0 & 0 & 0 & 0 & 0 & 0 \\ 
        0 & 0 & 0 & 0 & 0 & 0 & 0 & 0 & 0 & 0 & 0 & 0 & 0 & 0 & 0 \\ 
        0 & 0 & 0 & 0 & 0 & 0 & 0 & 0 & 0 & 0 & 0 & 0 & 0 & 0 & 0 \\ 
        0 & 0 & 0 & 0 & 0 & 0 & 0 & 0 & 0 & 0 & 0 & 0 & 0 & 0 & 0 \\ 
        0 & 0 & 0 & 0 & 0 & 0 & 0 & 0 & 0 & 0 & 0 & 0 & 0 & 0 & 0 \\ 
        0 & 0 & 0 & 0 & 0 & 0 & 0 & 0 & 0 & 0 & 0 & 0 & 0 & 0 & 0 \\ 
        0 & 0 & 0 & 0 & 0 & 0 & 0 & 0 & 0 & 0 & 0 & 0 & 0 & 0 & 0 \\ 
        0 & 0 & 0 & 0 & 0 & 0 & 0 & 0 & 0 & 0 & 0 & 0 & 0 & 0 & 0 
  \end{bmatrix}
 \end{equation}
\end{widetext}

We see that the above matrix has the expected structure of Eq.~\eqref{gamma_struc},
\begin{equation} 
	\Gamma=\left[ \begin{array}{ c | c  }
		\Gamma_{\alpha,{\tilde{\alpha}} < 4} & \Gamma_{\alpha < 4,{\tilde{\alpha}} \geq 4}\\ 
		\hline
		\Gamma_{\alpha \geq 4, {\tilde{\alpha}} < 4} & 0 \\
	\end{array} \right]
\end{equation}
and crucially, $ \Gamma_{\alpha < 4,{\tilde{\alpha}} \geq 4} \neq 0$. Therefore, as per the GKSL theorem, this RE will not preserve complete  positivity. This example was computed using QuTiP \cite{qutip_1,qutip_2}.

\section{Effective Lindblad equation satisfying local conservation laws}
\label{appendixc}
In this appendix, we will show that Eq.~\eqref{cont_conjecture_2} implies Eq.~\eqref{eq:rigorous_condition}. The condition for a QME preserving complete positivity and obeying local conservation laws is given by Eq.~\eqref{cont_conjecture_2}, which we recall to be
\begin{equation}
	\begin{aligned}\label{eq:cont_conjecture_3}
	& -i  [O_M,H^\prime] + \sum_{\alpha_M, {\tilde{\alpha}}_M=1}^{d_M^2-1} \widetilde{\Lambda}_{\alpha_M,{\tilde{\alpha}}_M} 
		 \Big(  g_{\alpha_M}^\dagger    O_M   g_{{\tilde{\alpha}}_M} \\
		 &- \frac{1}{2}  O_M  g_{\alpha_M}^\dagger g_{{\tilde{\alpha}}_M}  - \frac{1}{2}  g_{\alpha_M}^\dagger  g_{{\tilde{\alpha}}_M}   O_M  \Big)=0, \qquad \forall O_M,
	\end{aligned}
\end{equation}
where we write
\begin{equation}
H^\prime = \sum_{\alpha_M=1}^{d_M^2} \nu_{d_L^2, \alpha_M} g_{\alpha_M},
\end{equation}
for convenience. To move forward we will use the operator vector correspondence from Ref.~\onlinecite{watrous_2018}, where the vectorized version of the operator $X$ is given by $\superket{X}$, and can be constructed using linearity and
\begin{equation} \label{eq:vec_def_app}
\superket{\ket{i} \bra{j}} = \ket{i} \otimes \ket{j}^*,    
\end{equation}
where $\ket{j}^*$ denotes the complex conjugate of $\ket{j}$. We will apply Eq.~\eqref{eq:vec_def_app} to Eq.~\eqref{eq:cont_conjecture_3}, using the identity (Eq. 1.132 of Ref.~\onlinecite{watrous_2018})
\begin{equation}
    \label{eq:vec_identity}
    \superket{A_0 B A_1^T} = (A_0 \otimes A_1) \superket{B}.
\end{equation}

Applying the vec operation on both sides of Eq.~\eqref{eq:cont_conjecture_3}, we obtain

\begin{equation} \label{eq:vec1_app}
	\begin{aligned} 
		&-i\, \superket{I_M O_M H^\prime } +i\, \superket{ H^\prime O_M I_M} + \squash \squash \sum_{\alpha_M, {\tilde{\alpha}}_M=1}^{d_M^2-1}  \squash \widetilde{\Lambda}_{\alpha_M,{\tilde{\alpha}}_M} 	\\
		&\Big(  \superket{ g_{\alpha_M}^\dagger   O_M g_{{\tilde{\alpha}}_M}  } - \frac{1}{2} \superket{I_M O_M  g_{\alpha_M}^\dagger g_{{\tilde{\alpha}}_M} }  \\
		&-\frac{1}{2}  \superket{ g_{\alpha_M}^\dagger  g_{{\tilde{\alpha}}_M}   O_M I_M}  \Big)=0,	\qquad	\forall O_M.
	\end{aligned}
\end{equation}
Eq.~\eqref{eq:vec1_app} can be simplified using Eq.~\eqref{eq:vec_identity} to obtain, 

\begin{equation} \label{eq:vec2_app}
	\begin{aligned}
		& \bigg\{ -i I_M \otimes (H^{\prime} )^T + i H^\prime \otimes I_M    \sum_{\alpha_M, {\tilde{\alpha}}_M=1}^{d_M^2-1} \widetilde{\Lambda}_{\alpha_M,{\tilde{\alpha}}_M}     \\
		& \Big( g^\dagger_{\alpha_M} \otimes g^T_{{\tilde{\alpha}}_M} - \frac{1}{2} I_M \otimes (g^\dagger_{\alpha_M} g_{{\tilde{\alpha}}_M})^T \\
		&- \frac{1}{2}  (g^\dagger_{\alpha_M} g_{{\tilde{\alpha}}_M}) \otimes I_M \Big)
		\bigg\} \superket{O_M} = 0, \qquad 	\forall O_M.
	\end{aligned}
\end{equation}
Eq.~\eqref{eq:vec2_app} is of the form $\mathcal{M} \, \superket{O_M} = 0$ for all hermitian $O_M$. Since hermitian matrices (such as $O_M$) form a basis for the entire space of operators, this implies $\mathcal{M} \, \superket{X} = 0$ for all operators $X$. This is because one can expand $X$ as a linear combination of hermitian operators (such as $O_M$).
Now, $\mathcal{M} \, \superket{X} = 0 \, \forall \, X$ implies $\mathcal{M}=0$. Therefore, Eq.~\eqref{eq:vec2_app} implies
\begin{equation} \label{eq:vec3_app}
	\begin{aligned}
		\mathcal{M} = & -i I_M \otimes (H^{\prime} )^T +i H^{\prime} \otimes I_M   + \sum_{\alpha_M, {\tilde{\alpha}}_M=1}^{d_M^2-1} \widetilde{\Lambda}_{\alpha_M,{\tilde{\alpha}}_M}     \\
		& \Big( g^\dagger_{\alpha_M} \otimes g^T_{{\tilde{\alpha}}_M} - \frac{1}{2} I_M \otimes (g^\dagger_{\alpha_M} g_{{\tilde{\alpha}}_M})^T \\
		&- \frac{1}{2}  (g^\dagger_{\alpha_M} g_{{\tilde{\alpha}}_M}) \otimes I_M \Big)
		 = 0.
	\end{aligned}
\end{equation}
If $\mathcal{M} =0$, then $\text{Tr} (\mathcal{M}) = 0$. Taking the trace of Eq.~\eqref{eq:vec3_app}, and using the orthonormality of $\{ g_i \}$ along with the fact that $\text{Tr}(g_i) = \delta_{i, d^2_M} $, we obtain
\begin{equation}
		\text{Tr} (\mathcal{M}) =- \sum_{\alpha_M = 1}^{d_M^2-1} \widetilde{\Lambda}_{\alpha_M,\alpha_M}     d_M = 0
\end{equation}
which implies 
\begin{equation}
    \begin{aligned}
     \sum_{\alpha_M = 1}^{d_M^2-1} \widetilde{\Lambda}_{\alpha_M,\alpha_M} = \sum_{\alpha_M = 1}^{d_M^2-1} \sum_{\alpha_L=1}^{d_L^2} \widetilde{\Gamma}_{(\alpha_L, \alpha_M), (\alpha_L,\alpha_M)} = 0.
    \end{aligned}
\end{equation}
which is Eq.~\eqref{eq:rigorous_condition} in the main text.
 
\section{The condition for thermalization}
\label{appendix_thermal}
From fundamental principles of quantum statistical mechanics, we expect the system to thermalize when coupled to baths at equal temperatures. The exact condition that QME's must obey to satisfy thermalization has been derived in Ref.~\onlinecite{Tupkary_2022}. For the sake of completeness, we recall that discussion here. 

Let the total system Hamiltonian be given by $H =  H_S + \epsilon H_{SB} + H_{B}$, where $\epsilon$ is a dimensionless parameter controlling the strength of the system bath coupling, and $H_{SB}$ is the system bath coupling Hamiltonian. We proceed by obtaining an order-by-order solution to the steady state of our QME. Any QME describing our setup can be expanded in the so-called time-convolution-less form \cite{breuer_book}, 
\begin{equation}\label{eq:tcl}
    	\frac{\partial {\rho} (t)}{\partial t} = \sum_{m=0}^\infty \epsilon^{2m} \mathcal{L}_{2m} (t) [\rho(t)],
\end{equation}
where $\mathcal{L}$ could in general be time-dependent operators and $L_0(t) [\rho(t)] = i[\rho(t), H_S]$. For quantum master equations written to second-order in system-bath coupling, the above summation can be truncated at second order. Denoting $\mathcal{L}_{2m} \equiv \lim_{t\rightarrow \infty} \mathcal{L}_{2m}(t)$, the steady state $\rho_{SS}$ can be given by 
\begin{equation}
    \rho_{SS}=\lim_{t \to \infty} e^{t (\mathcal{L}_0 + \epsilon^2 \mathcal{L}_2)} \rho(0),
\end{equation}
which is assumed to be unique. The steady state satisfies
\begin{equation} \label{eq:steadystate1}
    0 = \sum_{m=0}^\infty\epsilon^{2m} \mathcal{L}_{2m} [\rho_{SS}].
\end{equation}
We can then perform an expansion of $\rho_{SS}$ in the even powers of $\epsilon$ as 
\begin{equation} \label{eq:rho_expansion}
    \rho_{SS} = \sum_{m=0}^\infty\epsilon^{2m} \rho^{(2m)}_{SS}
\end{equation}
Using Eq.~\eqref{eq:rho_expansion} in Eq.~\eqref{eq:steadystate1}, we can obtain an order by order solution of $\rho_{SS}$.
At the zeroth order in $\epsilon$, we obtain 
\begin{equation} \label{eq:zeroth_order_app}
[\rho^{(0)}_{SS}, H_S]=0.
\end{equation}
Assuming that the Hamiltonian has no degeneracies, Eq.~\eqref{eq:zeroth_order_app} implies that $\rho^{(0)}_{SS}$ is diagonal in the energy eigenbasis,
\begin{equation}
    \rho^{(0)}_{SS} = \sum_i p_i \ket{E_i} \bra{E_i}.
\end{equation}
where $\ket{E_i}$ is an eigenstate of the system. At second order in $\epsilon$ ($m=1$), we obtain the following two equations,
\begin{equation} \label{eq:order21}
    \bra{E_i} \mathcal{L}_2 [\rho^{(0)}_{SS} ] \ket{E_i} = 0, \quad \forall i
    \end{equation}
    \begin{equation} \label{eq:order22}
    \begin{aligned}
    i (E_i-E_j) \bra{E_i} \rho^{(2)}_{SS} \ket{E_j} & \\
    +\epsilon^2\bra{E_i} \mathcal{L}_2 [\rho^{(0)}_{SS} ] \ket{E_j} &= 0, \quad \forall i \neq j
    \end{aligned}
\end{equation}
Since $\rho^{(0)}_{SS}$ is diagonal in the energy eigenbasis, Eq.~\eqref{eq:order21} determines the diagonal elements of $\rho^{(0)}_{SS}$. Having obtained $\rho^{(0)}_{SS}$, Eq.~\eqref{eq:order22} then determines the off-diagonal elements of $\rho^{(2)}_{SS}$. Note from above equations that the leading order diagonal elements of $\rho_{SS}$ are independent of $\epsilon$. It can also be shown that the leading order off-diagonal elements of $\rho_{SS}$ in the energy eigenbasis of the system scale as $\epsilon^2$.
As discussed in the main text, the QME thermalizes if 
\begin{equation} \label{eq:rhoss_app}
    \lim_{\epsilon\to 0}\rho_{SS} = \rho_{th} 
\end{equation}
where $\rho_{th}$ is the Gibbs state of the system given by
\begin{equation} 
 \rho_{th}  = \frac{e^{-\beta H_S}}{\text{Tr} [e^{-\beta H_S}]}.
\end{equation}
We then conclude that the thermalization in this sense is a statement about leading order diagonal elements of $\rho_{SS}$.
Substituting Eq.~\eqref{eq:rhoss_app} in Eq.~\eqref{eq:order21}, we obtain the following condition on $\mathcal{L}_2$ for the system to thermalize,  
\begin{equation} \label{eq:thermal_cond_app}
    \bra{E_i} \mathcal{L}_2 [\rho_{th} ] \ket{E_i} = 0  \quad \forall i.
\end{equation}


\section{Semidefinite Programming (SDP)}
\subsection{ Basic Theory} \label{subsec:sdp}

In this section, we present the theoretical framework of semidefinite programming (SDP). We follow the definition of SDPs given in page 57 of Ref.~\onlinecite{watrous_2018}. In what follows, we will use $\Phi$ and $\Psi$ to denote hermitian preserving linear maps. We will also use $\Phi^\dagger$ to denote the ``adjoint map'' \cite{watrous_2018}, which is defined as the unique linear map that satisfies
\begin{equation} \label{eq:dagger_map}
    \braket{A,\Phi(B)} = \braket{\Phi^\dagger(A),B}
\end{equation}
where 
\begin{equation}
\label{eq_hs_app}
\braket{A,B} = \text{Tr}(A^\dagger B)
\end{equation}
denotes the Hilbert Schmidt inner product. 

An SDP is defined by the tuple ($\Phi$, $\Psi$, $A$, $B$, $C$), where $\Phi, \Psi$ are hermitian-preserving linear maps, and $A,B,C$ are hermitian operators. The ``primal'' problem of the SDP is given by
\begin{equation} \label{eq:SDP_definition_primal}
    \begin{aligned} 
    \text{maximize}: \quad & \braket{A,X} \quad w.r.t. \hspace{0.5em}  X \\
    \text{subject to}:  \quad  &\Phi(X) = B,~\Psi(X) \leq C,~X \geq 0,
    \end{aligned}
\end{equation}
where the inequalities represent matrix inequalities. I.e, $A \geq B$ is equivalent to $A-B \geq 0$ and implies $A-B$ is positive semidefinite. We will use the notation $X_{f}$ to denote any ``feasible'' value of $X$ that satisfies the three constraints in Eq.~\eqref{eq:SDP_definition_primal}, and $\mathcal{P}$ to denote the maximum value of $\braket{A,X}$ attained in Eq.~\eqref{eq:SDP_definition_primal} (assuming there is atleast one $X$ which satisfies constraints). 

For every ``primal'' problem, there exists a ``dual'' problem given by
\begin{equation} \label{eq:SDP_definition_dual}
    \begin{aligned} 
    \text{minimize}: \quad & \braket{B,Y} + \braket{C,Z} \quad w.r.t. \hspace{0.5em} Y, Z \\
    \text{subject to}:  \quad  &\Phi^\dagger(Y) + \Psi^\dagger(Z) \geq A, \\
    & Y \text{is hermitian},  Z \geq 0.
    \end{aligned}
\end{equation}
We will use the notation $(Y_{f}, Z_{f})$ to denote any ``feasible'' value of $Y,Z$ that satisfies the three constraints in Eq.~\eqref{eq:SDP_definition_dual}, and $\mathcal{D}$ to denote the minimum value of $\braket{B,Y} + \braket{C,Z}$ attained in Eqs.~\eqref{eq:SDP_definition_dual} (assuming atleast some $(Y,Z)$ satisfies constraints). 

Semidefinite programs have a notion of \textit{duality} associated with them, which relates properties of the primal and the dual problems. In particular, it can be shown that 
\begin{equation} \label{eq:weak_duality}
    \mathcal{P} \leq \mathcal{D}
\end{equation}
a property known as ``weak duality''.  In most situations, it can be shown that $\mathcal{P} = \mathcal{D}$, i.e, equality holds in Eq.~\eqref{eq:weak_duality}. This condition is known as ``strong duality''. 

\begin{figure}
    \centering
    \includegraphics[width =0.8\linewidth]{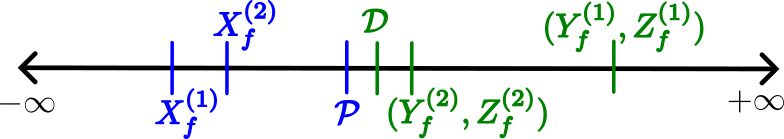}
    \caption{Schematic representing weak duality for the SDP given in Eqs.~\eqref{eq:SDP_definition_primal} and \eqref{eq:SDP_definition_dual}, according to Eq.~\eqref{eq:sdp_solution}. $X_f^{(j)}$ and $(Y_f^{(j)},Z_f^{(j)})$ represents any feasible input to the primal and dual problems respectively. Any such inputs yield upper and lower bounds on the solutions of the primal and dual problems. } 
    \label{fig_scheme1}
    \end{figure}

By weak duality and the definition of our primal and dual problems, using Eq.~\eqref{eq:SDP_definition_primal},\eqref{eq:SDP_definition_dual}, and \eqref{eq:weak_duality}, we obtain
\begin{equation} \label{eq:sdp_solution}
\braket{A, X_f} \leq \mathcal{P} \leq \mathcal{D} \leq \braket{B,Y_f} + \braket{C,Z_f}.
\end{equation}
From Eq.~\eqref{eq:sdp_solution}, any feasible choice of inputs to the primal and dual problem ($X_f,Y_f,Z_f$) leads to lower and upper bounds on the optimal values of the primal and dual problems [see Fig.~\eqref{fig_scheme1}].  In particular, if we obtain $\braket{A, X_f} =   \braket{B,Y_f} + \braket{C,Z_f}$, equality holds throughout in Eq.~\eqref{eq:sdp_solution}. This property can therefore be exploited to obtain exact solutions to the primal problem of an SDP. 
    
We will show in Sec.~\ref{subsec:reduction} that the thermal optimization problem (TOP) in Eq.~\eqref{eq:thermal_optimization} can be reduced to the standard form of SDP [Eq.~\eqref{eq:SDP_definition_primal}]. 

\subsection{ Reducing the thermal optimization problem (TOP) to standard form } \label{subsec:reduction}
\begin{figure}
    \centering
    \includegraphics[width =0.8\linewidth]{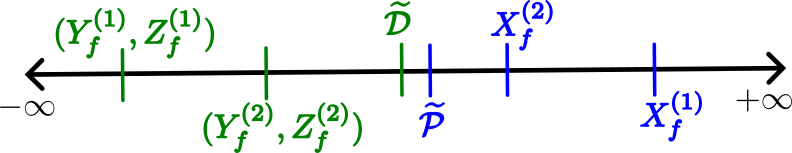}
    \caption{Schematic representing weak duality for the SDP given in Eqs.~\eqref{eq:SDP_definition_primal_v2} and \eqref{eq:SDP_definition_dual_v2}, according to Eq.~\eqref{eq:sdp_solution_new}. $X_f^{(j)}$ and $(Y_f^{(j)},Z_f^{(j)})$ represents any feasible input to the primal and dual problems respectively. Any such inputs yield upper and lower bounds on the solutions of the primal and dual problems. } 
    \label{fig_scheme2}
    \end{figure}

 Recall that the TOP was given by [Eq.\eqref{eq:thermal_optimization}]
\begin{equation} \label{eq:thermal_optimization_app}
    \begin{aligned}
    \text{minimize : }  &\tau \\
    \text{subject to : } & H^{(L)}_{LS} \text{ is hermitian,}~\text{Tr}(\Gamma^{(L)}) = 1, ~\Gamma^{(L)} \geq 0. \\
    \end{aligned}
\end{equation}
See Eq.~\eqref{eq:thermal_condition_tau} for definition of $\tau$ and Eq.~\eqref{eq:local_lindblad_for_TOP} for definition of $H^{(L)}_{LS}$, $\Gamma^{(L)}$.
In this subsection, we will show how the TOP from Eq.~\eqref{eq:thermal_optimization_app} can be reduced to the standard form of an SDP. We note that the standard form of SDP in Eq.~\eqref{eq:SDP_definition_primal} is not yet suitable for this purpose. Therefore, we replace $A \rightarrow -A,  C \rightarrow -C, \Psi \rightarrow -\Psi, Y \rightarrow -Y, $ in Eq.~\eqref{eq:SDP_definition_primal} and Eq.~\eqref{eq:SDP_definition_dual}, leaving $\Phi$, $B$, $X$ and $Z$ unchanged. Since maximizing any function is the same as minimizing its negative, we obtain the new ``primal'' form as 
\begin{equation} \label{eq:SDP_definition_primal_v2}
    \begin{aligned} 
    \text{minimize}: \quad & \braket{A,X} \\
    \text{subject to}:  \quad  &\Phi(X) = B,~ \Psi(X) \geq C,~X \geq 0,
    \end{aligned}
\end{equation}
where we use $\widetilde{P}$ to denote the minimum value of $\braket{A,X}$ obtained in Eq.~\eqref{eq:SDP_definition_primal_v2}. The new ``dual'' form is written as 
\begin{equation} \label{eq:SDP_definition_dual_v2}
    \begin{aligned} 
    \text{maximize}: \quad & \braket{B,Y} + \braket{C,Z} \\
    \text{subject to}:  \quad  &\Phi^\dagger(Y) + \Psi^\dagger(Z) \leq A, \\
    & Y \text{is hermitian},~ Z \geq 0,
    \end{aligned}
\end{equation}
where we use $\widetilde{D}$ to denote the minimum value of $\braket{B,Y}+\braket{C,Z}$ obtained in Eq.~\eqref{eq:SDP_definition_dual_v2}. Eq.~\eqref{eq:sdp_solution} is then transformed into [see Fig~\eqref{fig_scheme2}], 
\begin{equation} \label{eq:sdp_solution_new}
\braket{A, X_f} \geq \widetilde{\mathcal{P}} \geq \widetilde{\mathcal{D}} \geq \braket{B,Y_f} + \braket{C,Z_f}.
\end{equation}

We will now show how to reduce the TOP from Eq.~\eqref{eq:thermal_optimization_app} to Eq.~\eqref{eq:SDP_definition_primal_v2}, via a series of changes to the optimization problem in Eq.~\eqref{eq:SDP_definition_primal_v2}. We do so in three steps. 

\textit{Step 1}: Our first step is to write down a primal optimization problem whose solution (minimum value attained) is equal to 
\begin{equation}
\label{eq:trkdk}
||K ||_1 = \text{Tr}(\sqrt{K^\dagger K}).
\end{equation}
for any hermitian matrix $K$. Let $\Pi_p$ and $\Pi_n$ be projectors onto the positive and negative eigenspaces of $K$. In this case, 
\begin{equation}
||K ||_1 = \text{Tr} (\Pi_p K \Pi_p) - \text{Tr}(\Pi_n K \Pi_n).    
\end{equation} 
Let us now consider the optimization problem given by,
\begin{equation} \label{eq:SDP_primal1}
    \begin{aligned} 
    \text{minimize}: \quad & \left\langle \begin{pmatrix} I & 0 \\ 0 & I \end{pmatrix} ,\begin{pmatrix} P & . \\ . & Q \end{pmatrix} \right\rangle \\
    \text{subject to}: \quad & \Psi_1 \begin{pmatrix} P & . \\ . & Q \end{pmatrix}  = \begin{pmatrix} P & 0 \\ 0 & Q \end{pmatrix}  \geq \begin{pmatrix} K & 0 \\ 0 & -K \end{pmatrix} , \\
    & \begin{pmatrix} P & . \\ . & Q \end{pmatrix} \geq 0,
    \end{aligned}
\end{equation}
where we use dots to represent arbitrary blocks of the matrices which can always be set to zero without affecting the objective function or constraints.  Note that $\Psi_1$ in Eq.~\eqref{eq:SDP_primal1} is a map that replaces the off-diagonal blocks with null matrices, leaving the diagonal blocks unchanged. It is easy to see that Eq.~\eqref{eq:SDP_primal1} is of the form Eq.~\eqref{eq:SDP_definition_primal_v2} (with $\Phi$ and $B$ omitted i.e., no equality constraint).  Thus Eq.~\eqref{eq:SDP_primal1} is an SDP.

The dual problem to the primal problem in Eq.~\eqref{eq:SDP_primal1} is  given by
\begin{equation} \label{eq:SDP_dual1}
    \begin{aligned} 
    \text{maximize}: \quad &  \left\langle \begin{pmatrix} K & 0 \\ 0 & -K \end{pmatrix} ,  \begin{pmatrix} \bar{P} & . \\ . & \bar{Q} \end{pmatrix} \right\rangle  \\
    \text{subject to}:  \quad  &\Psi_1^\dagger  \begin{pmatrix} \bar{P} & . \\ . & \bar{Q} \end{pmatrix}  =  \begin{pmatrix} \bar{P} & 0 \\ 0 & \bar{Q} \end{pmatrix}  \leq  \begin{pmatrix} I & 0 \\ 0 & I \end{pmatrix} \\
    & \begin{pmatrix} \bar{P} & . \\ . & \bar{Q} \end{pmatrix} \geq 0.
    \end{aligned}
\end{equation}
where $\Psi_1^\dagger$ turns out to be the same map as $\Psi_1$ using Eq.~\eqref{eq:dagger_map}. It can be seen that 
Eq.~\eqref{eq:SDP_dual1} is of the form Eq.~\eqref{eq:SDP_definition_dual_v2} (again with $\Phi$ and $B$ omitted).

We will now show that the optimal values attained in both the primal and dual problems in Eqs.~\eqref{eq:SDP_primal1} and \eqref{eq:SDP_dual1} is equal to $||K||_1$.  To show this, note that setting
\begin{equation} \label{eq:PQ_sub_app}
    P_f =\Pi_p K \Pi_p, \quad Q_f = - \Pi_n K \Pi_n 
\end{equation}
(where $P_f$  and $Q_f$ denote `feasible' choices of $P$ and $Q$ respectively) in Eq.~\eqref{eq:SDP_primal1}
allows us to obtain $ ||K ||_1 $ in the primal objective function. Furthermore, setting
\begin{equation}
    \bar{P}_f = \Pi_p, \quad \bar{Q}_f = \Pi_n
    \end{equation}
in Eq.~\eqref{eq:SDP_dual1} allows us to obtain $ ||K ||_1 $ in the dual objective function. 
Thus, we have explicitly constructed feasible choices of inputs to the primal and dual problems of Eqs.~\eqref{eq:SDP_primal1} and \eqref{eq:SDP_dual1} respectively, that yield $||K ||_1 $ in the primal and dual objective functions. Therefore, according to Eq.~\eqref{eq:sdp_solution_new}, the optimal values attained in the primal and dual problems are both exactly equal to $||K ||_1 $.

\textit{Step 2}: In the first step we constructed an SDP whose solution is equal to $ ||K ||_1 $, for a fixed $K$. We will now construct an SDP which computes the minimum value of  $ ||K ||_1 $, subject to some constraints on $K$. We will first recast  the problem in Eq.~\eqref{eq:SDP_primal1} as 
\begin{equation} \label{eq:SDP_primal2}
    \begin{aligned} 
    \text{minimize}: \quad & \text{Tr}(P) + \text{Tr}(Q) \\
    \text{subject to}: \quad & P \geq K,~ Q \geq -K,~ P, Q \geq 0.
\end{aligned}
\end{equation}
Eq.~\eqref{eq:SDP_primal1} computes 
\begin{equation}
  ||K||_1 =   \sum_i | K_{ii}|. 
\end{equation}
when $K$ is diagonal. Let $\mathcal{G}$ be a linear, hermitian preserving map that always outputs a diagonal matrix. Then, the optimization problem given by  
\begin{equation} \label{eq:SDP_primal3}
    \begin{aligned} 
    \text{minimize}: \quad & \text{Tr}(P) + \text{Tr}(Q) \\
    \text{subject to}: \quad & P \geq \mathcal{G}(R),~ Q \geq -\mathcal{G}(R), \\
                       & \Phi(R) = B,~~ P, Q, R \geq 0,
\end{aligned}
\end{equation}
computes $\min_{R \geq 0, \Phi(R) = B} \sum_i | \mathcal{G}(R)_{ii} |$.
We will now begin to connect the above formalism to TOP from Eq.~\eqref{eq:thermal_optimization_app}. We will show how $R$ can be chosen to reflect the optimization over $H^{(L)}_{LS}$ and $\Gamma^{(L)}$, and $\Phi$ can be chosen to reflect the trace constraint on $\Gamma^{(L)}$. 
Then, we will specify a map $\mathcal{G}$ that takes $R$ as input (i.e, $H^{(L)}_{LS}$ and $\Gamma^{(L)}_{LS}$), and outputs a diagonal matrix such that the objective function computes 
\begin{equation} 
    \tau = \sum_i \left\vert \bra{E_i}  \mathcal{L}_2 (\rho_{th} ) \ket{E_i}\right\vert.
\end{equation}

\textit{Step 3}: In the thermal optimization problem [Eq.~\eqref{eq:thermal_optimization_app}], we have an optimization over $\Gamma^{(L)} \geq 0$, and hermitian $H^{(L)}_{LS}$. We use the fact that any hermitian matrix $H^{(L)}_{LS}$ can be written as a $H^{(L)}_{LS} =  S- T$, where $S,T \geq 0$. Moreover $S-T$ for any $S,T \geq 0$ is always hermitian. We will now replace the hermitian $H^{(L)}_{LS}$  with the difference of positive matrices $S-T$. This is needed, since semidefinite programs can only handle optimization over positive semidefinite variables. Furthermore, let us identify $\Gamma^{(L)}$ with some matrix $U$. Now consider the map $\mathcal{G}$ that acts as follows :
\begin{widetext}
\begin{equation} \label{eq:Gmap_definition_app}
    \mathcal{G} \begin{pmatrix} S & . & .\\ .& T & .\\ .& .& U \end{pmatrix} \equiv \mathcal{G}(S,T,U) \equiv \begin{pmatrix}  \bra{E_1} \mathcal{L}_2 (\rho_{th} ) \ket{E_1} & 0  & \hdots & 0 \\
    0 & \bra{E_2} \mathcal{L}_2 (\rho_{th} ) \ket{E_2} & \hdots & 0 \\
    \vdots & \hdots  & \ddots & \vdots \\
    0 & \hdots & 0 & \bra{E_d} \mathcal{L}_2 (\rho_{th} ) \ket{E_d} ,
    \end{pmatrix}
\end{equation}
\end{widetext}
where the map constructs $\mathcal{L}_2$ according to Eq.~\eqref{eq:local_lindblad_for_TOP} after setting $H^{(L)}_{LS} = S-T$, and $\Gamma^{(L)} = U$. Now, we consider a specific case of the optimization problem in Eq.~\eqref{eq:SDP_primal3}, for the choice of $\mathcal{G}$ in Eq.~\eqref{eq:Gmap_definition_app}. We obtain,
\begin{equation} \label{eq:SDP_primal4}
    \begin{aligned} 
    \text{minimize}: \quad & \text{Tr}(P) + \text{Tr}(Q) \\
    \text{subject to}: \quad & P \geq \mathcal{G} \begin{pmatrix} S &. &. \\. & T &. \\. & .& U \end{pmatrix} , ~Q \geq -\mathcal{G} \begin{pmatrix} S &. &. \\ .& T &. \\ .& .& U \end{pmatrix}, \\
                        \Phi_1 \begin{pmatrix} S &. & .\\ .& T & .\\ .&. & U \end{pmatrix} & = \text{Tr}(U) = 1,  P, Q, S, T, U \geq 0.
\end{aligned}
\end{equation}
Since $\mathcal{G}$ always outputs a diagonal matrix [see Eq.~\eqref{eq:Gmap_definition_app}], Eq.~\eqref{eq:SDP_primal4} computes 
\begin{equation} \label{eq:SDP_primal5}
    \begin{aligned} 
    \min \quad &\sum_i \left\vert  \mathcal{G} \begin{pmatrix} S & & \\ & T & \\ & & U \end{pmatrix}_{ii} \right\vert \\
     \quad \text{subject to : }\quad & S,T,U \geq 0, \text{Tr}(U) = 1
    \end{aligned}
\end{equation}
Since $H^{(L)}_{LS}$ can always be written as $S-T$, and $\Gamma^{(L)}$  as $U$, Eq.~\eqref{eq:SDP_primal5}  [and therefore Eq.~\eqref{eq:SDP_primal4} ] is identical to the thermal optimization problem in Eq.~\eqref{eq:thermal_optimization_app}. Therefore, Eq.~\eqref{eq:SDP_primal4} computes $\tau_{opt}$.

All that remains is converting Eq.~\eqref{eq:SDP_primal4} to the standard form Eq.~\eqref{eq:SDP_definition_primal_v2}. 
 Eq.~\eqref{eq:SDP_primal4} can be obtained from
 Eq.~\eqref{eq:SDP_primal4} after choosing
\begin{widetext} 
\begin{equation}
    \begin{aligned}
    &A  = \begin{pmatrix} I & & & & \\
    & I & &  & \\
    & &0 & & \\
    & & & 0 & \\
    & & & & 0 \\
    \end{pmatrix}, \quad    B  = 1, \quad
    X  = \begin{pmatrix}P &. &. &. &. \\
    .& Q &. &.  &. \\
    .&. &S &. &. \\
    .& .& .& T &. \\
    .&. &. &. & U \\
    \end{pmatrix}, \quad     C  = 0, \\
    &\Psi \begin{pmatrix}P &. &. &. &. \\
    .& Q &. &.  &. \\
    .&. &S &. &. \\
    .& .& .& T &. \\
    .&. &. &. & U \\
    \end{pmatrix} = \begin{pmatrix} P-\mathcal{G}(S,T,U) & 0 \\
    0 & Q+\mathcal{G}(S,T,U) \\ 
    \end{pmatrix},  \quad 
    \Phi\begin{pmatrix}P &. &. &. &. \\
    .& Q &. &.  &. \\
    .&. &S &. &. \\
    .& .& .& T &. \\
    .&. &. &. & U \\
    \end{pmatrix}  = \text{Tr}(U)
    \end{aligned}
\end{equation}
\end{widetext}
Recall that it is helpful to think of $P,Q$ as variables needed to compute the objective  function $\tau$ [Eq.~\eqref{eq:thermal_optimization_app}], $S,T$ are variables that give rise to $H^{(L)}_{LS} =  S-T$, and $U$ is a variable that encodes $\Gamma^{(L)}$. We have therefore shown that the TOP [Eq.~\eqref{eq:thermal_optimization_app}] is an SDP. 

It is to be noted that it is not necessary to reduce the TOP to the standard form of an SDP in order to use CVX \cite{cvx}. Infact, the TOP from Eq.~\eqref{eq:thermal_optimization_app} can be directly implemented into CVX, which itself handles the construction of the dual problem automatically in the background.

\bibliography{bibliography}

\begin{thebibliography}{74}%
\makeatletter
\providecommand \@ifxundefined [1]{%
 \@ifx{#1\undefined}
}%
\providecommand \@ifnum [1]{%
 \ifnum #1\expandafter \@firstoftwo
 \else \expandafter \@secondoftwo
 \fi
}%
\providecommand \@ifx [1]{%
 \ifx #1\expandafter \@firstoftwo
 \else \expandafter \@secondoftwo
 \fi
}%
\providecommand \natexlab [1]{#1}%
\providecommand \enquote  [1]{``#1''}%
\providecommand \bibnamefont  [1]{#1}%
\providecommand \bibfnamefont [1]{#1}%
\providecommand \citenamefont [1]{#1}%
\providecommand \href@noop [0]{\@secondoftwo}%
\providecommand \href [0]{\begingroup \@sanitize@url \@href}%
\providecommand \@href[1]{\@@startlink{#1}\@@href}%
\providecommand \@@href[1]{\endgroup#1\@@endlink}%
\providecommand \@sanitize@url [0]{\catcode `\\12\catcode `\$12\catcode
  `\&12\catcode `\#12\catcode `\^12\catcode `\_12\catcode `\%12\relax}%
\providecommand \@@startlink[1]{}%
\providecommand \@@endlink[0]{}%
\providecommand \url  [0]{\begingroup\@sanitize@url \@url }%
\providecommand \@url [1]{\endgroup\@href {#1}{\urlprefix }}%
\providecommand \urlprefix  [0]{URL }%
\providecommand \Eprint [0]{\href }%
\providecommand \doibase [0]{http://dx.doi.org/}%
\providecommand \selectlanguage [0]{\@gobble}%
\providecommand \bibinfo  [0]{\@secondoftwo}%
\providecommand \bibfield  [0]{\@secondoftwo}%
\providecommand \translation [1]{[#1]}%
\providecommand \BibitemOpen [0]{}%
\providecommand \bibitemStop [0]{}%
\providecommand \bibitemNoStop [0]{.\EOS\space}%
\providecommand \EOS [0]{\spacefactor3000\relax}%
\providecommand \BibitemShut  [1]{\csname bibitem#1\endcsname}%
\let\auto@bib@innerbib\@empty
\bibitem [{\citenamefont {Binder}\ \emph {et~al.}(2018)\citenamefont {Binder},
  \citenamefont {Correa}, \citenamefont {Gogolin}, \citenamefont {Anders},\
  and\ \citenamefont {Adesso}}]{quantum_thermodynamics}%
  \BibitemOpen
  \bibinfo {editor} {\bibfnamefont {F.}~\bibnamefont {Binder}}, \bibinfo
  {editor} {\bibfnamefont {L.~A.}\ \bibnamefont {Correa}}, \bibinfo {editor}
  {\bibfnamefont {C.}~\bibnamefont {Gogolin}}, \bibinfo {editor} {\bibfnamefont
  {J.}~\bibnamefont {Anders}}, \ and\ \bibinfo {editor} {\bibfnamefont
  {G.}~\bibnamefont {Adesso}},\ eds.,\ \href@noop {} {\emph {\bibinfo {title}
  {Thermodynamics in the Quantum Regime}}}\ (\bibinfo  {publisher} {Springer,
  Cham},\ \bibinfo {year} {2018})\BibitemShut {NoStop}%
\bibitem [{\citenamefont {Hur}\ \emph {et~al.}(2016)\citenamefont {Hur},
  \citenamefont {Henriet}, \citenamefont {Petrescu}, \citenamefont {Plekhanov},
  \citenamefont {Roux},\ and\ \citenamefont {Schir{\'{o}}}}]{LeHur2016}%
  \BibitemOpen
  \bibfield  {author} {\bibinfo {author} {\bibfnamefont {K.~L.}\ \bibnamefont
  {Hur}}, \bibinfo {author} {\bibfnamefont {L.}~\bibnamefont {Henriet}},
  \bibinfo {author} {\bibfnamefont {A.}~\bibnamefont {Petrescu}}, \bibinfo
  {author} {\bibfnamefont {K.}~\bibnamefont {Plekhanov}}, \bibinfo {author}
  {\bibfnamefont {G.}~\bibnamefont {Roux}}, \ and\ \bibinfo {author}
  {\bibfnamefont {M.}~\bibnamefont {Schir{\'{o}}}},\ }\href {\doibase
  10.1016/j.crhy.2016.05.003} {\bibfield  {journal} {\bibinfo  {journal}
  {Comptes Rendus Physique}\ }\textbf {\bibinfo {volume} {17}},\ \bibinfo
  {pages} {808} (\bibinfo {year} {2016})}\BibitemShut {NoStop}%
\bibitem [{\citenamefont {Cao}\ \emph {et~al.}(2019)\citenamefont {Cao},
  \citenamefont {Romero}, \citenamefont {Olson}, \citenamefont {Degroote},
  \citenamefont {Johnson}, \citenamefont {Kieferová}, \citenamefont
  {Kivlichan}, \citenamefont {Menke}, \citenamefont {Peropadre}, \citenamefont
  {Sawaya},\ and\ \citenamefont {et~al.}}]{quantum_chemistry}%
  \BibitemOpen
  \bibfield  {author} {\bibinfo {author} {\bibfnamefont {Y.}~\bibnamefont
  {Cao}}, \bibinfo {author} {\bibfnamefont {J.}~\bibnamefont {Romero}},
  \bibinfo {author} {\bibfnamefont {J.~P.}\ \bibnamefont {Olson}}, \bibinfo
  {author} {\bibfnamefont {M.}~\bibnamefont {Degroote}}, \bibinfo {author}
  {\bibfnamefont {P.~D.}\ \bibnamefont {Johnson}}, \bibinfo {author}
  {\bibfnamefont {M.}~\bibnamefont {Kieferová}}, \bibinfo {author}
  {\bibfnamefont {I.~D.}\ \bibnamefont {Kivlichan}}, \bibinfo {author}
  {\bibfnamefont {T.}~\bibnamefont {Menke}}, \bibinfo {author} {\bibfnamefont
  {B.}~\bibnamefont {Peropadre}}, \bibinfo {author} {\bibfnamefont {N.~P.~D.}\
  \bibnamefont {Sawaya}}, \ and\ \bibinfo {author} {\bibnamefont {et~al.}},\
  }\href {http://dx.doi.org/10.1021/acs.chemrev.8b00803} {\bibfield  {journal}
  {\bibinfo  {journal} {Chemical Reviews}\ }\textbf {\bibinfo {volume} {119}},\
  \bibinfo {pages} {10856–10915} (\bibinfo {year} {2019})}\BibitemShut
  {NoStop}%
\bibitem [{\citenamefont {Awschalom}\ \emph {et~al.}(2021)\citenamefont
  {Awschalom}, \citenamefont {Du}, \citenamefont {He}, \citenamefont
  {Heremans}, \citenamefont {Hoffmann}, \citenamefont {Hou}, \citenamefont
  {Kurebayashi}, \citenamefont {Li}, \citenamefont {Liu}, \citenamefont
  {Novosad}, \citenamefont {Sklenar}, \citenamefont {Sullivan}, \citenamefont
  {Sun}, \citenamefont {Tang}, \citenamefont {Tyberkevych}, \citenamefont
  {Trevillian}, \citenamefont {Tsen}, \citenamefont {Weiss}, \citenamefont
  {Zhang}, \citenamefont {Zhang}, \citenamefont {Zhao},\ and\ \citenamefont
  {Zollitsch}}]{quantum_engineering}%
  \BibitemOpen
  \bibfield  {author} {\bibinfo {author} {\bibfnamefont {D.~D.}\ \bibnamefont
  {Awschalom}}, \bibinfo {author} {\bibfnamefont {C.~H.~R.}\ \bibnamefont
  {Du}}, \bibinfo {author} {\bibfnamefont {R.}~\bibnamefont {He}}, \bibinfo
  {author} {\bibfnamefont {J.}~\bibnamefont {Heremans}}, \bibinfo {author}
  {\bibfnamefont {A.}~\bibnamefont {Hoffmann}}, \bibinfo {author}
  {\bibfnamefont {J.}~\bibnamefont {Hou}}, \bibinfo {author} {\bibfnamefont
  {H.}~\bibnamefont {Kurebayashi}}, \bibinfo {author} {\bibfnamefont
  {Y.}~\bibnamefont {Li}}, \bibinfo {author} {\bibfnamefont {L.}~\bibnamefont
  {Liu}}, \bibinfo {author} {\bibfnamefont {V.}~\bibnamefont {Novosad}},
  \bibinfo {author} {\bibfnamefont {J.}~\bibnamefont {Sklenar}}, \bibinfo
  {author} {\bibfnamefont {S.}~\bibnamefont {Sullivan}}, \bibinfo {author}
  {\bibfnamefont {D.}~\bibnamefont {Sun}}, \bibinfo {author} {\bibfnamefont
  {H.}~\bibnamefont {Tang}}, \bibinfo {author} {\bibfnamefont {V.}~\bibnamefont
  {Tyberkevych}}, \bibinfo {author} {\bibfnamefont {C.}~\bibnamefont
  {Trevillian}}, \bibinfo {author} {\bibfnamefont {A.~W.}\ \bibnamefont
  {Tsen}}, \bibinfo {author} {\bibfnamefont {L.}~\bibnamefont {Weiss}},
  \bibinfo {author} {\bibfnamefont {W.}~\bibnamefont {Zhang}}, \bibinfo
  {author} {\bibfnamefont {X.}~\bibnamefont {Zhang}}, \bibinfo {author}
  {\bibfnamefont {L.}~\bibnamefont {Zhao}}, \ and\ \bibinfo {author}
  {\bibfnamefont {C.~W.}\ \bibnamefont {Zollitsch}},\ }\href {\doibase
  10.1109/TQE.2021.3057799} {\bibfield  {journal} {\bibinfo  {journal} {IEEE
  Transactions on Quantum Engineering}\ ,\ \bibinfo {pages} {1}} (\bibinfo
  {year} {2021})}\BibitemShut {NoStop}%
\bibitem [{\citenamefont {Cao}\ \emph {et~al.}(2020)\citenamefont {Cao},
  \citenamefont {Cogdell}, \citenamefont {Coker}, \citenamefont {Duan},
  \citenamefont {Hauer}, \citenamefont {Kleinekath{\"o}fer}, \citenamefont
  {Jansen}, \citenamefont {Man{\v c}al}, \citenamefont {Miller}, \citenamefont
  {Ogilvie}, \citenamefont {Prokhorenko}, \citenamefont {Renger}, \citenamefont
  {Tan}, \citenamefont {Tempelaar}, \citenamefont {Thorwart}, \citenamefont
  {Thyrhaug}, \citenamefont {Westenhoff},\ and\ \citenamefont
  {Zigmantas}}]{quantum_biology}%
  \BibitemOpen
  \bibfield  {author} {\bibinfo {author} {\bibfnamefont {J.}~\bibnamefont
  {Cao}}, \bibinfo {author} {\bibfnamefont {R.~J.}\ \bibnamefont {Cogdell}},
  \bibinfo {author} {\bibfnamefont {D.~F.}\ \bibnamefont {Coker}}, \bibinfo
  {author} {\bibfnamefont {H.-G.}\ \bibnamefont {Duan}}, \bibinfo {author}
  {\bibfnamefont {J.}~\bibnamefont {Hauer}}, \bibinfo {author} {\bibfnamefont
  {U.}~\bibnamefont {Kleinekath{\"o}fer}}, \bibinfo {author} {\bibfnamefont
  {T.~L.~C.}\ \bibnamefont {Jansen}}, \bibinfo {author} {\bibfnamefont
  {T.}~\bibnamefont {Man{\v c}al}}, \bibinfo {author} {\bibfnamefont
  {R.~J.~D.}\ \bibnamefont {Miller}}, \bibinfo {author} {\bibfnamefont {J.~P.}\
  \bibnamefont {Ogilvie}}, \bibinfo {author} {\bibfnamefont {V.~I.}\
  \bibnamefont {Prokhorenko}}, \bibinfo {author} {\bibfnamefont
  {T.}~\bibnamefont {Renger}}, \bibinfo {author} {\bibfnamefont {H.-S.}\
  \bibnamefont {Tan}}, \bibinfo {author} {\bibfnamefont {R.}~\bibnamefont
  {Tempelaar}}, \bibinfo {author} {\bibfnamefont {M.}~\bibnamefont {Thorwart}},
  \bibinfo {author} {\bibfnamefont {E.}~\bibnamefont {Thyrhaug}}, \bibinfo
  {author} {\bibfnamefont {S.}~\bibnamefont {Westenhoff}}, \ and\ \bibinfo
  {author} {\bibfnamefont {D.}~\bibnamefont {Zigmantas}},\ }\href
  {https://advances.sciencemag.org/content/6/14/eaaz4888} {\bibfield  {journal}
  {\bibinfo  {journal} {Science Advances}\ }\textbf {\bibinfo {volume} {6}}
  (\bibinfo {year} {2020})}\BibitemShut {NoStop}%
\bibitem [{\citenamefont {Breuer}\ and\ \citenamefont
  {Petruccione}(2006)}]{breuer_book}%
  \BibitemOpen
  \bibfield  {author} {\bibinfo {author} {\bibfnamefont {H.-P.}\ \bibnamefont
  {Breuer}}\ and\ \bibinfo {author} {\bibfnamefont {F.}~\bibnamefont
  {Petruccione}},\ }\href@noop {} {\emph {\bibinfo {title} {The Theory of Open
  Quantum Systems}}}\ (\bibinfo  {publisher} {Oxford University Press,
  Oxford},\ \bibinfo {year} {2006})\BibitemShut {NoStop}%
\bibitem [{\citenamefont {Rivas}\ and\ \citenamefont
  {Huelga}(2012)}]{Rivas_2012}%
  \BibitemOpen
  \bibfield  {author} {\bibinfo {author} {\bibfnamefont {{\'A}.}~\bibnamefont
  {Rivas}}\ and\ \bibinfo {author} {\bibfnamefont {S.~F.}\ \bibnamefont
  {Huelga}},\ }\href {http://dx.doi.org/10.1007/978-3-642-23354-8} {\bibfield
  {journal} {\bibinfo  {journal} {SpringerBriefs in Physics}\ } (\bibinfo
  {year} {2012})}\BibitemShut {NoStop}%
\bibitem [{\citenamefont {Carmichael}(2002)}]{carmichael_book}%
  \BibitemOpen
  \bibfield  {author} {\bibinfo {author} {\bibfnamefont {H.}~\bibnamefont
  {Carmichael}},\ }\href@noop {} {\emph {\bibinfo {title} {Statistical Methods
  in Quantum Optics 1. Master Equations and Fokker–Planck Equations}}}\
  (\bibinfo  {publisher} {Springer-Verlag Berlin Heidelberg},\ \bibinfo {year}
  {2002})\BibitemShut {NoStop}%
\bibitem [{\citenamefont {Watrous}(2018)}]{watrous_2018}%
  \BibitemOpen
  \bibfield  {author} {\bibinfo {author} {\bibfnamefont {J.}~\bibnamefont
  {Watrous}},\ }\href {\doibase 10.1017/9781316848142} {\emph {\bibinfo {title}
  {The Theory of Quantum Information}}}\ (\bibinfo  {publisher} {Cambridge
  University Press},\ \bibinfo {year} {2018})\BibitemShut {NoStop}%
\bibitem [{\citenamefont {Siddhu}\ and\ \citenamefont
  {Tayur}(2022)}]{Siddhu_2022}%
  \BibitemOpen
  \bibfield  {author} {\bibinfo {author} {\bibfnamefont {V.}~\bibnamefont
  {Siddhu}}\ and\ \bibinfo {author} {\bibfnamefont {S.}~\bibnamefont {Tayur}},\
  }\enquote {\bibinfo {title} {Five starter pieces: Quantum information science
  via semidefinite programs},}\ in\ \href {\doibase 10.1287/educ.2022.0243}
  {\emph {\bibinfo {booktitle} {Tutorials in Operations Research: Emerging and
  Impactful Topics in Operations}}}\ (\bibinfo {year} {2022})\ Chap.~\bibinfo
  {chapter} {3}, pp.\ \bibinfo {pages} {59--92}\BibitemShut {NoStop}%
\bibitem [{\citenamefont {Mazziotti}(2004)}]{Mazziotti_2004}%
  \BibitemOpen
  \bibfield  {author} {\bibinfo {author} {\bibfnamefont {D.~A.}\ \bibnamefont
  {Mazziotti}},\ }\href {\doibase 10.1103/PhysRevLett.93.213001} {\bibfield
  {journal} {\bibinfo  {journal} {Phys. Rev. Lett.}\ }\textbf {\bibinfo
  {volume} {93}},\ \bibinfo {pages} {213001} (\bibinfo {year}
  {2004})}\BibitemShut {NoStop}%
\bibitem [{\citenamefont {Mazziotti}(2011)}]{Mazziotti_2011}%
  \BibitemOpen
  \bibfield  {author} {\bibinfo {author} {\bibfnamefont {D.~A.}\ \bibnamefont
  {Mazziotti}},\ }\href {\doibase 10.1103/PhysRevLett.106.083001} {\bibfield
  {journal} {\bibinfo  {journal} {Phys. Rev. Lett.}\ }\textbf {\bibinfo
  {volume} {106}},\ \bibinfo {pages} {083001} (\bibinfo {year}
  {2011})}\BibitemShut {NoStop}%
\bibitem [{\citenamefont {Wolf}\ \emph {et~al.}(2008)\citenamefont {Wolf},
  \citenamefont {Eisert}, \citenamefont {Cubitt},\ and\ \citenamefont
  {Cirac}}]{Wolf_2008}%
  \BibitemOpen
  \bibfield  {author} {\bibinfo {author} {\bibfnamefont {M.~M.}\ \bibnamefont
  {Wolf}}, \bibinfo {author} {\bibfnamefont {J.}~\bibnamefont {Eisert}},
  \bibinfo {author} {\bibfnamefont {T.~S.}\ \bibnamefont {Cubitt}}, \ and\
  \bibinfo {author} {\bibfnamefont {J.~I.}\ \bibnamefont {Cirac}},\ }\href
  {\doibase 10.1103/PhysRevLett.101.150402} {\bibfield  {journal} {\bibinfo
  {journal} {Phys. Rev. Lett.}\ }\textbf {\bibinfo {volume} {101}},\ \bibinfo
  {pages} {150402} (\bibinfo {year} {2008})}\BibitemShut {NoStop}%
\bibitem [{\citenamefont {Cubitt}\ \emph {et~al.}(2012)\citenamefont {Cubitt},
  \citenamefont {Eisert},\ and\ \citenamefont {Wolf}}]{Cubitt_2012}%
  \BibitemOpen
  \bibfield  {author} {\bibinfo {author} {\bibfnamefont {T.~S.}\ \bibnamefont
  {Cubitt}}, \bibinfo {author} {\bibfnamefont {J.}~\bibnamefont {Eisert}}, \
  and\ \bibinfo {author} {\bibfnamefont {M.~M.}\ \bibnamefont {Wolf}},\ }\href
  {\doibase 10.1007/s00220-011-1402-y} {\bibfield  {journal} {\bibinfo
  {journal} {Communications in Mathematical Physics}\ }\textbf {\bibinfo
  {volume} {310}},\ \bibinfo {pages} {383} (\bibinfo {year}
  {2012})}\BibitemShut {NoStop}%
\bibitem [{\citenamefont {Kiilerich}\ and\ \citenamefont
  {M\o{}lmer}(2018)}]{Kiilerich_2018}%
  \BibitemOpen
  \bibfield  {author} {\bibinfo {author} {\bibfnamefont {A.~H.}\ \bibnamefont
  {Kiilerich}}\ and\ \bibinfo {author} {\bibfnamefont {K.}~\bibnamefont
  {M\o{}lmer}},\ }\href {\doibase 10.1103/PhysRevA.97.052113} {\bibfield
  {journal} {\bibinfo  {journal} {Phys. Rev. A}\ }\textbf {\bibinfo {volume}
  {97}},\ \bibinfo {pages} {052113} (\bibinfo {year} {2018})}\BibitemShut
  {NoStop}%
\bibitem [{\citenamefont {Gorini}\ \emph {et~al.}(1976)\citenamefont {Gorini},
  \citenamefont {Kossakowski},\ and\ \citenamefont {Sudarshan}}]{GKS1976}%
  \BibitemOpen
  \bibfield  {author} {\bibinfo {author} {\bibfnamefont {V.}~\bibnamefont
  {Gorini}}, \bibinfo {author} {\bibfnamefont {A.}~\bibnamefont {Kossakowski}},
  \ and\ \bibinfo {author} {\bibfnamefont {E.~C.~G.}\ \bibnamefont
  {Sudarshan}},\ }\href {\doibase 10.1063/1.522979} {\bibfield  {journal}
  {\bibinfo  {journal} {Journal of Mathematical Physics}\ }\textbf {\bibinfo
  {volume} {17}},\ \bibinfo {pages} {821} (\bibinfo {year} {1976})}\BibitemShut
  {NoStop}%
\bibitem [{\citenamefont {Gorini}\ \emph {et~al.}(1978)\citenamefont {Gorini},
  \citenamefont {Frigerio}, \citenamefont {Verri}, \citenamefont
  {Kossakowski},\ and\ \citenamefont {Sudarshan}}]{Gorini_1978}%
  \BibitemOpen
  \bibfield  {author} {\bibinfo {author} {\bibfnamefont {V.}~\bibnamefont
  {Gorini}}, \bibinfo {author} {\bibfnamefont {A.}~\bibnamefont {Frigerio}},
  \bibinfo {author} {\bibfnamefont {M.}~\bibnamefont {Verri}}, \bibinfo
  {author} {\bibfnamefont {A.}~\bibnamefont {Kossakowski}}, \ and\ \bibinfo
  {author} {\bibfnamefont {E.}~\bibnamefont {Sudarshan}},\ }\href {\doibase
  https://doi.org/10.1016/0034-4877(78)90050-2} {\bibfield  {journal} {\bibinfo
   {journal} {Reports on Mathematical Physics}\ }\textbf {\bibinfo {volume}
  {13}},\ \bibinfo {pages} {149} (\bibinfo {year} {1978})}\BibitemShut
  {NoStop}%
\bibitem [{\citenamefont {Lindblad}(1976)}]{lindblad1976}%
  \BibitemOpen
  \bibfield  {author} {\bibinfo {author} {\bibfnamefont {G.}~\bibnamefont
  {Lindblad}},\ }\href {\doibase 10.1007/bf01608499} {\bibfield  {journal}
  {\bibinfo  {journal} {Communications in Mathematical Physics}\ }\textbf
  {\bibinfo {volume} {48}},\ \bibinfo {pages} {119} (\bibinfo {year}
  {1976})}\BibitemShut {NoStop}%
\bibitem [{\citenamefont {Weimer}\ \emph {et~al.}(2021)\citenamefont {Weimer},
  \citenamefont {Kshetrimayum},\ and\ \citenamefont {Or\'us}}]{Weimer_2021}%
  \BibitemOpen
  \bibfield  {author} {\bibinfo {author} {\bibfnamefont {H.}~\bibnamefont
  {Weimer}}, \bibinfo {author} {\bibfnamefont {A.}~\bibnamefont
  {Kshetrimayum}}, \ and\ \bibinfo {author} {\bibfnamefont {R.}~\bibnamefont
  {Or\'us}},\ }\href {\doibase 10.1103/RevModPhys.93.015008} {\bibfield
  {journal} {\bibinfo  {journal} {Rev. Mod. Phys.}\ }\textbf {\bibinfo {volume}
  {93}},\ \bibinfo {pages} {015008} (\bibinfo {year} {2021})}\BibitemShut
  {NoStop}%
\bibitem [{\citenamefont {Plenio}\ and\ \citenamefont
  {Knight}(1998)}]{Plenio_1998}%
  \BibitemOpen
  \bibfield  {author} {\bibinfo {author} {\bibfnamefont {M.~B.}\ \bibnamefont
  {Plenio}}\ and\ \bibinfo {author} {\bibfnamefont {P.~L.}\ \bibnamefont
  {Knight}},\ }\href {\doibase 10.1103/RevModPhys.70.101} {\bibfield  {journal}
  {\bibinfo  {journal} {Rev. Mod. Phys.}\ }\textbf {\bibinfo {volume} {70}},\
  \bibinfo {pages} {101} (\bibinfo {year} {1998})}\BibitemShut {NoStop}%
\bibitem [{\citenamefont {Dalibard}\ \emph {et~al.}(1992)\citenamefont
  {Dalibard}, \citenamefont {Castin},\ and\ \citenamefont
  {M{\o}lmer}}]{Dalibard1992}%
  \BibitemOpen
  \bibfield  {author} {\bibinfo {author} {\bibfnamefont {J.}~\bibnamefont
  {Dalibard}}, \bibinfo {author} {\bibfnamefont {Y.}~\bibnamefont {Castin}}, \
  and\ \bibinfo {author} {\bibfnamefont {K.}~\bibnamefont {M{\o}lmer}},\ }\href
  {\doibase 10.1103/physrevlett.68.580} {\bibfield  {journal} {\bibinfo
  {journal} {Phys. Rev. Lett.}\ }\textbf {\bibinfo {volume} {68}},\ \bibinfo
  {pages} {580} (\bibinfo {year} {1992})}\BibitemShut {NoStop}%
\bibitem [{\citenamefont {M{\o}lmer}\ \emph {et~al.}(1993)\citenamefont
  {M{\o}lmer}, \citenamefont {Castin},\ and\ \citenamefont
  {Dalibard}}]{Mlmer1993}%
  \BibitemOpen
  \bibfield  {author} {\bibinfo {author} {\bibfnamefont {K.}~\bibnamefont
  {M{\o}lmer}}, \bibinfo {author} {\bibfnamefont {Y.}~\bibnamefont {Castin}}, \
  and\ \bibinfo {author} {\bibfnamefont {J.}~\bibnamefont {Dalibard}},\ }\href
  {\doibase 10.1364/josab.10.000524} {\bibfield  {journal} {\bibinfo  {journal}
  {Journal of the Optical Society of America B}\ }\textbf {\bibinfo {volume}
  {10}},\ \bibinfo {pages} {524} (\bibinfo {year} {1993})}\BibitemShut
  {NoStop}%
\bibitem [{\citenamefont {Tupkary}\ \emph {et~al.}(2022)\citenamefont
  {Tupkary}, \citenamefont {Dhar}, \citenamefont {Kulkarni},\ and\
  \citenamefont {Purkayastha}}]{Tupkary_2022}%
  \BibitemOpen
  \bibfield  {author} {\bibinfo {author} {\bibfnamefont {D.}~\bibnamefont
  {Tupkary}}, \bibinfo {author} {\bibfnamefont {A.}~\bibnamefont {Dhar}},
  \bibinfo {author} {\bibfnamefont {M.}~\bibnamefont {Kulkarni}}, \ and\
  \bibinfo {author} {\bibfnamefont {A.}~\bibnamefont {Purkayastha}},\ }\href
  {\doibase 10.1103/PhysRevA.105.032208} {\bibfield  {journal} {\bibinfo
  {journal} {Phys. Rev. A}\ }\textbf {\bibinfo {volume} {105}},\ \bibinfo
  {pages} {032208} (\bibinfo {year} {2022})}\BibitemShut {NoStop}%
\bibitem [{\citenamefont {Redfield}(1957)}]{redfield_1957}%
  \BibitemOpen
  \bibfield  {author} {\bibinfo {author} {\bibfnamefont {A.~G.}\ \bibnamefont
  {Redfield}},\ }\href {\doibase 10.1147/rd.11.0019} {\bibfield  {journal}
  {\bibinfo  {journal} {IBM Journal of Research and Development}\ }\textbf
  {\bibinfo {volume} {1}},\ \bibinfo {pages} {19} (\bibinfo {year}
  {1957})}\BibitemShut {NoStop}%
\bibitem [{\citenamefont {Bloch}(1957)}]{block_1957}%
  \BibitemOpen
  \bibfield  {author} {\bibinfo {author} {\bibfnamefont {F.}~\bibnamefont
  {Bloch}},\ }\href {\doibase 10.1103/PhysRev.105.1206} {\bibfield  {journal}
  {\bibinfo  {journal} {Phys. Rev.}\ }\textbf {\bibinfo {volume} {105}},\
  \bibinfo {pages} {1206} (\bibinfo {year} {1957})}\BibitemShut {NoStop}%
\bibitem [{\citenamefont {Hartmann}\ and\ \citenamefont
  {Strunz}(2020)}]{Hartmann_2020_1}%
  \BibitemOpen
  \bibfield  {author} {\bibinfo {author} {\bibfnamefont {R.}~\bibnamefont
  {Hartmann}}\ and\ \bibinfo {author} {\bibfnamefont {W.~T.}\ \bibnamefont
  {Strunz}},\ }\href {\doibase 10.1103/PhysRevA.101.012103} {\bibfield
  {journal} {\bibinfo  {journal} {Phys. Rev. A}\ }\textbf {\bibinfo {volume}
  {101}},\ \bibinfo {pages} {012103} (\bibinfo {year} {2020})}\BibitemShut
  {NoStop}%
\bibitem [{\citenamefont {Eastham}\ \emph {et~al.}(2016)\citenamefont
  {Eastham}, \citenamefont {Kirton}, \citenamefont {Cammack}, \citenamefont
  {Lovett},\ and\ \citenamefont {Keeling}}]{Eastham_2016}%
  \BibitemOpen
  \bibfield  {author} {\bibinfo {author} {\bibfnamefont {P.~R.}\ \bibnamefont
  {Eastham}}, \bibinfo {author} {\bibfnamefont {P.}~\bibnamefont {Kirton}},
  \bibinfo {author} {\bibfnamefont {H.~M.}\ \bibnamefont {Cammack}}, \bibinfo
  {author} {\bibfnamefont {B.~W.}\ \bibnamefont {Lovett}}, \ and\ \bibinfo
  {author} {\bibfnamefont {J.}~\bibnamefont {Keeling}},\ }\href {\doibase
  10.1103/PhysRevA.94.012110} {\bibfield  {journal} {\bibinfo  {journal} {Phys.
  Rev. A}\ }\textbf {\bibinfo {volume} {94}},\ \bibinfo {pages} {012110}
  (\bibinfo {year} {2016})}\BibitemShut {NoStop}%
\bibitem [{\citenamefont {Anderloni}\ \emph {et~al.}(2007)\citenamefont
  {Anderloni}, \citenamefont {Benatti},\ and\ \citenamefont
  {Floreanini}}]{anderloni_2007}%
  \BibitemOpen
  \bibfield  {author} {\bibinfo {author} {\bibfnamefont {S.}~\bibnamefont
  {Anderloni}}, \bibinfo {author} {\bibfnamefont {F.}~\bibnamefont {Benatti}},
  \ and\ \bibinfo {author} {\bibfnamefont {R.}~\bibnamefont {Floreanini}},\
  }\href {\doibase 10.1088/1751-8113/40/7/013} {\bibfield  {journal} {\bibinfo
  {journal} {Journal of Physics A: Mathematical and Theoretical}\ }\textbf
  {\bibinfo {volume} {40}},\ \bibinfo {pages} {1625} (\bibinfo {year}
  {2007})}\BibitemShut {NoStop}%
\bibitem [{\citenamefont {Gaspard}\ and\ \citenamefont
  {Nagaoka}(1999)}]{Gaspard_Nagaoka_1999}%
  \BibitemOpen
  \bibfield  {author} {\bibinfo {author} {\bibfnamefont {P.}~\bibnamefont
  {Gaspard}}\ and\ \bibinfo {author} {\bibfnamefont {M.}~\bibnamefont
  {Nagaoka}},\ }\href {\doibase 10.1063/1.479867} {\bibfield  {journal}
  {\bibinfo  {journal} {The Journal of Chemical Physics}\ }\textbf {\bibinfo
  {volume} {111}},\ \bibinfo {pages} {5668} (\bibinfo {year}
  {1999})}\BibitemShut {NoStop}%
\bibitem [{\citenamefont {Kohen}\ \emph {et~al.}(1997)\citenamefont {Kohen},
  \citenamefont {Marston},\ and\ \citenamefont {Tannor}}]{Kohen_1997}%
  \BibitemOpen
  \bibfield  {author} {\bibinfo {author} {\bibfnamefont {D.}~\bibnamefont
  {Kohen}}, \bibinfo {author} {\bibfnamefont {C.~C.}\ \bibnamefont {Marston}},
  \ and\ \bibinfo {author} {\bibfnamefont {D.~J.}\ \bibnamefont {Tannor}},\
  }\href {\doibase 10.1063/1.474887} {\bibfield  {journal} {\bibinfo  {journal}
  {The Journal of Chemical Physics}\ }\textbf {\bibinfo {volume} {107}},\
  \bibinfo {pages} {5236} (\bibinfo {year} {1997})}\BibitemShut {NoStop}%
\bibitem [{\citenamefont {Gnutzmann}\ and\ \citenamefont
  {Haake}(1996)}]{Gnutzmann_1996}%
  \BibitemOpen
  \bibfield  {author} {\bibinfo {author} {\bibfnamefont {S.}~\bibnamefont
  {Gnutzmann}}\ and\ \bibinfo {author} {\bibfnamefont {F.}~\bibnamefont
  {Haake}},\ }\href {\doibase 10.1007/s002570050208} {\bibfield  {journal}
  {\bibinfo  {journal} {Zeitschrift f{\"u}r Physik B Condensed Matter}\
  }\textbf {\bibinfo {volume} {101}},\ \bibinfo {pages} {263} (\bibinfo {year}
  {1996})}\BibitemShut {NoStop}%
\bibitem [{\citenamefont {Su{\'{a}}rez}\ \emph {et~al.}(1992)\citenamefont
  {Su{\'{a}}rez}, \citenamefont {Silbey},\ and\ \citenamefont
  {Oppenheim}}]{Suarez_1992}%
  \BibitemOpen
  \bibfield  {author} {\bibinfo {author} {\bibfnamefont {A.}~\bibnamefont
  {Su{\'{a}}rez}}, \bibinfo {author} {\bibfnamefont {R.}~\bibnamefont
  {Silbey}}, \ and\ \bibinfo {author} {\bibfnamefont {I.}~\bibnamefont
  {Oppenheim}},\ }\href {\doibase 10.1063/1.463831} {\bibfield  {journal}
  {\bibinfo  {journal} {The Journal of Chemical Physics}\ }\textbf {\bibinfo
  {volume} {97}},\ \bibinfo {pages} {5101} (\bibinfo {year}
  {1992})}\BibitemShut {NoStop}%
\bibitem [{\citenamefont {Whitney}(2008)}]{Whitney_2008}%
  \BibitemOpen
  \bibfield  {author} {\bibinfo {author} {\bibfnamefont {R.~S.}\ \bibnamefont
  {Whitney}},\ }\href {\doibase 10.1088/1751-8113/41/17/175304} {\bibfield
  {journal} {\bibinfo  {journal} {Journal of Physics A: Mathematical and
  Theoretical}\ }\textbf {\bibinfo {volume} {41}},\ \bibinfo {pages} {175304}
  (\bibinfo {year} {2008})}\BibitemShut {NoStop}%
\bibitem [{\citenamefont {Nathan}\ and\ \citenamefont {Rudner}(2020)}]{ule}%
  \BibitemOpen
  \bibfield  {author} {\bibinfo {author} {\bibfnamefont {F.}~\bibnamefont
  {Nathan}}\ and\ \bibinfo {author} {\bibfnamefont {M.~S.}\ \bibnamefont
  {Rudner}},\ }\href {\doibase 10.1103/PhysRevB.102.115109} {\bibfield
  {journal} {\bibinfo  {journal} {Phys. Rev. B}\ }\textbf {\bibinfo {volume}
  {102}},\ \bibinfo {pages} {115109} (\bibinfo {year} {2020})}\BibitemShut
  {NoStop}%
\bibitem [{\citenamefont {Spohn}(1980)}]{Spohn_1980}%
  \BibitemOpen
  \bibfield  {author} {\bibinfo {author} {\bibfnamefont {H.}~\bibnamefont
  {Spohn}},\ }\href {\doibase 10.1103/RevModPhys.52.569} {\bibfield  {journal}
  {\bibinfo  {journal} {Rev. Mod. Phys.}\ }\textbf {\bibinfo {volume} {52}},\
  \bibinfo {pages} {569} (\bibinfo {year} {1980})}\BibitemShut {NoStop}%
\bibitem [{\citenamefont {Grant}\ and\ \citenamefont {Boyd}(2014)}]{cvx}%
  \BibitemOpen
  \bibfield  {author} {\bibinfo {author} {\bibfnamefont {M.}~\bibnamefont
  {Grant}}\ and\ \bibinfo {author} {\bibfnamefont {S.}~\bibnamefont {Boyd}},\
  }\href@noop {} {\enquote {\bibinfo {title} {{CVX}: Matlab software for
  disciplined convex programming, version 2.1},}\ }\bibinfo {howpublished}
  {\url{http://cvxr.com/cvx}} (\bibinfo {year} {2014})\BibitemShut {NoStop}%
\bibitem [{\citenamefont {Walls}\ and\ \citenamefont
  {Milburn}(2008)}]{Milburn_book1}%
  \BibitemOpen
  \bibfield  {author} {\bibinfo {author} {\bibfnamefont {D.}~\bibnamefont
  {Walls}}\ and\ \bibinfo {author} {\bibfnamefont {G.~J.}\ \bibnamefont
  {Milburn}},\ }\href@noop {} {\emph {\bibinfo {title} {Quantum optics}}}\
  (\bibinfo  {publisher} {Springer-Verlag Berlin Heidelberg},\ \bibinfo {year}
  {2008})\BibitemShut {NoStop}%
\bibitem [{\citenamefont {Becker}\ \emph {et~al.}(2021)\citenamefont {Becker},
  \citenamefont {Wu},\ and\ \citenamefont {Eckardt}}]{Becker_2021}%
  \BibitemOpen
  \bibfield  {author} {\bibinfo {author} {\bibfnamefont {T.}~\bibnamefont
  {Becker}}, \bibinfo {author} {\bibfnamefont {L.-N.}\ \bibnamefont {Wu}}, \
  and\ \bibinfo {author} {\bibfnamefont {A.}~\bibnamefont {Eckardt}},\ }\href
  {\doibase 10.1103/PhysRevE.104.014110} {\bibfield  {journal} {\bibinfo
  {journal} {Phys. Rev. E}\ }\textbf {\bibinfo {volume} {104}},\ \bibinfo
  {pages} {014110} (\bibinfo {year} {2021})}\BibitemShut {NoStop}%
\bibitem [{\citenamefont {Prosen}\ and\ \citenamefont
  {Žnidarič}(2009)}]{Prosen_2009}%
  \BibitemOpen
  \bibfield  {author} {\bibinfo {author} {\bibfnamefont {T.}~\bibnamefont
  {Prosen}}\ and\ \bibinfo {author} {\bibfnamefont {M.}~\bibnamefont
  {Žnidarič}},\ }\href {\doibase 10.1088/1742-5468/2009/02/P02035} {\bibfield
   {journal} {\bibinfo  {journal} {Journal of Statistical Mechanics: Theory and
  Experiment}\ }\textbf {\bibinfo {volume} {2009}},\ \bibinfo {pages} {P02035}
  (\bibinfo {year} {2009})}\BibitemShut {NoStop}%
\bibitem [{\citenamefont {Mendoza-Arenas}\ \emph {et~al.}(2015)\citenamefont
  {Mendoza-Arenas}, \citenamefont {Clark},\ and\ \citenamefont
  {Jaksch}}]{Mendoza-Arenas_2015}%
  \BibitemOpen
  \bibfield  {author} {\bibinfo {author} {\bibfnamefont {J.~J.}\ \bibnamefont
  {Mendoza-Arenas}}, \bibinfo {author} {\bibfnamefont {S.~R.}\ \bibnamefont
  {Clark}}, \ and\ \bibinfo {author} {\bibfnamefont {D.}~\bibnamefont
  {Jaksch}},\ }\href {\doibase 10.1103/PhysRevE.91.042129} {\bibfield
  {journal} {\bibinfo  {journal} {Phys. Rev. E}\ }\textbf {\bibinfo {volume}
  {91}},\ \bibinfo {pages} {042129} (\bibinfo {year} {2015})}\BibitemShut
  {NoStop}%
\bibitem [{\citenamefont {\ifmmode \check{Z}\else
  \v{Z}\fi{}nidari\ifmmode~\check{c}\else \v{c}\fi{}}\ \emph
  {et~al.}(2010)\citenamefont {\ifmmode \check{Z}\else
  \v{Z}\fi{}nidari\ifmmode~\check{c}\else \v{c}\fi{}}, \citenamefont {Prosen},
  \citenamefont {Benenti}, \citenamefont {Casati},\ and\ \citenamefont
  {Rossini}}]{thermalization_2010_znidaric}%
  \BibitemOpen
  \bibfield  {author} {\bibinfo {author} {\bibfnamefont {M.}~\bibnamefont
  {\ifmmode \check{Z}\else \v{Z}\fi{}nidari\ifmmode~\check{c}\else
  \v{c}\fi{}}}, \bibinfo {author} {\bibfnamefont {T.~c.~v.}\ \bibnamefont
  {Prosen}}, \bibinfo {author} {\bibfnamefont {G.}~\bibnamefont {Benenti}},
  \bibinfo {author} {\bibfnamefont {G.}~\bibnamefont {Casati}}, \ and\ \bibinfo
  {author} {\bibfnamefont {D.}~\bibnamefont {Rossini}},\ }\href {\doibase
  10.1103/PhysRevE.81.051135} {\bibfield  {journal} {\bibinfo  {journal} {Phys.
  Rev. E}\ }\textbf {\bibinfo {volume} {81}},\ \bibinfo {pages} {051135}
  (\bibinfo {year} {2010})}\BibitemShut {NoStop}%
\bibitem [{\citenamefont {Barreiro}\ \emph {et~al.}(2011)\citenamefont
  {Barreiro}, \citenamefont {M{\"u}ller}, \citenamefont {Schindler},
  \citenamefont {Nigg}, \citenamefont {Monz}, \citenamefont {Chwalla},
  \citenamefont {Hennrich}, \citenamefont {Roos}, \citenamefont {Zoller},\ and\
  \citenamefont {Blatt}}]{Barreiro_2011}%
  \BibitemOpen
  \bibfield  {author} {\bibinfo {author} {\bibfnamefont {J.~T.}\ \bibnamefont
  {Barreiro}}, \bibinfo {author} {\bibfnamefont {M.}~\bibnamefont
  {M{\"u}ller}}, \bibinfo {author} {\bibfnamefont {P.}~\bibnamefont
  {Schindler}}, \bibinfo {author} {\bibfnamefont {D.}~\bibnamefont {Nigg}},
  \bibinfo {author} {\bibfnamefont {T.}~\bibnamefont {Monz}}, \bibinfo {author}
  {\bibfnamefont {M.}~\bibnamefont {Chwalla}}, \bibinfo {author} {\bibfnamefont
  {M.}~\bibnamefont {Hennrich}}, \bibinfo {author} {\bibfnamefont {C.~F.}\
  \bibnamefont {Roos}}, \bibinfo {author} {\bibfnamefont {P.}~\bibnamefont
  {Zoller}}, \ and\ \bibinfo {author} {\bibfnamefont {R.}~\bibnamefont
  {Blatt}},\ }\href {\doibase 10.1038/nature09801} {\bibfield  {journal}
  {\bibinfo  {journal} {Nature}\ }\textbf {\bibinfo {volume} {470}},\ \bibinfo
  {pages} {486} (\bibinfo {year} {2011})}\BibitemShut {NoStop}%
\bibitem [{\citenamefont {Schindler}\ \emph {et~al.}(2013)\citenamefont
  {Schindler}, \citenamefont {M{\"u}ller}, \citenamefont {Nigg}, \citenamefont
  {Barreiro}, \citenamefont {Martinez}, \citenamefont {Hennrich}, \citenamefont
  {Monz}, \citenamefont {Diehl}, \citenamefont {Zoller},\ and\ \citenamefont
  {Blatt}}]{Schindler_2013}%
  \BibitemOpen
  \bibfield  {author} {\bibinfo {author} {\bibfnamefont {P.}~\bibnamefont
  {Schindler}}, \bibinfo {author} {\bibfnamefont {M.}~\bibnamefont
  {M{\"u}ller}}, \bibinfo {author} {\bibfnamefont {D.}~\bibnamefont {Nigg}},
  \bibinfo {author} {\bibfnamefont {J.~T.}\ \bibnamefont {Barreiro}}, \bibinfo
  {author} {\bibfnamefont {E.~A.}\ \bibnamefont {Martinez}}, \bibinfo {author}
  {\bibfnamefont {M.}~\bibnamefont {Hennrich}}, \bibinfo {author}
  {\bibfnamefont {T.}~\bibnamefont {Monz}}, \bibinfo {author} {\bibfnamefont
  {S.}~\bibnamefont {Diehl}}, \bibinfo {author} {\bibfnamefont
  {P.}~\bibnamefont {Zoller}}, \ and\ \bibinfo {author} {\bibfnamefont
  {R.}~\bibnamefont {Blatt}},\ }\href {\doibase 10.1038/nphys2630} {\bibfield
  {journal} {\bibinfo  {journal} {Nature Physics}\ }\textbf {\bibinfo {volume}
  {9}},\ \bibinfo {pages} {361} (\bibinfo {year} {2013})}\BibitemShut {NoStop}%
\bibitem [{\citenamefont {Weimer}\ \emph {et~al.}(2010)\citenamefont {Weimer},
  \citenamefont {M{\"u}ller}, \citenamefont {Lesanovsky}, \citenamefont
  {Zoller},\ and\ \citenamefont {B{\"u}chler}}]{Weimer_2010}%
  \BibitemOpen
  \bibfield  {author} {\bibinfo {author} {\bibfnamefont {H.}~\bibnamefont
  {Weimer}}, \bibinfo {author} {\bibfnamefont {M.}~\bibnamefont {M{\"u}ller}},
  \bibinfo {author} {\bibfnamefont {I.}~\bibnamefont {Lesanovsky}}, \bibinfo
  {author} {\bibfnamefont {P.}~\bibnamefont {Zoller}}, \ and\ \bibinfo {author}
  {\bibfnamefont {H.~P.}\ \bibnamefont {B{\"u}chler}},\ }\href {\doibase
  10.1038/nphys1614} {\bibfield  {journal} {\bibinfo  {journal} {Nature
  Physics}\ }\textbf {\bibinfo {volume} {6}},\ \bibinfo {pages} {382} (\bibinfo
  {year} {2010})}\BibitemShut {NoStop}%
\bibitem [{\citenamefont {Nguyen}\ \emph {et~al.}(2018)\citenamefont {Nguyen},
  \citenamefont {Raimond}, \citenamefont {Sayrin}, \citenamefont {Corti\~nas},
  \citenamefont {Cantat-Moltrecht}, \citenamefont {Assemat}, \citenamefont
  {Dotsenko}, \citenamefont {Gleyzes}, \citenamefont {Haroche}, \citenamefont
  {Roux}, \citenamefont {Jolicoeur},\ and\ \citenamefont
  {Brune}}]{Nguyen_2018}%
  \BibitemOpen
  \bibfield  {author} {\bibinfo {author} {\bibfnamefont {T.~L.}\ \bibnamefont
  {Nguyen}}, \bibinfo {author} {\bibfnamefont {J.~M.}\ \bibnamefont {Raimond}},
  \bibinfo {author} {\bibfnamefont {C.}~\bibnamefont {Sayrin}}, \bibinfo
  {author} {\bibfnamefont {R.}~\bibnamefont {Corti\~nas}}, \bibinfo {author}
  {\bibfnamefont {T.}~\bibnamefont {Cantat-Moltrecht}}, \bibinfo {author}
  {\bibfnamefont {F.}~\bibnamefont {Assemat}}, \bibinfo {author} {\bibfnamefont
  {I.}~\bibnamefont {Dotsenko}}, \bibinfo {author} {\bibfnamefont
  {S.}~\bibnamefont {Gleyzes}}, \bibinfo {author} {\bibfnamefont
  {S.}~\bibnamefont {Haroche}}, \bibinfo {author} {\bibfnamefont
  {G.}~\bibnamefont {Roux}}, \bibinfo {author} {\bibfnamefont {T.}~\bibnamefont
  {Jolicoeur}}, \ and\ \bibinfo {author} {\bibfnamefont {M.}~\bibnamefont
  {Brune}},\ }\href {\doibase 10.1103/PhysRevX.8.011032} {\bibfield  {journal}
  {\bibinfo  {journal} {Phys. Rev. X}\ }\textbf {\bibinfo {volume} {8}},\
  \bibinfo {pages} {011032} (\bibinfo {year} {2018})}\BibitemShut {NoStop}%
\bibitem [{\citenamefont {Garc{\'i}a-P{\'e}rez}\ \emph
  {et~al.}(2020)\citenamefont {Garc{\'i}a-P{\'e}rez}, \citenamefont {Rossi},\
  and\ \citenamefont {Maniscalco}}]{Garcia_2020}%
  \BibitemOpen
  \bibfield  {author} {\bibinfo {author} {\bibfnamefont {G.}~\bibnamefont
  {Garc{\'i}a-P{\'e}rez}}, \bibinfo {author} {\bibfnamefont {M.~A.~C.}\
  \bibnamefont {Rossi}}, \ and\ \bibinfo {author} {\bibfnamefont
  {S.}~\bibnamefont {Maniscalco}},\ }\href {\doibase 10.1038/s41534-019-0235-y}
  {\bibfield  {journal} {\bibinfo  {journal} {npj Quantum Information}\
  }\textbf {\bibinfo {volume} {6}},\ \bibinfo {pages} {1} (\bibinfo {year}
  {2020})}\BibitemShut {NoStop}%
\bibitem [{\citenamefont {Kim}\ \emph {et~al.}(2022)\citenamefont {Kim},
  \citenamefont {Nichol}, \citenamefont {Jordan},\ and\ \citenamefont
  {Franco}}]{Kim_2022}%
  \BibitemOpen
  \bibfield  {author} {\bibinfo {author} {\bibfnamefont {C.~W.}\ \bibnamefont
  {Kim}}, \bibinfo {author} {\bibfnamefont {J.~M.}\ \bibnamefont {Nichol}},
  \bibinfo {author} {\bibfnamefont {A.~N.}\ \bibnamefont {Jordan}}, \ and\
  \bibinfo {author} {\bibfnamefont {I.}~\bibnamefont {Franco}},\ }\href
  {\doibase 10.1103/PRXQuantum.3.040308} {\bibfield  {journal} {\bibinfo
  {journal} {PRX Quantum}\ }\textbf {\bibinfo {volume} {3}},\ \bibinfo {pages}
  {040308} (\bibinfo {year} {2022})}\BibitemShut {NoStop}%
\bibitem [{\citenamefont {Becker}\ \emph {et~al.}(2022)\citenamefont {Becker},
  \citenamefont {Schnell},\ and\ \citenamefont {Thingna}}]{Becker2022}%
  \BibitemOpen
  \bibfield  {author} {\bibinfo {author} {\bibfnamefont {T.}~\bibnamefont
  {Becker}}, \bibinfo {author} {\bibfnamefont {A.}~\bibnamefont {Schnell}}, \
  and\ \bibinfo {author} {\bibfnamefont {J.}~\bibnamefont {Thingna}},\ }\href
  {\doibase 10.1103/PhysRevLett.129.200403} {\bibfield  {journal} {\bibinfo
  {journal} {Phys. Rev. Lett.}\ }\textbf {\bibinfo {volume} {129}},\ \bibinfo
  {pages} {200403} (\bibinfo {year} {2022})}\BibitemShut {NoStop}%
\bibitem [{\citenamefont {McCauley}\ \emph {et~al.}(2020)\citenamefont
  {McCauley}, \citenamefont {Cruikshank}, \citenamefont {Bondar},\ and\
  \citenamefont {Jacobs}}]{mccauley2020}%
  \BibitemOpen
  \bibfield  {author} {\bibinfo {author} {\bibfnamefont {G.}~\bibnamefont
  {McCauley}}, \bibinfo {author} {\bibfnamefont {B.}~\bibnamefont
  {Cruikshank}}, \bibinfo {author} {\bibfnamefont {D.~I.}\ \bibnamefont
  {Bondar}}, \ and\ \bibinfo {author} {\bibfnamefont {K.}~\bibnamefont
  {Jacobs}},\ }\href {http://dx.doi.org/10.1038/s41534-020-00299-6} {\bibfield
  {journal} {\bibinfo  {journal} {npj Quantum Information}\ }\textbf {\bibinfo
  {volume} {6}} (\bibinfo {year} {2020})}\BibitemShut {NoStop}%
\bibitem [{\citenamefont {Harbola}\ \emph {et~al.}(2006)\citenamefont
  {Harbola}, \citenamefont {Esposito},\ and\ \citenamefont
  {Mukamel}}]{Harbola2006}%
  \BibitemOpen
  \bibfield  {author} {\bibinfo {author} {\bibfnamefont {U.}~\bibnamefont
  {Harbola}}, \bibinfo {author} {\bibfnamefont {M.}~\bibnamefont {Esposito}}, \
  and\ \bibinfo {author} {\bibfnamefont {S.}~\bibnamefont {Mukamel}},\ }\href
  {\doibase 10.1103/PhysRevB.74.235309} {\bibfield  {journal} {\bibinfo
  {journal} {Phys. Rev. B}\ }\textbf {\bibinfo {volume} {74}},\ \bibinfo
  {pages} {235309} (\bibinfo {year} {2006})}\BibitemShut {NoStop}%
\bibitem [{\citenamefont {Nathan}\ and\ \citenamefont {Rudner}(2022)}]{ULE2}%
  \BibitemOpen
  \bibfield  {author} {\bibinfo {author} {\bibfnamefont {F.}~\bibnamefont
  {Nathan}}\ and\ \bibinfo {author} {\bibfnamefont {M.~S.}\ \bibnamefont
  {Rudner}},\ }\href {\doibase 10.48550/ARXIV.2206.02917} {\  (\bibinfo {year}
  {2022}),\ 10.48550/ARXIV.2206.02917}\BibitemShut {NoStop}%
\bibitem [{\citenamefont {Trushechkin}\ and\ \citenamefont
  {Volovich}(2016)}]{Trushechkin_2016}%
  \BibitemOpen
  \bibfield  {author} {\bibinfo {author} {\bibfnamefont {A.~S.}\ \bibnamefont
  {Trushechkin}}\ and\ \bibinfo {author} {\bibfnamefont {I.~V.}\ \bibnamefont
  {Volovich}},\ }\href {\doibase 10.1209/0295-5075/113/30005} {\bibfield
  {journal} {\bibinfo  {journal} {{EPL} (Europhysics Letters)}\ }\textbf
  {\bibinfo {volume} {113}},\ \bibinfo {pages} {30005} (\bibinfo {year}
  {2016})}\BibitemShut {NoStop}%
\bibitem [{\citenamefont {Kleinherbers}\ \emph {et~al.}(2020)\citenamefont
  {Kleinherbers}, \citenamefont {Szpak}, \citenamefont {K\"onig},\ and\
  \citenamefont {Sch\"utzhold}}]{Kleinherbers_2020}%
  \BibitemOpen
  \bibfield  {author} {\bibinfo {author} {\bibfnamefont {E.}~\bibnamefont
  {Kleinherbers}}, \bibinfo {author} {\bibfnamefont {N.}~\bibnamefont {Szpak}},
  \bibinfo {author} {\bibfnamefont {J.}~\bibnamefont {K\"onig}}, \ and\
  \bibinfo {author} {\bibfnamefont {R.}~\bibnamefont {Sch\"utzhold}},\ }\href
  {\doibase 10.1103/PhysRevB.101.125131} {\bibfield  {journal} {\bibinfo
  {journal} {Phys. Rev. B}\ }\textbf {\bibinfo {volume} {101}},\ \bibinfo
  {pages} {125131} (\bibinfo {year} {2020})}\BibitemShut {NoStop}%
\bibitem [{\citenamefont {Davidovi{\'c}}(2020)}]{Davidovic_2020}%
  \BibitemOpen
  \bibfield  {author} {\bibinfo {author} {\bibfnamefont {D.}~\bibnamefont
  {Davidovi{\'c}}},\ }\href {\doibase 10.22331/q-2020-09-21-326} {\bibfield
  {journal} {\bibinfo  {journal} {Quantum}\ }\textbf {\bibinfo {volume} {4}},\
  \bibinfo {pages} {326} (\bibinfo {year} {2020})}\BibitemShut {NoStop}%
\bibitem [{\citenamefont {Mozgunov}\ and\ \citenamefont
  {Lidar}(2020)}]{mozgunov2020}%
  \BibitemOpen
  \bibfield  {author} {\bibinfo {author} {\bibfnamefont {E.}~\bibnamefont
  {Mozgunov}}\ and\ \bibinfo {author} {\bibfnamefont {D.}~\bibnamefont
  {Lidar}},\ }\href {http://dx.doi.org/10.22331/q-2020-02-06-227} {\bibfield
  {journal} {\bibinfo  {journal} {Quantum}\ }\textbf {\bibinfo {volume} {4}},\
  \bibinfo {pages} {227} (\bibinfo {year} {2020})}\BibitemShut {NoStop}%
\bibitem [{\citenamefont {Kir\ifmmode~\check{s}\else \v{s}\fi{}anskas}\ \emph
  {et~al.}(2018)\citenamefont {Kir\ifmmode~\check{s}\else \v{s}\fi{}anskas},
  \citenamefont {Francki\'e},\ and\ \citenamefont {Wacker}}]{kirvsanskas2018}%
  \BibitemOpen
  \bibfield  {author} {\bibinfo {author} {\bibfnamefont {G.}~\bibnamefont
  {Kir\ifmmode~\check{s}\else \v{s}\fi{}anskas}}, \bibinfo {author}
  {\bibfnamefont {M.}~\bibnamefont {Francki\'e}}, \ and\ \bibinfo {author}
  {\bibfnamefont {A.}~\bibnamefont {Wacker}},\ }\href {\doibase
  10.1103/PhysRevB.97.035432} {\bibfield  {journal} {\bibinfo  {journal} {Phys.
  Rev. B}\ }\textbf {\bibinfo {volume} {97}},\ \bibinfo {pages} {035432}
  (\bibinfo {year} {2018})}\BibitemShut {NoStop}%
\bibitem [{\citenamefont {Gerry}\ and\ \citenamefont
  {Segal}(2022)}]{segal_2022}%
  \BibitemOpen
  \bibfield  {author} {\bibinfo {author} {\bibfnamefont {M.}~\bibnamefont
  {Gerry}}\ and\ \bibinfo {author} {\bibfnamefont {D.}~\bibnamefont {Segal}},\
  }\href@noop {} {\  (\bibinfo {year} {2022})},\ \Eprint
  {http://arxiv.org/abs/arXiv:2212.11307} {arXiv:2212.11307} \BibitemShut
  {NoStop}%
\bibitem [{\citenamefont {Ostilli}\ and\ \citenamefont
  {Presilla}(2017)}]{massimo_2017}%
  \BibitemOpen
  \bibfield  {author} {\bibinfo {author} {\bibfnamefont {M.}~\bibnamefont
  {Ostilli}}\ and\ \bibinfo {author} {\bibfnamefont {C.}~\bibnamefont
  {Presilla}},\ }\href {\doibase 10.1103/PhysRevA.95.062112} {\bibfield
  {journal} {\bibinfo  {journal} {Phys. Rev. A}\ }\textbf {\bibinfo {volume}
  {95}},\ \bibinfo {pages} {062112} (\bibinfo {year} {2017})}\BibitemShut
  {NoStop}%
\bibitem [{\citenamefont {Macr\`{\i}}\ \emph {et~al.}(2017)\citenamefont
  {Macr\`{\i}}, \citenamefont {Ostilli},\ and\ \citenamefont
  {Presilla}}]{massimo_2017_2}%
  \BibitemOpen
  \bibfield  {author} {\bibinfo {author} {\bibfnamefont {T.}~\bibnamefont
  {Macr\`{\i}}}, \bibinfo {author} {\bibfnamefont {M.}~\bibnamefont {Ostilli}},
  \ and\ \bibinfo {author} {\bibfnamefont {C.}~\bibnamefont {Presilla}},\
  }\href {\doibase 10.1103/PhysRevA.95.042107} {\bibfield  {journal} {\bibinfo
  {journal} {Phys. Rev. A}\ }\textbf {\bibinfo {volume} {95}},\ \bibinfo
  {pages} {042107} (\bibinfo {year} {2017})}\BibitemShut {NoStop}%
\bibitem [{\citenamefont {Purkayastha}\ \emph {et~al.}(2020)\citenamefont
  {Purkayastha}, \citenamefont {Guarnieri}, \citenamefont {Mitchison},
  \citenamefont {Filip},\ and\ \citenamefont {Goold}}]{Archak_2020}%
  \BibitemOpen
  \bibfield  {author} {\bibinfo {author} {\bibfnamefont {A.}~\bibnamefont
  {Purkayastha}}, \bibinfo {author} {\bibfnamefont {G.}~\bibnamefont
  {Guarnieri}}, \bibinfo {author} {\bibfnamefont {M.~T.}\ \bibnamefont
  {Mitchison}}, \bibinfo {author} {\bibfnamefont {R.}~\bibnamefont {Filip}}, \
  and\ \bibinfo {author} {\bibfnamefont {J.}~\bibnamefont {Goold}},\ }\href
  {http://dx.doi.org/10.1038/s41534-020-0256-6} {\bibfield  {journal} {\bibinfo
   {journal} {npj Quantum Information}\ }\textbf {\bibinfo {volume} {6}}
  (\bibinfo {year} {2020})}\BibitemShut {NoStop}%
\bibitem [{\citenamefont {Fux}\ \emph {et~al.}(2022)\citenamefont {Fux},
  \citenamefont {Kilda}, \citenamefont {Lovett},\ and\ \citenamefont
  {Keeling}}]{Fux_2022}%
  \BibitemOpen
  \bibfield  {author} {\bibinfo {author} {\bibfnamefont {G.~E.}\ \bibnamefont
  {Fux}}, \bibinfo {author} {\bibfnamefont {D.}~\bibnamefont {Kilda}}, \bibinfo
  {author} {\bibfnamefont {B.~W.}\ \bibnamefont {Lovett}}, \ and\ \bibinfo
  {author} {\bibfnamefont {J.}~\bibnamefont {Keeling}},\ }\href {\doibase
  10.48550/ARXIV.2201.05529} {\  (\bibinfo {year} {2022}),\
  10.48550/ARXIV.2201.05529}\BibitemShut {NoStop}%
\bibitem [{\citenamefont {Purkayastha}\ \emph {et~al.}(2021)\citenamefont
  {Purkayastha}, \citenamefont {Guarnieri}, \citenamefont {Campbell},
  \citenamefont {Prior},\ and\ \citenamefont {Goold}}]{PReB_2021}%
  \BibitemOpen
  \bibfield  {author} {\bibinfo {author} {\bibfnamefont {A.}~\bibnamefont
  {Purkayastha}}, \bibinfo {author} {\bibfnamefont {G.}~\bibnamefont
  {Guarnieri}}, \bibinfo {author} {\bibfnamefont {S.}~\bibnamefont {Campbell}},
  \bibinfo {author} {\bibfnamefont {J.}~\bibnamefont {Prior}}, \ and\ \bibinfo
  {author} {\bibfnamefont {J.}~\bibnamefont {Goold}},\ }\href {\doibase
  10.1103/PhysRevB.104.045417} {\bibfield  {journal} {\bibinfo  {journal}
  {Phys. Rev. B}\ }\textbf {\bibinfo {volume} {104}},\ \bibinfo {pages}
  {045417} (\bibinfo {year} {2021})}\BibitemShut {NoStop}%
\bibitem [{\citenamefont {Lacerda}\ \emph {et~al.}(2022)\citenamefont
  {Lacerda}, \citenamefont {Purkayastha}, \citenamefont {Kewming},
  \citenamefont {Landi},\ and\ \citenamefont {Goold}}]{Lacerda_2022}%
  \BibitemOpen
  \bibfield  {author} {\bibinfo {author} {\bibfnamefont {A.~M.}\ \bibnamefont
  {Lacerda}}, \bibinfo {author} {\bibfnamefont {A.}~\bibnamefont
  {Purkayastha}}, \bibinfo {author} {\bibfnamefont {M.}~\bibnamefont
  {Kewming}}, \bibinfo {author} {\bibfnamefont {G.~T.}\ \bibnamefont {Landi}},
  \ and\ \bibinfo {author} {\bibfnamefont {J.}~\bibnamefont {Goold}},\ }\href
  {\doibase 10.48550/ARXIV.2206.01090} {\  (\bibinfo {year} {2022}),\
  10.48550/ARXIV.2206.01090}\BibitemShut {NoStop}%
\bibitem [{\citenamefont {Majumdar}\ \emph {et~al.}(2020)\citenamefont
  {Majumdar}, \citenamefont {Hall},\ and\ \citenamefont
  {Ahmadi}}]{SDP_scalability_2020}%
  \BibitemOpen
  \bibfield  {author} {\bibinfo {author} {\bibfnamefont {A.}~\bibnamefont
  {Majumdar}}, \bibinfo {author} {\bibfnamefont {G.}~\bibnamefont {Hall}}, \
  and\ \bibinfo {author} {\bibfnamefont {A.~A.}\ \bibnamefont {Ahmadi}},\
  }\href {\doibase 10.1146/annurev-control-091819-074326} {\bibfield  {journal}
  {\bibinfo  {journal} {Annual Review of Control, Robotics, and Autonomous
  Systems}\ }\textbf {\bibinfo {volume} {3}},\ \bibinfo {pages} {331} (\bibinfo
  {year} {2020})}\BibitemShut {NoStop}%
\bibitem [{\citenamefont {Guarnieri}\ \emph {et~al.}(2018)\citenamefont
  {Guarnieri}, \citenamefont {Kol\'a\ifmmode~\check{r}\else \v{r}\fi{}},\ and\
  \citenamefont {Filip}}]{Guarnieri_2018}%
  \BibitemOpen
  \bibfield  {author} {\bibinfo {author} {\bibfnamefont {G.}~\bibnamefont
  {Guarnieri}}, \bibinfo {author} {\bibfnamefont {M.}~\bibnamefont
  {Kol\'a\ifmmode~\check{r}\else \v{r}\fi{}}}, \ and\ \bibinfo {author}
  {\bibfnamefont {R.}~\bibnamefont {Filip}},\ }\href {\doibase
  10.1103/PhysRevLett.121.070401} {\bibfield  {journal} {\bibinfo  {journal}
  {Phys. Rev. Lett.}\ }\textbf {\bibinfo {volume} {121}},\ \bibinfo {pages}
  {070401} (\bibinfo {year} {2018})}\BibitemShut {NoStop}%
\bibitem [{\citenamefont {Trushechkin}\ \emph {et~al.}(2022)\citenamefont
  {Trushechkin}, \citenamefont {Merkli}, \citenamefont {Cresser},\ and\
  \citenamefont {Anders}}]{Trushechkin_2022}%
  \BibitemOpen
  \bibfield  {author} {\bibinfo {author} {\bibfnamefont {A.~S.}\ \bibnamefont
  {Trushechkin}}, \bibinfo {author} {\bibfnamefont {M.}~\bibnamefont {Merkli}},
  \bibinfo {author} {\bibfnamefont {J.~D.}\ \bibnamefont {Cresser}}, \ and\
  \bibinfo {author} {\bibfnamefont {J.}~\bibnamefont {Anders}},\ }\href
  {\doibase 10.1116/5.0073853} {\bibfield  {journal} {\bibinfo  {journal} {AVS
  Quantum Science}\ }\textbf {\bibinfo {volume} {4}},\ \bibinfo {pages}
  {012301} (\bibinfo {year} {2022})}\BibitemShut {NoStop}%
\bibitem [{\citenamefont {Cresser}\ and\ \citenamefont
  {Anders}(2021)}]{Cresser_2021}%
  \BibitemOpen
  \bibfield  {author} {\bibinfo {author} {\bibfnamefont {J.~D.}\ \bibnamefont
  {Cresser}}\ and\ \bibinfo {author} {\bibfnamefont {J.}~\bibnamefont
  {Anders}},\ }\href {\doibase 10.1103/PhysRevLett.127.250601} {\bibfield
  {journal} {\bibinfo  {journal} {Phys. Rev. Lett.}\ }\textbf {\bibinfo
  {volume} {127}},\ \bibinfo {pages} {250601} (\bibinfo {year}
  {2021})}\BibitemShut {NoStop}%
\bibitem [{\citenamefont {Streltsov}\ \emph {et~al.}(2017)\citenamefont
  {Streltsov}, \citenamefont {Adesso},\ and\ \citenamefont
  {Plenio}}]{Streltsov_2017}%
  \BibitemOpen
  \bibfield  {author} {\bibinfo {author} {\bibfnamefont {A.}~\bibnamefont
  {Streltsov}}, \bibinfo {author} {\bibfnamefont {G.}~\bibnamefont {Adesso}}, \
  and\ \bibinfo {author} {\bibfnamefont {M.~B.}\ \bibnamefont {Plenio}},\
  }\href {\doibase 10.1103/RevModPhys.89.041003} {\bibfield  {journal}
  {\bibinfo  {journal} {Rev. Mod. Phys.}\ }\textbf {\bibinfo {volume} {89}},\
  \bibinfo {pages} {041003} (\bibinfo {year} {2017})}\BibitemShut {NoStop}%
\bibitem [{\citenamefont {Francica}\ \emph {et~al.}(2019)\citenamefont
  {Francica}, \citenamefont {Goold},\ and\ \citenamefont
  {Plastina}}]{Francica_2019}%
  \BibitemOpen
  \bibfield  {author} {\bibinfo {author} {\bibfnamefont {G.}~\bibnamefont
  {Francica}}, \bibinfo {author} {\bibfnamefont {J.}~\bibnamefont {Goold}}, \
  and\ \bibinfo {author} {\bibfnamefont {F.}~\bibnamefont {Plastina}},\ }\href
  {\doibase 10.1103/PhysRevE.99.042105} {\bibfield  {journal} {\bibinfo
  {journal} {Phys. Rev. E}\ }\textbf {\bibinfo {volume} {99}},\ \bibinfo
  {pages} {042105} (\bibinfo {year} {2019})}\BibitemShut {NoStop}%
\bibitem [{\citenamefont {Santos}\ \emph {et~al.}(2019)\citenamefont {Santos},
  \citenamefont {Céleri}, \citenamefont {Landi},\ and\ \citenamefont
  {Paternostro}}]{Santos_2019}%
  \BibitemOpen
  \bibfield  {author} {\bibinfo {author} {\bibfnamefont {J.~P.}\ \bibnamefont
  {Santos}}, \bibinfo {author} {\bibfnamefont {L.~C.}\ \bibnamefont {Céleri}},
  \bibinfo {author} {\bibfnamefont {G.~T.}\ \bibnamefont {Landi}}, \ and\
  \bibinfo {author} {\bibfnamefont {M.}~\bibnamefont {Paternostro}},\ }\href
  {http://dx.doi.org/10.1038/s41534-019-0138-y} {\bibfield  {journal} {\bibinfo
   {journal} {npj Quantum Information}\ }\textbf {\bibinfo {volume} {5}}
  (\bibinfo {year} {2019})}\BibitemShut {NoStop}%
\bibitem [{\citenamefont {Narasimhachar}\ and\ \citenamefont
  {Gour}(2015)}]{Narasimhachar_2015}%
  \BibitemOpen
  \bibfield  {author} {\bibinfo {author} {\bibfnamefont {V.}~\bibnamefont
  {Narasimhachar}}\ and\ \bibinfo {author} {\bibfnamefont {G.}~\bibnamefont
  {Gour}},\ }\href {http://dx.doi.org/10.1038/ncomms8689} {\bibfield  {journal}
  {\bibinfo  {journal} {Nature Communications}\ }\textbf {\bibinfo {volume}
  {6}} (\bibinfo {year} {2015})}\BibitemShut {NoStop}%
\bibitem [{\citenamefont {Tupkary}(2022)}]{devashishcodegithub}%
  \BibitemOpen
  \bibfield  {author} {\bibinfo {author} {\bibfnamefont {D.}~\bibnamefont
  {Tupkary}},\ }\href@noop {} {}\bibinfo {howpublished}
  {\url{https://github.com/dtupkary/SearchingLindbladians}} (\bibinfo {year}
  {2022})\BibitemShut {NoStop}%
\bibitem [{\citenamefont {Johansson}\ \emph {et~al.}(2012)\citenamefont
  {Johansson}, \citenamefont {Nation},\ and\ \citenamefont {Nori}}]{qutip_1}%
  \BibitemOpen
  \bibfield  {author} {\bibinfo {author} {\bibfnamefont {J.}~\bibnamefont
  {Johansson}}, \bibinfo {author} {\bibfnamefont {P.}~\bibnamefont {Nation}}, \
  and\ \bibinfo {author} {\bibfnamefont {F.}~\bibnamefont {Nori}},\ }\href
  {https://www.sciencedirect.com/science/article/pii/S0010465512000835}
  {\bibfield  {journal} {\bibinfo  {journal} {Computer Physics Communications}\
  }\textbf {\bibinfo {volume} {183}},\ \bibinfo {pages} {1760} (\bibinfo {year}
  {2012})}\BibitemShut {NoStop}%
\bibitem [{\citenamefont {Johansson}\ \emph {et~al.}(2013)\citenamefont
  {Johansson}, \citenamefont {Nation},\ and\ \citenamefont {Nori}}]{qutip_2}%
  \BibitemOpen
  \bibfield  {author} {\bibinfo {author} {\bibfnamefont {J.}~\bibnamefont
  {Johansson}}, \bibinfo {author} {\bibfnamefont {P.}~\bibnamefont {Nation}}, \
  and\ \bibinfo {author} {\bibfnamefont {F.}~\bibnamefont {Nori}},\ }\href
  {\doibase https://doi.org/10.1016/j.cpc.2012.11.019} {\bibfield  {journal}
  {\bibinfo  {journal} {Computer Physics Communications}\ }\textbf {\bibinfo
  {volume} {184}},\ \bibinfo {pages} {1234} (\bibinfo {year}
  {2013})}\BibitemShut {NoStop}%
\end{thebibliography}%

\end{document}